\documentclass[%
 a4paper,
 prl,
 preprint,%
 reprint,%
 superscriptaddress,
 nofootinbib
]{revtex4-1}

\usepackage[utf8]{inputenc}
\usepackage[T1]{fontenc}
\usepackage{mathptmx}

\usepackage{amsmath,amssymb}
\usepackage{lipsum,soul}
\usepackage{bm}
\usepackage{graphicx}
\usepackage{graphics}
\usepackage{amsmath}
\usepackage{array}
\usepackage[caption=false]{subfig}
\usepackage{xcolor}
\usepackage{subfig}
\usepackage{hyperref}
\usepackage[capitalise]{cleveref}
\usepackage{textcomp} % regestered trademark
\usepackage{epstopdf}
\usepackage{cases}
\usepackage{amssymb}
\usepackage{multirow}
\usepackage{siunitx}
\usepackage{cases}
\usepackage{tabularx}
\usepackage{nomencl}
\usepackage{verbatim}
\usepackage{tikz}
\usepackage{comment}
\usepackage{lineno,hyperref}
\begin{document}

\title{Perturbation theory of thermal rectification}
\author{Chuang Zhang}
\email{zhangcmzt@hust.edu.cn}
\affiliation{%
 State Key Laboratory of Coal Combustion, School of Energy and Power Engineering, Huazhong University of Science and Technology, Wuhan 430074, China}%
\author{Meng An}
\email{anmeng@sust.edu.cn}
\affiliation{%
College of Mechanical and Electrical Engineering, Shaanxi University of Science and Technology, 6 Xuefuzhong Road, Weiyangdaxueyuan, Xi’an 710021, China}
\author{Zhaoli Guo}%
\email{zlguo@hust.edu.cn}
\affiliation{%
 State Key Laboratory of Coal Combustion, School of Energy and Power Engineering, Huazhong University of Science and Technology, Wuhan 430074, China}%
\author{Songze Chen}
\email{Corresponding author: jacksongze@hust.edu.cn}
\affiliation{%
 State Key Laboratory of Coal Combustion, School of Energy and Power Engineering, Huazhong University of Science and Technology, Wuhan 430074, China}%
\date{\today}

\begin{abstract}

Thermal rectification which is a diode-like behavior of heat flux has been studied over a long time.
However, a universal and systematic physical description is still lacking.
{\color{black}{In this letter, a perturbation theory of thermal rectification is developed, which provides an analytical formula of the thermal rectification ratio.
It reveals the linear relationship between the thermal rectification ratio and temperature difference.
Furthermore, the size-dependence of the thermal rectification relies on the specific form of the thermal conductivity.
In addition, several experimental and numerical observations in previous literatures are well explained.}}
This theory can be applicable to any system in which a differentiable effective thermal conductivity can be derived, and is helpful to unveil general principle for thermal rectification.

\end{abstract}

\maketitle

Thermal rectification~\cite{starr1936,terraneo2002a,li_thermal_2004,RevModPhysLibaowen,ZHANG2020} is a diode-like behavior of heat flux.
It plays an important role on thermal management and engineering in solid-state devices or materials.

In the past decades, much attention has been paid to identify the underlying physics and to enhance the thermal rectification ratio~\cite{terraneo2002a,li_thermal_2004,chang_solid-state_2006,eckmann2006,RevModPhysLibaowen,yang2012a,liu2019a,PhysRevLett.124.075903}.
Many studies show that the thermal rectification between two-segment bulk materials can be realized by selecting materials with suitable properties or different temperature dependent thermal conductivities~\cite{kobayashi2009,dames2009,peyrard2006,PhysRevE.98.042131}.
%A simple algebraic expression of the thermal rectification is also given in the common case of low thermal bias and thermal conductivities with power-law temperature dependencies~\cite{dames2009}.
Meanwhile, a general conclusion was made, i.e., thermal rectification is impossible if the thermal conductivity $\kappa (x,T)$ is separable~\cite{go2010}.
For asymmetric nanoscale materials or systems, many accessible strategies have been proposed to realize the thermal rectification, such as asymmetric shape~\cite{yang_thermal_2009,hu2009,wang2019,yang_carbon_2008,wang2017,wang2014}, mass graded~\cite{yang2007,wu2008}, porous or inhomogeneous materials~\cite{criado-sancho2013,wu_thermal_2007,hu_series_2017}.
Some physical mechanisms~\cite{RevModPhysLibaowen,liu2019a,wang2014} were identified to explain the thermal rectification, such as the different phonon spectra overlap by switching the direction of the temperature gradient~\cite{yang_thermal_2009,liu2019a,roberts2011a}, asymmetric phonon ballistic or edge scattering~\cite{ouyang2010,wang2017,ma2018}, nonseparable dependence of the thermal conductivity $\kappa (T,x)$ on temperature $T$ and spatial position $x$~\cite{wang2017,wang2014,zhu2014}.
In addition, some theoretical work was also made based on some simplified microscopic models to identify the essential conditions~\cite{pereira2006,pereira2017,wehmeyer2017} for thermal rectification, such as the unusual temperature-dependent potential~\cite{terraneo2002a,li_thermal_2004}, nonuniform or graded mass distribution~\cite{pereira2010,pereira2011}.

However, to the best of our knowledge, previous studies of thermal rectification are based on specific physical problems or theoretical models.
In other words, a universal and systematic physical description of the thermal rectification is still lacking.

In this letter, a perturbation theory of thermal rectification is established and three dimensionless parameters are identified for the first time.
The theory is not limited by system length or material properties and presents a clear physical picture of  thermal rectification based on explicit physical assumptions and rigorous theoretical derivations.
Several experimental and numerical observations in literatures are well explained based on this theory.

Let's introduce the main assumptions of this study.
{\color{black}{Given a thermal conduction system satisfying the local thermal equilibrium~\cite{Kubo1991statistical,Wang_2019LTE}, the local temperature $T(x)$ or other local physical properties can be defined well.
Suppose that an effective thermal conductivity $\kappa_e$ can be identified so that the Fourier law is satisfied formally,
\begin{align}
q = -\kappa_e(W,T, L) \frac{dT}{dx}, \quad x \in [x_0 -L/2, x_0 +L/2], \label{eq:conductivity}
\end{align}
where $q$ is the heat flux, $x$ is the spatial position. $x_0$, $L$ and $W$ are the central position, total system length and the local physical quantity in this system, respectively.
The physical quantities except for the temperature that influence the effective thermal conductivity are grouped into two categories.
The first category stands for the local physical quantity varying with position, i.e., $W=W(x)$, for instance, the characteristic length in other directions~\cite{wu2008,hu2009,wang2014,yang_carbon_2008,yang_thermal_2009,sawaki2011}, mass~\cite{chang_solid-state_2006,yang2007}, porosity~\cite{criado-sancho2013}.
The other one represents a kind of global physical quantity like the system length $L$.}}
The expression, $\kappa_e(W,T, L)$, also requires assumptions that the representative variable $W$ is independent of the temperature, and neglecting the dependence of higher order derivative of temperature.
{\color{black}{In this study, we assume that the effective thermal conductivity changes smoothly and slightly in the whole system.
In addition, the temperature gradient inside the thermal system should be non-zero and finite.}}
Note that Eq.~\eqref{eq:conductivity} is valid for any thermal conduction systems regardless of system length and material properties if the local thermal equilibrium~\cite{Kubo1991statistical,Wang_2019LTE} is satisfied and an effective thermal conductivity could be identified.

Next, the theoretical derivations of our theory is introduced.
Given a (quasi) one-dimensional thermal conduction system inside $[x_0 -L/2, x_0 +L/2 ]$, two temperatures ($T_0-\Delta T/2, T_0+\Delta T/2$) are imposed at the two boundaries, where $T_0$ and $\Delta T$ are the average temperature and temperature difference, respectively.
At steady state, based on energy conservation, the heat conduction satisfies
\begin{align}
\frac{\partial q}{\partial x} = 0. \label{eq:laplacian}
\end{align}
According to the assumptions we made, the spatial distribution of $W(x)$ is fixed in the system.
Therefore, the effective thermal conductivity in Eq.~\eqref{eq:conductivity} can be formally taken as a function of the position $x$, temperature $T$ and system length $L$,
\begin{align}
\kappa_e = \kappa_e( W(x),T, L) = \kappa_e(x,T, L). \label{eq:conductivity-xT}
\end{align}
%{\color{black}{Note that as the system length $L$ of thermal system is comparable to the phonon mean free path, $\kappa_e(L)$ may change as the system length $L$ changes~\cite{ZHANG2020,bae2013ballistic,xu_length-dependent_2014,RevModPhysLibaowen,yang2012a,wang2017,wang2014}. }}

{\color{black}{Given that the effective thermal conductivity $\kappa_e(x,T, L)$ changes smoothly and slightly in the whole system,}} it can be approximated by the Taylor expansion~\cite{rudin1964principles} based on $(x_0,T_0)$, i.e.,
\begin{align}
%&\kappa_e = \kappa_0 + \left.\ \frac{\partial \kappa_e}{\partial x} \right|_{(x,T)=(x_0,T_0)}  (x-x_0)
%+  \left.\ \frac{\partial \kappa_e}{\partial T} \right|_{(x,T)=(x_0,T_0)} (T-T_0)  \notag \\
%&+ \left.\ \frac{\partial^2 \kappa_e}{ 2 \partial x^2} \right|_{(x,T)=(x_0,T_0)} (x-x_0)^2 +  \left.\  \frac{\partial^2 \kappa_e}{2 \partial T^2} \right|_{(x,T)=(x_0,T_0)} (T-T_0)^2 \notag \\
%& + \left.\ \frac{\partial^2 \kappa_e}{\partial x\partial T} \right|_{(x,T)=(x_0,T_0)} (x-x_0)(T-T_0), \label{eq:taylorsm}
\kappa_e =& \kappa_0 + \frac{\partial \kappa_e}{\partial x} (x-x_0)  + \frac{\partial \kappa_e}{\partial T} (T-T_0) + \frac{1}{2}\frac{\partial^2 \kappa_e}{\partial x^2} (x-x_0)^2 \notag \\
&+ \frac{1}{2}\frac{\partial^2 \kappa_e}{\partial T^2} (T-T_0)^2+ \frac{\partial^2 \kappa_e}{\partial x\partial T} (x-x_0)(T-T_0). \label{eq:taylorsm}
\end{align}
{\color{black}{Note that all the partial derivatives of the effective thermal conductivity in this work are calculated at $(x_0,T_0)$ and $\kappa_0 =\kappa_e (x_0,T_0,L) \neq 0$~\cite{rudin1964principles}. }}
The higher order terms are assumed to be negligible.
Choosing $\kappa_0, L, \Delta T$ as reference variables to normalize the equations (Eqs.~(\ref{eq:conductivity},\ref{eq:laplacian},\ref{eq:taylorsm}),
we can get the dimensionless equations as follows,
\begin{align}
&\frac{\partial q^*}{\partial x^*} = 0, \quad q^* = -\kappa_e^*\frac{d T^*}{d x^*},
\end{align}
where,
\begin{align}
& x^* =\frac{ x-x_0}{L }, \quad \ T^* =\frac{ T-T_0 }{ \Delta T }, \quad q^* =\frac{q L }{\kappa_0 \Delta T},\\
&\kappa_e^* = 1+\alpha_x x^* + \alpha_T T^* + \alpha_{xT} x^*T^* + \alpha_{x^2}x^{*2}+\alpha_{T^2}T^{*2},
\label{eq:dimensionlesskappa}
\end{align}
and the associated dimensionless parameters are
\begin{align}
&\alpha_x = \frac{L}{\kappa_0}\frac{\partial \kappa_e}{\partial x}, \quad
\alpha_T = \frac{\Delta T}{\kappa_0}\frac{\partial \kappa_e}{\partial T}, \quad\alpha_{xT} = \frac{L\Delta T}{\kappa_0}\frac{\partial^2 \kappa_e}{\partial x\partial T}, \label{eq:alphathree} \\
&\alpha_{x^2} = \frac{L^2}{2\kappa_0}\frac{\partial^2 \kappa_e}{\partial x^2}, \quad
\alpha_{T^2} = \frac{\Delta T^2}{2\kappa_0}\frac{\partial^2 \kappa_e}{\partial T^2}.
%\alpha_x &= \frac{L}{\kappa_0} \left.\ \frac{\partial \kappa_e}{\partial x} \right|_{(x,T)=(x_0,T_0)}, \label{eq:alphax} \\
%\alpha_T &= \frac{\Delta T}{\kappa_0} \left.\ \frac{\partial \kappa_e}{\partial T} \right|_{(x,T)=(x_0,T_0)}, \label{eq:alphaT} \\
%\alpha_{xT} &= \frac{L\Delta T}{\kappa_0} \left.\ \frac{\partial^2 \kappa_e}{\partial x\partial T} \right|_{(x,T)=(x_0,T_0)}, \label{eq:alphaxT} \\
%\alpha_{x^2} &= \frac{L^2}{2\kappa_0} \left.\ \frac{\partial^2 \kappa_e}{\partial x^2} \right|_{(x,T)=(x_0,T_0)} , \label{eq:alphaxx} \\
%\alpha_{T^2} &= \frac{\Delta T^2}{2\kappa_0} \left.\ \frac{\partial^2 \kappa_e}{\partial T^2} \right|_{(x,T)=(x_0,T_0)}. \label{eq:alphaTT}
\end{align}
Then, the dimensionless equations are solved with two sets of boundary conditions respectively,
%\mbox{Eqs}.~(\ref{eq:conductivity},\ref{eq:conductivity-xT},\ref{eq:taylorsm}) and
\begin{align}
\text{forward (`+')}&:\quad T^*(-\frac{1}{2}) = -\frac{1}{2}, & T^*(\frac{1}{2}) = \frac{1}{2} ,\label{eq:forwardBc} \\
%\text{forward ('+')}&:\ T(-1/2) = -1/2,\quad  T(1/2) = 1/2,\label{eq:forwardBc} \\
\text{backward (`-')}&:\quad T^*(-\frac{1}{2}) = \frac{1}{2}, & T^*(\frac{1}{2}) = -\frac{1}{2} . \label{eq:backwardBc}
%\text{backward ('-')}&:\ T(-1/2) = 1/2, \quad\ \ T(1/2) = -1/2. \label{eq:backwardBc}
\end{align}
%\text{backward ('-')}&:\quad T(x_1) = T_2, \quad T(x_2) = T_1 . \label{eq:backwardBc}
%\end{align}
%as shown in~\cref{rectification}(b).
%\begin{align}
%\beta_T &= a \Delta T /\kappa_0, &\quad \beta_W = b L /\kappa_0. \label{eq:relativeChanges}
%\end{align}

{\color{black}{With all these assumptions, the thermal rectification ratio of the whole thermal system can be deduced from \mbox{Eqs}.~(\ref{eq:conductivity},\ref{eq:laplacian},\ref{eq:conductivity-xT},\ref{eq:taylorsm}) based on perturbation method~\cite{bender2013advanced} or also direct Taylor expansion~\cite{rudin1964principles} (detailed derivations and numerical validations can be found in Sec. I-V in Supplemental Material),
\begin{align}
q^*_{+}&\approx -1 + \frac{1}{12} (\alpha_x \alpha_T - \alpha_{xT} -\alpha_{x^2 }-\alpha_{T^2} -\frac{1}{10} \alpha_T^2 \alpha_{x^2} ), \label{eq:qzheng} \\
q^*_{-}&\approx 1 + \frac{1}{12} (\alpha_x \alpha_T - \alpha_{xT} + \alpha_{x^2 } + \alpha_{T^2}  + \frac{1}{10} \alpha_T^2 \alpha_{x^2}  ), \label{eq:qfu} \\
\beta &= \frac{q_{+} +q_{-} }{q_{+} - q_{-} } \approx  \frac{1}{12}(\alpha_{xT} -\alpha_x \alpha_T) \notag \\
 &= \frac{L\Delta T}{12} \left.\ \left( \frac{1}{\kappa_0 }\frac{\partial^2 \kappa_e}{\partial x \partial T}-\frac{1}{\kappa_0^2}\frac{\partial \kappa_e}{\partial x}\frac{\partial \kappa_e}{\partial T} \right)  \right|_{(x,T)=(x_0,T_0)},    \label{eq:CVMsq15}
\end{align}
where $q_+$ ($q_+^*$) is the forward (dimensionless) heat flux, $q_-$ ($q_{-}^*$) is the backward (dimensionless) heat flux, and $\beta$ is the thermal rectification ratio predicted by perturbation theory, which is a function of the total system length $L$, temperature difference $\Delta T$ and the effective thermal conductivity $\kappa_e (x, T, L)$.
Based on theoretical constraints and the numerical validations, Eq.~\eqref{eq:CVMsq15} is valid as the effective thermal conductivity $\kappa_e(x,T,L)$ changes smoothly and slightly in the whole system or these dimensionless parameters are small.
Note that small $\alpha_x$ ($\alpha_T$) are not equivalent to small $L$ ($\Delta T$).
}}

{\color{black}{Equation~\eqref{eq:CVMsq15} is the central result of the present study.}}
This theoretical formula reveals the essential condition to realize the  thermal rectification, i.e.,
\begin{align}
(\alpha_x\alpha_T-\alpha_{xT}) L \Delta T \neq 0,
\label{eq:formularec}
\end{align}
which is a subset of the nonseparable condition~\cite{go2010}.
And three dimensionless parameters in Eq.~\eqref{eq:CVMsq15} denote the relative change of the effective thermal conductivity throughout the whole system due to the temperature change and the heterogeneity of the other physical properties.
In addition, Eq.~\eqref{eq:CVMsq15} gives rigorous theoretical supports to the linear relationship between the thermal rectification ratio and temperature difference $\Delta T$, which has been widely observed in previous literature~\cite{dames2009,sawaki2011,zhu2014,wang2019,wang2014}.

{\color{black}{Furthermore, the size-dependent thermal rectification is associated with the size-dependent thermal conductivity~\cite{ZHANG2020}.
According to Eq.~\eqref{eq:CVMsq15}, the thermal rectification ratio is proportional to the system length provided that the terms in the brackets are independent of system length, such as $x_0$ and the effective thermal conductivity $\kappa_e$ are independent of the system length $L$.
Otherwise, the thermal rectification ratio depends on the specific formulas of thermal conductivity or materials properties.

Actually, the size-dependent thermal rectification phenomena have been observed in previous studies~\cite{wang2014,wang2017,wang2019,yang_experimental_2020,ZHANG2020}.
For example, as the system length increases, the thermal rectification ratio increases proportionally with system length in 2D Lorentz gas model~\cite{wang2019}, but decreases gradually in trapezoid graphene nanoribbons~\cite{wang2014,wang2017}.
However, the exploration of underlying mechanisms is still lacking, especially for its relationship with the size-dependent thermal conductivity as mentioned in a latest review~\cite{ZHANG2020}.
Based on the present study, the different size-dependent thermal rectification behaviors in 2D Lorentz gas model~\cite{wang2019} and trapezoid graphene nanoribbons~\cite{wang2014} are related to different size-dependent thermal conductivity.
In 2D Lorentz gas model, the thermal conductivity almost keeps a constant as the length of the rectangular space changes~\cite{wang2019} (see Sec. VI in Supplemental Material).
But the thermal conductivity of graphene nanoribbons increases with system length as the characteristic length is comparable to phonon mean free path~\cite{bae2013ballistic,xu_length-dependent_2014}.
%In the future, the present theory can be taken as a guideline for experimental studies on the size dependent thermal rectification~\cite{wang2017,ZHANG2020,yang_experimental_2020}.
}}

\begin{figure}
 \centering
 \includegraphics[scale=0.33,viewport=60 520 850 1380,clip=true]{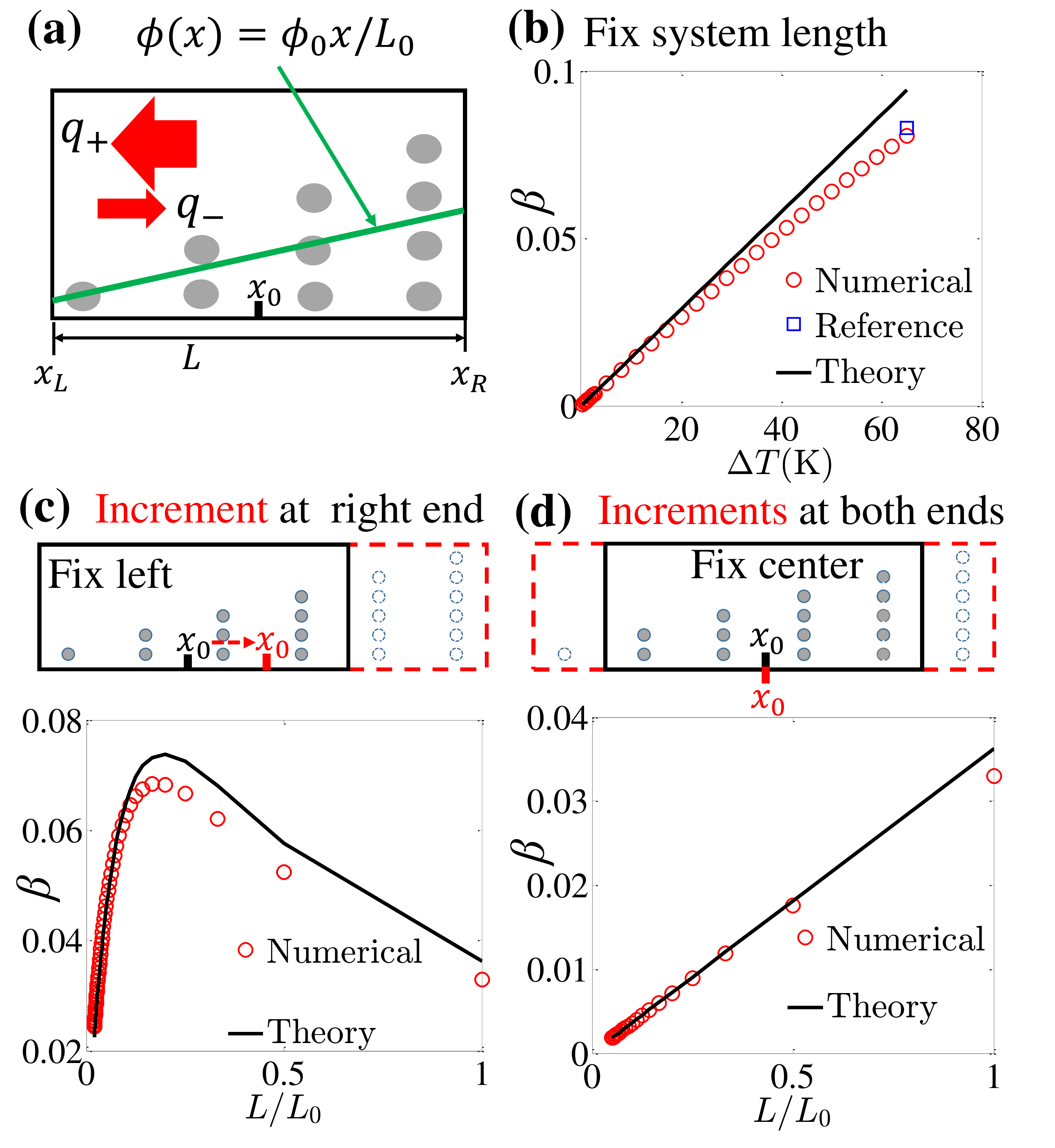}
 \caption{ Thermal rectification of inhomogeneous porous silicon device~\cite{criado-sancho2013}. Black line: Analytical solutions (Eq.~\eqref{eq:CVMsq15}). Red circle: numerical simulations. Blue square: the data obtained in reference~\cite{criado-sancho2013}. (a) A sketch for graded inhomogeneous porous silicon device. The spatial porosity distributions satisfy $\phi (x)=\phi_0 x/L_0$, $\phi_0 =0.1$, $L_0 =100$ mm is the reference length, $x$ is the spatial position. $L$ is the system length, $x_0$ is the central position. $x_L=x_0 -L/2 $ and $x_R=x_0 +L/2$ are the position of left and right boundaries, respectively. (b) The linear relationship between the thermal rectification ratio $\beta$ and temperature difference $\Delta T$, where $x_0 =L/2=L_0/2$, $T_0=92.5$ K. (c)(d) Two different ways to change system length $L$ (see FIG. S6 more clearly in Supplemental Material). And the associated results of the distributions of the thermal rectification ratio with different system length $L$, where $T_0=92.5$ K, $|\Delta T|=25$ K. (c) First, the left end of the system is fixed, and the other end changes, so that $x_0=L/2$ changes with system length. (d) Second, The central position $x_0=L_0 /2$ is anchored to a fixed point and the system length is changed along both two ends symmetrically, so that $x_0$ is independent of system length. }
 \label{porousND}
\end{figure}

To better show the relationship between the thermal rectification and system length $L$ (or temperature difference $\Delta T$), the thermal rectification coefficient of a quasi-one dimensional graded inhomogeneous porous silicon device is studied~\cite{criado-sancho2013}.
As shown in~\cref{porousND}(a), a lot of nanopores are distributed in a bulk silicon bar.
The spatial porosity distributions satisfy $\phi (x)=\phi_0 x/L_0$, where $\phi$ is the porosity (ratio of the volume of the pores divided by the total volume), $x$ is the spatial position, $\phi_0= 0.10$, $L_0=100$ mm is the reference length.
$L$ is the system length, $x_0$ is the central position.
$x_L=x_0 -L/2 $ and $x_R=x_0 +L/2$ are the position of left and right boundaries, respectively.
Two temperatures are imposed at the left and right boundaries, i.e., $T_L=T_0 +\Delta T/2$, $T_R=T_0 - \Delta T/2$.
According to previous studies~\cite{hydrodynamicporeousAFX2010,criado-sancho2013}, an effective thermal conductivity $\kappa_e(x,T)=\kappa_e(\phi,\lambda/r ,T )$ can be identified as,
\begin{align}
\kappa_e(\phi,\lambda/r ,T )= \frac{ \kappa_{\text{bulk}} (1- \phi)^3    }{ 1+ \frac{9}{2} \phi (1- \phi)^3  (1+ \frac{3 \sqrt{\phi}}{\sqrt{2}} ) \frac{ (\lambda /r)^2 }{ 1+A'( \lambda/r)}},
\label{eq:kappaporoussi}
\end{align}
where $r=150$ nm is pore radius, $A'(\lambda/r)= 0.864+ 0.290 \times \exp( -1.25 r/ \lambda )$.
$\kappa_{\text{bulk}}$ and $\lambda$ are the thermal conductivity and phonon mean free path of bulk silicon (see Sec. VII in Supplemental Material).
It can be observed that the effective thermal conductivity (Eq.~\eqref{eq:kappaporoussi}) is independent of system length.

In this system, the linear relationship between the thermal rectification ratio and temperature difference can be easily observed.
Given $T_0=92.5$ K, $x_0= L/2 =L_0/2$, the thermal rectification ratio increases linearly with temperature difference $\Delta T$, as shown in~\cref{porousND}(b).
As $\Delta T=65$ K, the results predicted by present numerical simulations are in good agreement with the data obtained in reference~\cite{criado-sancho2013}.
Besides, the heat flux prefers to flow from high porosity to low porosity $(\beta > 0)$, which is consistent with those mentioned in previous studies~\cite{criado-sancho2013}.

The length-dependence thermal rectification is more complex.
Given $T_0=92.5$ K and $\Delta T=25$ K, we propose two different ways to change the system length, and investigate their influences to the thermal rectification.
The first way is to fix the left end of system and change the right end (\cref{porousND}(c)). In this case, $x_0 = L/2$, moves as the system length changes.
As shown in~\cref{porousND}(c), the thermal rectification ratio keeps linear relationship with $L$ only in a very narrow range.
That's because $x_0$ depending on $L$ leads to the change of the terms in the brackets of Eq.~\eqref{eq:CVMsq15}.
The second way is to anchor the central position $x_0$ to a fixed point and move left and right ends symmetrically (\cref{porousND}(d)), so that $x_0=L_0/2$ is independent of system length.
As shown in~\cref{porousND}(d), the linear relationship between the thermal rectification ratio and system length can be observed in a wide range.
Above results show that whether $x_0$ changes with $L$ affects the length-dependent thermal rectification phenomena a lot, which is consistent with our theory.
Besides, there is a linear relationship between the thermal rectification ratio and system length if both the effective thermal conductivity and $x_0$ are independent of system length.

\begin{figure}
 \centering
 \includegraphics[scale=0.28,viewport=20 230 950 1020,clip=true]{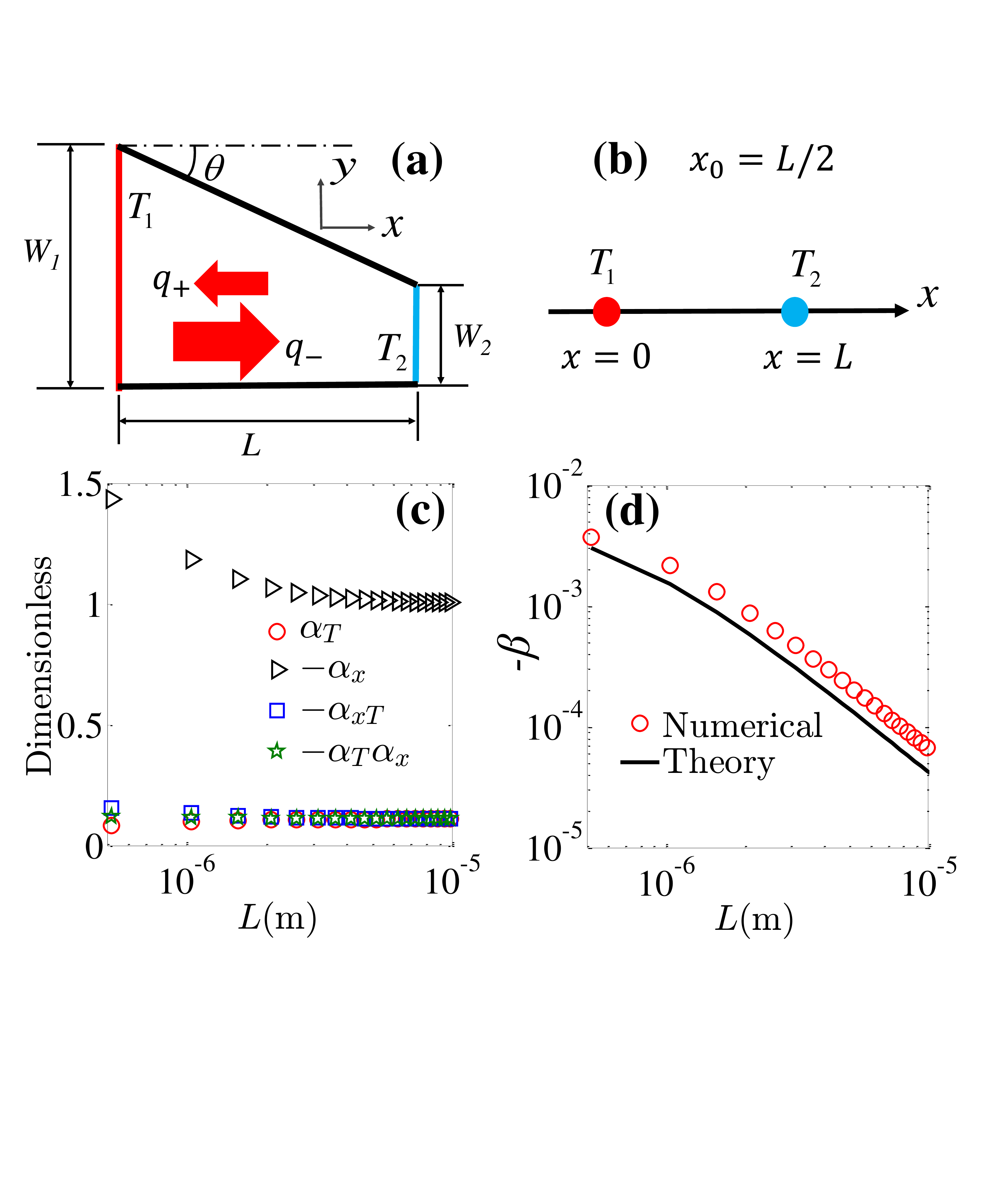}
\caption{(a) Geometrical definition of the length ($L$) and width ($W_1,W_2$) of the trapezoidal-graphene nanoribbons~\cite{wang2014}, where $x,~y$ are spatial coordinations, $T_{1}=T_0 +\Delta T/2$ and $T_2=T_0 -\Delta T/2$. (b) Reduced quasi-one dimensional heat conduction, where $x_0=L/2$ is the central position. $x=0$ and $x=L$ are the left and right boundaries, respectively. (c) The distributions of these dimensionless parameters $-\alpha_x$, $\alpha_T$, $-\alpha_{xT}$ and $-\alpha_x  \alpha_T$ as the system length increases, which are calculated directly from Eqs.~\eqref{eq:kappareduced} and~\eqref{eq:alphathree} for a given $(x_0,T_0)$ (see Sec. VIII in Supplemental Material). (d) Comparison of the thermal rectification ratio $\beta$ in trapezoid graphene between the numerical and analytical results (Eq.~\eqref{eq:CVMsq15}) based on Eq.~\eqref{eq:kappareduced}.
 }
 \label{GNRrectification}
\end{figure}

Different from inhomogeneous porous silicon materials~\cite{criado-sancho2013}, the thermal conductivity depends on system length in trapezoid suspended graphene~\cite{wang2014,wang2017} as characteristic length is comparable to phonon mean free path~\cite{bae2013ballistic,xu_length-dependent_2014}.
Considering a trapezoid suspended graphene, as shown in~\cref{GNRrectification}(a)~\cite{wang2014}, $T_{1}=T_0 +\Delta T/2$ and $T_2=T_0 -\Delta T/2$ are the temperatures at left and right boundaries.
The length of the geometry is $L$ and $\theta$ is the inclined angle.
The widthes of left and right boundaries are $W_1$ and $W_2$, respectively.
The two-dimensional heat conduction problem can be approximately reduced into one-dimensional problem (\cref{GNRrectification}(b)) and its heat flux along $x$ direction, i.e.,
\begin{align}
q  =  - \int \kappa(x,y,T) \frac{dT}{dx} dy= -\kappa_e  \frac{d \overline{T} }{dx},
\label{eq:fourier1DNC}
\end{align}
where
\begin{align}
\kappa_e \approx  W_3 (x)  \kappa, \quad W_3 (x)  = \frac{ W_2 -W_1 }{L} x +W_1, \label{eq:reducedW}
\end{align}
%\begin{align}
%q  =  - \int \kappa \frac{dT}{dx} dy,
%\label{eq:fourier2DNC}
%\end{align}
where $\kappa=\kappa(x,y,T)$ is the thermal conductivity of thermal system, $T=T(x,y)$ is the temperature. $\kappa_e$ and $\overline{T}$ are the reduced effective thermal conductivity and reduced temperature.
$W_3 (x)$ is the transverse length in $y$ direction, $x \in [0,L],~x_0=L/2$.

The thermal conductivity in graphene nanoribbons $(\kappa_{\text{GNR} } (L,T,W_g))$ is given by an experimental study~\cite{bae2013ballistic}, which is a function of temperature $T$, system length $L$ and width $W_g$ of nanoribbons, i.e.,
\begin{align}
%\frac{G_{\text{ball}} } {A} &= \left( \frac{1}{4.4 \times 10^5 T^{1.68}  } +\frac{1}{ 1.2 \times  10^{10} }   \right)^{-1}, \notag \\
\frac{1}{ \kappa_{1} (L,T) } &= \left( \frac{1}{4.4 \times 10^5 T^{1.68}  } +\frac{1}{ 1.2 \times  10^{10} }   \right) \left(\frac{1}{L} +\frac{1}{\pi \lambda /2 }     \right), \notag  \\
\kappa_{\text{GNR} } & = \left( \frac{1}{c_1}  \left(\frac{\Delta}{W_g } \right)^{m} +\frac{1}{ \kappa_{1} (L,T)  }   \right)^{-1},
\label{eq:conductivityncgraphene}
\end{align}
where $\Delta$ is the root-mean-square edge roughness, $\lambda$ is the phonon mean free path.
According to Eq.~\eqref{eq:conductivityncgraphene}, the reduced effective thermal conductivity $\kappa_e$ in Eq.~\eqref{eq:fourier1DNC} is
\begin{align}
\kappa_e (x , \overline{T}, L ) = \kappa_e (W_3 , \overline{T}, L ) \approx W_3 (x) \times \kappa_{\text{GNR} } (L, \overline{T} ,W_3).
\label{eq:kappareduced}
\end{align}

{\color{black}{Note that almost all parameters in Eqs.~\eqref{eq:kappareduced} and~\eqref{eq:conductivityncgraphene} will influence the thermal rectification.
Besides, it is very hard to identify an effective thermal conductivity $\kappa_e (x,\overline{T},L)$ with accurate parameters for the specific problems mentioned in previous studies~\cite{wang2014,wang2017} due to insufficient data, so that only a qualitative analysis of the size-dependent thermal rectification is made.}}
Based on previous experimental studies~\cite{bae2013ballistic,xu_length-dependent_2014}, here we set $\Delta =0.6$ nm, $c_1=0.04$ W/(mK), $T_0=300~\text{K}$, $m=1.8$, $ | \Delta T |=30~\text{K}$, $\lambda=240$ nm~\cite{xu_length-dependent_2014}.
In the smallest size, $W_1=450 $ nm, $W_2=150 $ nm, $L=520$nm, $\theta \approx \pi /6$ and both $L$ and $W_1$ are increased proportionally with fixed $\theta$.
The smallest size is selected by taking into account the validity of the thermal conductivity in graphene nanoribbon $\kappa_{ \text{GNR} }$~\cite{bae2013ballistic} (Eq.~\eqref{eq:conductivityncgraphene}).

Both numerical solutions and theoretical analysis are implemented based on Eqs.~(\ref{eq:conductivity},\ref{eq:laplacian},~\ref{eq:kappareduced}) and shown in~\cref{GNRrectification}(c)(d).
Although there are some deviations between the theoretical and numerical results, both of them show that the heat flux prefers to flow from the wider side to the narrow side ($\beta <0$,~\cref{GNRrectification}(d)).
In addition, as the system length increases, the thermal rectification ratio decreases.
Above two phenomena are consistent with those mentioned in previous studies~\cite{wang2017,wang2014} qualitatively.
Based on~\cref{GNRrectification}(c), it can be observed that $\alpha_x$ changes significantly with system length.
In other words, the length-dependent thermal rectification is greatly related to the size-dependent thermal conductivity~\cite{ZHANG2020,bae2013ballistic,xu_length-dependent_2014}.

Finally, some limitations and accessible extensions about the perturbation theory of thermal rectification (Eq.~\eqref{eq:CVMsq15}) are addressed.
{\color{black}{Firstly, Eq.~\eqref{eq:conductivity} or our theory is invalid if the local thermal equilibrium~\cite{Kubo1991statistical,Wang_2019LTE} is not satisfied or the local temperature, effective thermal conductivity cannot be well defined~\cite{Pandey17PEG,pandey_thermalization_2016}.
Secondly, Eq.~\eqref{eq:CVMsq15} is unapplicable if the effective thermal conductivity $\kappa_e (x,T, L)$ is discontinuous or singular/infinite~\cite{leitner_thermal_2013}, such as the thermal rectification with sharp interface~\cite{kobayashi2009,dames2009,peyrard2006}, phase-change materials~\cite{kobayashi2012,kang2018a}.
Thirdly, the present theory cannot describe the thermal rectification only caused by boundaries or heat bath~\cite{rectificationcontacts2020JAP,leitner_thermal_2013}.
Fourthly, for two-segment bulk materials~\cite{kobayashi2009,dames2009,peyrard2006,PhysRevE.98.042131}, although Eq.~\eqref{eq:CVMsq15} is unapplicable, Eqs.~\eqref{eq:qzheng}~\eqref{eq:qfu} and the perturbation method~\cite{bender2013advanced,rudin1964principles} can be directly applied to simultaneously solve the heat conduction equation for each segment so that analytical solutions of the thermal rectification ratio~\cite{dames2009,kobayashi2009} can also be obtained, which can be seen in Sec. IX in Supplemental Material.
}}

In conclusion, a systematic physical description of the thermal rectification is established through perturbation theory (Eq.~\eqref{eq:CVMsq15}).
{\color{black}{A physical relationship among the thermal rectification, system length, temperature difference and thermal conductivity is built.
It reveals the linear relationship between the thermal rectification ratio and the temperature difference.
Furthermore, the thermal rectification ratio is proportional to the system length provided that the terms in the brackets of Eq.~\eqref{eq:CVMsq15} are independent of system length.
Otherwise, the thermal rectification ratio depends on the specific formulas of thermal conductivity or materials properties.}}
Several previous experimental and numerical observations are also well explained.
In addition, the physical meanings, limitations and possible extensions of present theory are also discussed in details.
We believe that the proposed theory and these three dimensionless parameters therein will be taken
as a guideline for future experimental studies on the size dependent thermal rectification and shed light on the design of the thermal rectifier.

This study was supported by the National Science Foundation of China (51836003).
The authors acknowledge Shiqian Hu, Nuo Yang, Lifa Zhang, Xiulin Ruan, David Jou for useful communications of the thermal rectification in 2D Lorentz gas model, trapezoidal graphene nanoribbons and inhomogeneous porous silicon.

%\renewcommand{\theequation}{S\arabic{equation}}
%\renewcommand{\thefigure}{S\arabic{figure}}
%\setcounter{figure}{0}
%\setcounter{equation}{0}

%\twocolumngrid
%\bibliographystyle{IEEEtr}
\bibliography{phonon}

%merlin.mbs apsrev4-1.bst 2010-07-25 4.21a (PWD, AO, DPC) hacked
%Control: key (0)
%Control: author (8) initials jnrlst
%Control: editor formatted (1) identically to author
%Control: production of article title (-1) disabled
%Control: page (0) single
%Control: year (1) truncated
%Control: production of eprint (0) enabled
\begin{thebibliography}{50}%
\makeatletter
\providecommand \@ifxundefined [1]{%
 \@ifx{#1\undefined}
}%
\providecommand \@ifnum [1]{%
 \ifnum #1\expandafter \@firstoftwo
 \else \expandafter \@secondoftwo
 \fi
}%
\providecommand \@ifx [1]{%
 \ifx #1\expandafter \@firstoftwo
 \else \expandafter \@secondoftwo
 \fi
}%
\providecommand \natexlab [1]{#1}%
\providecommand \enquote  [1]{``#1''}%
\providecommand \bibnamefont  [1]{#1}%
\providecommand \bibfnamefont [1]{#1}%
\providecommand \citenamefont [1]{#1}%
\providecommand \href@noop [0]{\@secondoftwo}%
\providecommand \href [0]{\begingroup \@sanitize@url \@href}%
\providecommand \@href[1]{\@@startlink{#1}\@@href}%
\providecommand \@@href[1]{\endgroup#1\@@endlink}%
\providecommand \@sanitize@url [0]{\catcode `\\12\catcode `\$12\catcode
  `\&12\catcode `\#12\catcode `\^12\catcode `\_12\catcode `\%12\relax}%
\providecommand \@@startlink[1]{}%
\providecommand \@@endlink[0]{}%
\providecommand \url  [0]{\begingroup\@sanitize@url \@url }%
\providecommand \@url [1]{\endgroup\@href {#1}{\urlprefix }}%
\providecommand \urlprefix  [0]{URL }%
\providecommand \Eprint [0]{\href }%
\providecommand \doibase [0]{http://dx.doi.org/}%
\providecommand \selectlanguage [0]{\@gobble}%
\providecommand \bibinfo  [0]{\@secondoftwo}%
\providecommand \bibfield  [0]{\@secondoftwo}%
\providecommand \translation [1]{[#1]}%
\providecommand \BibitemOpen [0]{}%
\providecommand \bibitemStop [0]{}%
\providecommand \bibitemNoStop [0]{.\EOS\space}%
\providecommand \EOS [0]{\spacefactor3000\relax}%
\providecommand \BibitemShut  [1]{\csname bibitem#1\endcsname}%
\let\auto@bib@innerbib\@empty
%</preamble>
\bibitem [{\citenamefont {Starr}(1936)}]{starr1936}%
  \BibitemOpen
  \bibfield  {author} {\bibinfo {author} {\bibfnamefont {C.}~\bibnamefont
  {Starr}},\ }\href {\doibase 10.1063/1.1745338} {\bibfield  {journal}
  {\bibinfo  {journal} {Physics}\ }\textbf {\bibinfo {volume} {7}},\ \bibinfo
  {pages} {15} (\bibinfo {year} {1936})}\BibitemShut {NoStop}%
\bibitem [{\citenamefont {Terraneo}\ \emph {et~al.}(2002)\citenamefont
  {Terraneo}, \citenamefont {Peyrard},\ and\ \citenamefont
  {Casati}}]{terraneo2002a}%
  \BibitemOpen
  \bibfield  {author} {\bibinfo {author} {\bibfnamefont {M.}~\bibnamefont
  {Terraneo}}, \bibinfo {author} {\bibfnamefont {M.}~\bibnamefont {Peyrard}}, \
  and\ \bibinfo {author} {\bibfnamefont {G.}~\bibnamefont {Casati}},\ }\href
  {\doibase 10.1103/PhysRevLett.88.094302} {\bibfield  {journal} {\bibinfo
  {journal} {Phys. Rev. Lett.}\ }\textbf {\bibinfo {volume} {88}},\ \bibinfo
  {pages} {094302} (\bibinfo {year} {2002})}\BibitemShut {NoStop}%
\bibitem [{\citenamefont {Li}\ \emph {et~al.}(2004)\citenamefont {Li},
  \citenamefont {Wang},\ and\ \citenamefont {Casati}}]{li_thermal_2004}%
  \BibitemOpen
  \bibfield  {author} {\bibinfo {author} {\bibfnamefont {B.}~\bibnamefont
  {Li}}, \bibinfo {author} {\bibfnamefont {L.}~\bibnamefont {Wang}}, \ and\
  \bibinfo {author} {\bibfnamefont {G.}~\bibnamefont {Casati}},\ }\href
  {\doibase 10.1103/PhysRevLett.93.184301} {\bibfield  {journal} {\bibinfo
  {journal} {Phys. Rev. Lett.}\ }\textbf {\bibinfo {volume} {93}},\ \bibinfo
  {pages} {184301} (\bibinfo {year} {2004})}\BibitemShut {NoStop}%
\bibitem [{\citenamefont {Li}\ \emph {et~al.}(2012)\citenamefont {Li},
  \citenamefont {Ren}, \citenamefont {Wang}, \citenamefont {Zhang},
  \citenamefont {H\"anggi},\ and\ \citenamefont {Li}}]{RevModPhysLibaowen}%
  \BibitemOpen
  \bibfield  {author} {\bibinfo {author} {\bibfnamefont {N.}~\bibnamefont
  {Li}}, \bibinfo {author} {\bibfnamefont {J.}~\bibnamefont {Ren}}, \bibinfo
  {author} {\bibfnamefont {L.}~\bibnamefont {Wang}}, \bibinfo {author}
  {\bibfnamefont {G.}~\bibnamefont {Zhang}}, \bibinfo {author} {\bibfnamefont
  {P.}~\bibnamefont {H\"anggi}}, \ and\ \bibinfo {author} {\bibfnamefont
  {B.}~\bibnamefont {Li}},\ }\href {\doibase 10.1103/RevModPhys.84.1045}
  {\bibfield  {journal} {\bibinfo  {journal} {Rev. Mod. Phys.}\ }\textbf
  {\bibinfo {volume} {84}},\ \bibinfo {pages} {1045} (\bibinfo {year}
  {2012})}\BibitemShut {NoStop}%
\bibitem [{\citenamefont {Zhang}\ \emph {et~al.}(2020)\citenamefont {Zhang},
  \citenamefont {Ouyang}, \citenamefont {Cheng}, \citenamefont {Chen},
  \citenamefont {Li},\ and\ \citenamefont {Zhang}}]{ZHANG2020}%
  \BibitemOpen
  \bibfield  {author} {\bibinfo {author} {\bibfnamefont {Z.}~\bibnamefont
  {Zhang}}, \bibinfo {author} {\bibfnamefont {Y.}~\bibnamefont {Ouyang}},
  \bibinfo {author} {\bibfnamefont {Y.}~\bibnamefont {Cheng}}, \bibinfo
  {author} {\bibfnamefont {J.}~\bibnamefont {Chen}}, \bibinfo {author}
  {\bibfnamefont {N.}~\bibnamefont {Li}}, \ and\ \bibinfo {author}
  {\bibfnamefont {G.}~\bibnamefont {Zhang}},\ }\href {\doibase
  https://doi.org/10.1016/j.physrep.2020.03.001} {\bibfield  {journal}
  {\bibinfo  {journal} {Phys. Rep.}\ }\textbf {\bibinfo {volume} {860}},\
  \bibinfo {pages} {1 } (\bibinfo {year} {2020})}\BibitemShut {NoStop}%
\bibitem [{\citenamefont {Chang}\ \emph {et~al.}(2006)\citenamefont {Chang},
  \citenamefont {Okawa}, \citenamefont {Majumdar},\ and\ \citenamefont
  {Zettl}}]{chang_solid-state_2006}%
  \BibitemOpen
  \bibfield  {author} {\bibinfo {author} {\bibfnamefont {C.~W.}\ \bibnamefont
  {Chang}}, \bibinfo {author} {\bibfnamefont {D.}~\bibnamefont {Okawa}},
  \bibinfo {author} {\bibfnamefont {A.}~\bibnamefont {Majumdar}}, \ and\
  \bibinfo {author} {\bibfnamefont {A.}~\bibnamefont {Zettl}},\ }\href
  {\doibase 10.1126/science.1132898} {\bibfield  {journal} {\bibinfo  {journal}
  {Science}\ }\textbf {\bibinfo {volume} {314}},\ \bibinfo {pages} {1121}
  (\bibinfo {year} {2006})}\BibitemShut {NoStop}%
\bibitem [{\citenamefont {Eckmann}\ and\ \citenamefont
  {{Mej{\'i}a-Monasterio}}(2006)}]{eckmann2006}%
  \BibitemOpen
  \bibfield  {author} {\bibinfo {author} {\bibfnamefont {J.-P.}\ \bibnamefont
  {Eckmann}}\ and\ \bibinfo {author} {\bibfnamefont {C.}~\bibnamefont
  {{Mej{\'i}a-Monasterio}}},\ }\href {\doibase 10.1103/PhysRevLett.97.094301}
  {\bibfield  {journal} {\bibinfo  {journal} {Phys. Rev. Lett.}\ }\textbf
  {\bibinfo {volume} {97}},\ \bibinfo {pages} {094301} (\bibinfo {year}
  {2006})}\BibitemShut {NoStop}%
\bibitem [{\citenamefont {Yang}\ \emph {et~al.}(2012)\citenamefont {Yang},
  \citenamefont {Xu}, \citenamefont {Zhang},\ and\ \citenamefont
  {Li}}]{yang2012a}%
  \BibitemOpen
  \bibfield  {author} {\bibinfo {author} {\bibfnamefont {N.}~\bibnamefont
  {Yang}}, \bibinfo {author} {\bibfnamefont {X.}~\bibnamefont {Xu}}, \bibinfo
  {author} {\bibfnamefont {G.}~\bibnamefont {Zhang}}, \ and\ \bibinfo {author}
  {\bibfnamefont {B.}~\bibnamefont {Li}},\ }\href {\doibase 10.1063/1.4773462}
  {\bibfield  {journal} {\bibinfo  {journal} {AIP Adv.}\ }\textbf {\bibinfo
  {volume} {2}},\ \bibinfo {pages} {041410} (\bibinfo {year}
  {2012})}\BibitemShut {NoStop}%
\bibitem [{\citenamefont {Liu}\ \emph {et~al.}(2019)\citenamefont {Liu},
  \citenamefont {Wang},\ and\ \citenamefont {Zhang}}]{liu2019a}%
  \BibitemOpen
  \bibfield  {author} {\bibinfo {author} {\bibfnamefont {H.}~\bibnamefont
  {Liu}}, \bibinfo {author} {\bibfnamefont {H.}~\bibnamefont {Wang}}, \ and\
  \bibinfo {author} {\bibfnamefont {X.}~\bibnamefont {Zhang}},\ }\href
  {\doibase 10.3390/app9020344} {\bibfield  {journal} {\bibinfo  {journal}
  {Appl. Sci.}\ }\textbf {\bibinfo {volume} {9}},\ \bibinfo {pages} {344}
  (\bibinfo {year} {2019})}\BibitemShut {NoStop}%
\bibitem [{\citenamefont {Donovan}\ and\ \citenamefont
  {Warzoha}(2020)}]{PhysRevLett.124.075903}%
  \BibitemOpen
  \bibfield  {author} {\bibinfo {author} {\bibfnamefont {B.~F.}\ \bibnamefont
  {Donovan}}\ and\ \bibinfo {author} {\bibfnamefont {R.~J.}\ \bibnamefont
  {Warzoha}},\ }\href {\doibase 10.1103/PhysRevLett.124.075903} {\bibfield
  {journal} {\bibinfo  {journal} {Phys. Rev. Lett.}\ }\textbf {\bibinfo
  {volume} {124}},\ \bibinfo {pages} {075903} (\bibinfo {year}
  {2020})}\BibitemShut {NoStop}%
\bibitem [{\citenamefont {Kobayashi}\ \emph {et~al.}(2009)\citenamefont
  {Kobayashi}, \citenamefont {Teraoka},\ and\ \citenamefont
  {Terasaki}}]{kobayashi2009}%
  \BibitemOpen
  \bibfield  {author} {\bibinfo {author} {\bibfnamefont {W.}~\bibnamefont
  {Kobayashi}}, \bibinfo {author} {\bibfnamefont {Y.}~\bibnamefont {Teraoka}},
  \ and\ \bibinfo {author} {\bibfnamefont {I.}~\bibnamefont {Terasaki}},\
  }\href {\doibase 10.1063/1.3253712} {\bibfield  {journal} {\bibinfo
  {journal} {Appl. Phys. Lett.}\ }\textbf {\bibinfo {volume} {95}},\ \bibinfo
  {pages} {171905} (\bibinfo {year} {2009})}\BibitemShut {NoStop}%
\bibitem [{\citenamefont {Dames}(2009)}]{dames2009}%
  \BibitemOpen
  \bibfield  {author} {\bibinfo {author} {\bibfnamefont {C.}~\bibnamefont
  {Dames}},\ }\href {\doibase 10.1115/1.3089552} {\bibfield  {journal}
  {\bibinfo  {journal} {J. Heat Transfer}\ }\textbf {\bibinfo {volume} {131}},\
  \bibinfo {pages} {061301} (\bibinfo {year} {2009})}\BibitemShut {NoStop}%
\bibitem [{\citenamefont {Peyrard}(2006)}]{peyrard2006}%
  \BibitemOpen
  \bibfield  {author} {\bibinfo {author} {\bibfnamefont {M.}~\bibnamefont
  {Peyrard}},\ }\href {\doibase 10.1209/epl/i2006-10223-5} {\bibfield
  {journal} {\bibinfo  {journal} {EPL}\ }\textbf {\bibinfo {volume} {76}},\
  \bibinfo {pages} {49} (\bibinfo {year} {2006})}\BibitemShut {NoStop}%
\bibitem [{\citenamefont {Yang}\ \emph {et~al.}(2018)\citenamefont {Yang},
  \citenamefont {Chen}, \citenamefont {Wang}, \citenamefont {Li},\ and\
  \citenamefont {Zhang}}]{PhysRevE.98.042131}%
  \BibitemOpen
  \bibfield  {author} {\bibinfo {author} {\bibfnamefont {Y.}~\bibnamefont
  {Yang}}, \bibinfo {author} {\bibfnamefont {H.}~\bibnamefont {Chen}}, \bibinfo
  {author} {\bibfnamefont {H.}~\bibnamefont {Wang}}, \bibinfo {author}
  {\bibfnamefont {N.}~\bibnamefont {Li}}, \ and\ \bibinfo {author}
  {\bibfnamefont {L.}~\bibnamefont {Zhang}},\ }\href {\doibase
  10.1103/PhysRevE.98.042131} {\bibfield  {journal} {\bibinfo  {journal} {Phys.
  Rev. E}\ }\textbf {\bibinfo {volume} {98}},\ \bibinfo {pages} {042131}
  (\bibinfo {year} {2018})}\BibitemShut {NoStop}%
\bibitem [{\citenamefont {Go}\ and\ \citenamefont {Sen}(2010)}]{go2010}%
  \BibitemOpen
  \bibfield  {author} {\bibinfo {author} {\bibfnamefont {D.~B.}\ \bibnamefont
  {Go}}\ and\ \bibinfo {author} {\bibfnamefont {M.}~\bibnamefont {Sen}},\
  }\href {\doibase 10.1115/1.4002286} {\bibfield  {journal} {\bibinfo
  {journal} {J. Heat Transfer}\ }\textbf {\bibinfo {volume} {132}},\ \bibinfo
  {pages} {124502} (\bibinfo {year} {2010})}\BibitemShut {NoStop}%
\bibitem [{\citenamefont {Yang}\ \emph {et~al.}(2009)\citenamefont {Yang},
  \citenamefont {Zhang},\ and\ \citenamefont {Li}}]{yang_thermal_2009}%
  \BibitemOpen
  \bibfield  {author} {\bibinfo {author} {\bibfnamefont {N.}~\bibnamefont
  {Yang}}, \bibinfo {author} {\bibfnamefont {G.}~\bibnamefont {Zhang}}, \ and\
  \bibinfo {author} {\bibfnamefont {B.}~\bibnamefont {Li}},\ }\href {\doibase
  10.1063/1.3183587} {\bibfield  {journal} {\bibinfo  {journal} {Appl. Phys.
  Lett.}\ }\textbf {\bibinfo {volume} {95}},\ \bibinfo {pages} {033107}
  (\bibinfo {year} {2009})}\BibitemShut {NoStop}%
\bibitem [{\citenamefont {Hu}\ \emph {et~al.}(2009)\citenamefont {Hu},
  \citenamefont {Ruan},\ and\ \citenamefont {Chen}}]{hu2009}%
  \BibitemOpen
  \bibfield  {author} {\bibinfo {author} {\bibfnamefont {J.}~\bibnamefont
  {Hu}}, \bibinfo {author} {\bibfnamefont {X.}~\bibnamefont {Ruan}}, \ and\
  \bibinfo {author} {\bibfnamefont {Y.~P.}\ \bibnamefont {Chen}},\ }\href
  {\doibase 10.1021/nl901231s} {\bibfield  {journal} {\bibinfo  {journal} {Nano
  Lett.}\ }\textbf {\bibinfo {volume} {9}},\ \bibinfo {pages} {2730} (\bibinfo
  {year} {2009})}\BibitemShut {NoStop}%
\bibitem [{\citenamefont {Wang}\ \emph
  {et~al.}(2019{\natexlab{a}})\citenamefont {Wang}, \citenamefont {Yang},
  \citenamefont {Chen}, \citenamefont {Li},\ and\ \citenamefont
  {Zhang}}]{wang2019}%
  \BibitemOpen
  \bibfield  {author} {\bibinfo {author} {\bibfnamefont {H.}~\bibnamefont
  {Wang}}, \bibinfo {author} {\bibfnamefont {Y.}~\bibnamefont {Yang}}, \bibinfo
  {author} {\bibfnamefont {H.}~\bibnamefont {Chen}}, \bibinfo {author}
  {\bibfnamefont {N.}~\bibnamefont {Li}}, \ and\ \bibinfo {author}
  {\bibfnamefont {L.}~\bibnamefont {Zhang}},\ }\href {\doibase
  10.1103/PhysRevE.99.062111} {\bibfield  {journal} {\bibinfo  {journal} {Phys.
  Rev. E}\ }\textbf {\bibinfo {volume} {99}},\ \bibinfo {pages} {062111}
  (\bibinfo {year} {2019}{\natexlab{a}})}\BibitemShut {NoStop}%
\bibitem [{\citenamefont {Yang}\ \emph {et~al.}(2008)\citenamefont {Yang},
  \citenamefont {Zhang},\ and\ \citenamefont {Li}}]{yang_carbon_2008}%
  \BibitemOpen
  \bibfield  {author} {\bibinfo {author} {\bibfnamefont {N.}~\bibnamefont
  {Yang}}, \bibinfo {author} {\bibfnamefont {G.}~\bibnamefont {Zhang}}, \ and\
  \bibinfo {author} {\bibfnamefont {B.}~\bibnamefont {Li}},\ }\href {\doibase
  10.1063/1.3049603} {\bibfield  {journal} {\bibinfo  {journal} {Appl. Phys.
  Lett.}\ }\textbf {\bibinfo {volume} {93}},\ \bibinfo {pages} {243111}
  (\bibinfo {year} {2008})}\BibitemShut {NoStop}%
\bibitem [{\citenamefont {Wang}\ \emph {et~al.}(2017)\citenamefont {Wang},
  \citenamefont {Hu}, \citenamefont {Takahashi}, \citenamefont {Zhang},
  \citenamefont {Takamatsu},\ and\ \citenamefont {Chen}}]{wang2017}%
  \BibitemOpen
  \bibfield  {author} {\bibinfo {author} {\bibfnamefont {H.}~\bibnamefont
  {Wang}}, \bibinfo {author} {\bibfnamefont {S.}~\bibnamefont {Hu}}, \bibinfo
  {author} {\bibfnamefont {K.}~\bibnamefont {Takahashi}}, \bibinfo {author}
  {\bibfnamefont {X.}~\bibnamefont {Zhang}}, \bibinfo {author} {\bibfnamefont
  {H.}~\bibnamefont {Takamatsu}}, \ and\ \bibinfo {author} {\bibfnamefont
  {J.}~\bibnamefont {Chen}},\ }\href {\doibase 10.1038/ncomms15843} {\bibfield
  {journal} {\bibinfo  {journal} {Nat. Commun.}\ }\textbf {\bibinfo {volume}
  {8}},\ \bibinfo {pages} {15843} (\bibinfo {year} {2017})}\BibitemShut
  {NoStop}%
\bibitem [{\citenamefont {Wang}\ \emph {et~al.}(2014)\citenamefont {Wang},
  \citenamefont {Vallabhaneni}, \citenamefont {Hu}, \citenamefont {Qiu},
  \citenamefont {Chen},\ and\ \citenamefont {Ruan}}]{wang2014}%
  \BibitemOpen
  \bibfield  {author} {\bibinfo {author} {\bibfnamefont {Y.}~\bibnamefont
  {Wang}}, \bibinfo {author} {\bibfnamefont {A.}~\bibnamefont {Vallabhaneni}},
  \bibinfo {author} {\bibfnamefont {J.}~\bibnamefont {Hu}}, \bibinfo {author}
  {\bibfnamefont {B.}~\bibnamefont {Qiu}}, \bibinfo {author} {\bibfnamefont
  {Y.~P.}\ \bibnamefont {Chen}}, \ and\ \bibinfo {author} {\bibfnamefont
  {X.}~\bibnamefont {Ruan}},\ }\href {\doibase 10.1021/nl403773f} {\bibfield
  {journal} {\bibinfo  {journal} {Nano Lett.}\ }\textbf {\bibinfo {volume}
  {14}},\ \bibinfo {pages} {592} (\bibinfo {year} {2014})}\BibitemShut
  {NoStop}%
\bibitem [{\citenamefont {Yang}\ \emph {et~al.}(2007)\citenamefont {Yang},
  \citenamefont {Li}, \citenamefont {Wang},\ and\ \citenamefont
  {Li}}]{yang2007}%
  \BibitemOpen
  \bibfield  {author} {\bibinfo {author} {\bibfnamefont {N.}~\bibnamefont
  {Yang}}, \bibinfo {author} {\bibfnamefont {N.}~\bibnamefont {Li}}, \bibinfo
  {author} {\bibfnamefont {L.}~\bibnamefont {Wang}}, \ and\ \bibinfo {author}
  {\bibfnamefont {B.}~\bibnamefont {Li}},\ }\href {\doibase
  10.1103/PhysRevB.76.020301} {\bibfield  {journal} {\bibinfo  {journal} {Phys.
  Rev. B}\ }\textbf {\bibinfo {volume} {76}},\ \bibinfo {pages} {020301}
  (\bibinfo {year} {2007})}\BibitemShut {NoStop}%
\bibitem [{\citenamefont {Wu}\ and\ \citenamefont {Li}(2008)}]{wu2008}%
  \BibitemOpen
  \bibfield  {author} {\bibinfo {author} {\bibfnamefont {G.}~\bibnamefont
  {Wu}}\ and\ \bibinfo {author} {\bibfnamefont {B.}~\bibnamefont {Li}},\ }\href
  {\doibase 10.1088/0953-8984/20/17/175211} {\bibfield  {journal} {\bibinfo
  {journal} {J. Phys.: Condens. Matter}\ }\textbf {\bibinfo {volume} {20}},\
  \bibinfo {pages} {175211} (\bibinfo {year} {2008})}\BibitemShut {NoStop}%
\bibitem [{\citenamefont {{Criado-Sancho}}\ \emph {et~al.}(2013)\citenamefont
  {{Criado-Sancho}}, \citenamefont {Alvarez},\ and\ \citenamefont
  {Jou}}]{criado-sancho2013}%
  \BibitemOpen
  \bibfield  {author} {\bibinfo {author} {\bibfnamefont {M.}~\bibnamefont
  {{Criado-Sancho}}}, \bibinfo {author} {\bibfnamefont {F.~X.}\ \bibnamefont
  {Alvarez}}, \ and\ \bibinfo {author} {\bibfnamefont {D.}~\bibnamefont
  {Jou}},\ }\href {\doibase 10.1063/1.4816685} {\bibfield  {journal} {\bibinfo
  {journal} {J. Appl. Phys.}\ }\textbf {\bibinfo {volume} {114}},\ \bibinfo
  {pages} {053512} (\bibinfo {year} {2013})}\BibitemShut {NoStop}%
\bibitem [{\citenamefont {Wu}\ and\ \citenamefont
  {Li}(2007)}]{wu_thermal_2007}%
  \BibitemOpen
  \bibfield  {author} {\bibinfo {author} {\bibfnamefont {G.}~\bibnamefont
  {Wu}}\ and\ \bibinfo {author} {\bibfnamefont {B.}~\bibnamefont {Li}},\ }\href
  {\doibase 10.1103/PhysRevB.76.085424} {\bibfield  {journal} {\bibinfo
  {journal} {Phys. Rev. B}\ }\textbf {\bibinfo {volume} {76}},\ \bibinfo
  {pages} {085424} (\bibinfo {year} {2007})}\BibitemShut {NoStop}%
\bibitem [{\citenamefont {Hu}\ \emph {et~al.}(2017)\citenamefont {Hu},
  \citenamefont {An}, \citenamefont {Yang},\ and\ \citenamefont
  {Li}}]{hu_series_2017}%
  \BibitemOpen
  \bibfield  {author} {\bibinfo {author} {\bibfnamefont {S.}~\bibnamefont
  {Hu}}, \bibinfo {author} {\bibfnamefont {M.}~\bibnamefont {An}}, \bibinfo
  {author} {\bibfnamefont {N.}~\bibnamefont {Yang}}, \ and\ \bibinfo {author}
  {\bibfnamefont {B.}~\bibnamefont {Li}},\ }\href {\doibase
  10.1002/smll.201602726} {\bibfield  {journal} {\bibinfo  {journal} {Small}\
  }\textbf {\bibinfo {volume} {13}},\ \bibinfo {pages} {1602726} (\bibinfo
  {year} {2017})}\BibitemShut {NoStop}%
\bibitem [{\citenamefont {Roberts}\ and\ \citenamefont
  {Walker}(2011)}]{roberts2011a}%
  \BibitemOpen
  \bibfield  {author} {\bibinfo {author} {\bibfnamefont {N.~A.}\ \bibnamefont
  {Roberts}}\ and\ \bibinfo {author} {\bibfnamefont {D.~G.}\ \bibnamefont
  {Walker}},\ }\href {\doibase 10.1016/j.ijthermalsci.2010.12.004} {\bibfield
  {journal} {\bibinfo  {journal} {Int. J. Therm. Sci.}\ }\textbf {\bibinfo
  {volume} {50}},\ \bibinfo {pages} {648} (\bibinfo {year} {2011})}\BibitemShut
  {NoStop}%
\bibitem [{\citenamefont {Ouyang}\ \emph {et~al.}(2010)\citenamefont {Ouyang},
  \citenamefont {Chen}, \citenamefont {Xie}, \citenamefont {Wei}, \citenamefont
  {Yang}, \citenamefont {Yang},\ and\ \citenamefont {Zhong}}]{ouyang2010}%
  \BibitemOpen
  \bibfield  {author} {\bibinfo {author} {\bibfnamefont {T.}~\bibnamefont
  {Ouyang}}, \bibinfo {author} {\bibfnamefont {Y.}~\bibnamefont {Chen}},
  \bibinfo {author} {\bibfnamefont {Y.}~\bibnamefont {Xie}}, \bibinfo {author}
  {\bibfnamefont {X.~L.}\ \bibnamefont {Wei}}, \bibinfo {author} {\bibfnamefont
  {K.}~\bibnamefont {Yang}}, \bibinfo {author} {\bibfnamefont {P.}~\bibnamefont
  {Yang}}, \ and\ \bibinfo {author} {\bibfnamefont {J.}~\bibnamefont {Zhong}},\
  }\href {\doibase 10.1103/PhysRevB.82.245403} {\bibfield  {journal} {\bibinfo
  {journal} {Phys. Rev. B}\ }\textbf {\bibinfo {volume} {82}},\ \bibinfo
  {pages} {245403} (\bibinfo {year} {2010})}\BibitemShut {NoStop}%
\bibitem [{\citenamefont {Ma}\ and\ \citenamefont {Tian}(2018)}]{ma2018}%
  \BibitemOpen
  \bibfield  {author} {\bibinfo {author} {\bibfnamefont {H.}~\bibnamefont
  {Ma}}\ and\ \bibinfo {author} {\bibfnamefont {Z.}~\bibnamefont {Tian}},\
  }\href {\doibase 10.1021/acs.nanolett.7b02867} {\bibfield  {journal}
  {\bibinfo  {journal} {Nano Lett.}\ }\textbf {\bibinfo {volume} {18}},\
  \bibinfo {pages} {43} (\bibinfo {year} {2018})}\BibitemShut {NoStop}%
\bibitem [{\citenamefont {Zhu}\ \emph {et~al.}(2014)\citenamefont {Zhu},
  \citenamefont {Hippalgaonkar}, \citenamefont {Shen}, \citenamefont {Wang},
  \citenamefont {Abate}, \citenamefont {Lee}, \citenamefont {Wu}, \citenamefont
  {Yin}, \citenamefont {Majumdar},\ and\ \citenamefont {Zhang}}]{zhu2014}%
  \BibitemOpen
  \bibfield  {author} {\bibinfo {author} {\bibfnamefont {J.}~\bibnamefont
  {Zhu}}, \bibinfo {author} {\bibfnamefont {K.}~\bibnamefont {Hippalgaonkar}},
  \bibinfo {author} {\bibfnamefont {S.}~\bibnamefont {Shen}}, \bibinfo {author}
  {\bibfnamefont {K.}~\bibnamefont {Wang}}, \bibinfo {author} {\bibfnamefont
  {Y.}~\bibnamefont {Abate}}, \bibinfo {author} {\bibfnamefont
  {S.}~\bibnamefont {Lee}}, \bibinfo {author} {\bibfnamefont {J.}~\bibnamefont
  {Wu}}, \bibinfo {author} {\bibfnamefont {X.}~\bibnamefont {Yin}}, \bibinfo
  {author} {\bibfnamefont {A.}~\bibnamefont {Majumdar}}, \ and\ \bibinfo
  {author} {\bibfnamefont {X.}~\bibnamefont {Zhang}},\ }\href {\doibase
  10.1021/nl502261m} {\bibfield  {journal} {\bibinfo  {journal} {Nano Lett.}\
  }\textbf {\bibinfo {volume} {14}},\ \bibinfo {pages} {4867} (\bibinfo {year}
  {2014})}\BibitemShut {NoStop}%
\bibitem [{\citenamefont {Pereira}\ and\ \citenamefont
  {Falcao}(2006)}]{pereira2006}%
  \BibitemOpen
  \bibfield  {author} {\bibinfo {author} {\bibfnamefont {E.}~\bibnamefont
  {Pereira}}\ and\ \bibinfo {author} {\bibfnamefont {R.}~\bibnamefont
  {Falcao}},\ }\href {\doibase 10.1103/PhysRevLett.96.100601} {\bibfield
  {journal} {\bibinfo  {journal} {Phys. Rev. Lett.}\ }\textbf {\bibinfo
  {volume} {96}},\ \bibinfo {pages} {100601} (\bibinfo {year}
  {2006})}\BibitemShut {NoStop}%
\bibitem [{\citenamefont {Pereira}(2017)}]{pereira2017}%
  \BibitemOpen
  \bibfield  {author} {\bibinfo {author} {\bibfnamefont {E.}~\bibnamefont
  {Pereira}},\ }\href {\doibase 10.1103/PhysRevE.96.012114} {\bibfield
  {journal} {\bibinfo  {journal} {Phys. Rev. E}\ }\textbf {\bibinfo {volume}
  {96}},\ \bibinfo {pages} {012114} (\bibinfo {year} {2017})}\BibitemShut
  {NoStop}%
\bibitem [{\citenamefont {Wehmeyer}\ \emph {et~al.}(2017)\citenamefont
  {Wehmeyer}, \citenamefont {Yabuki}, \citenamefont {Monachon}, \citenamefont
  {Wu},\ and\ \citenamefont {Dames}}]{wehmeyer2017}%
  \BibitemOpen
  \bibfield  {author} {\bibinfo {author} {\bibfnamefont {G.}~\bibnamefont
  {Wehmeyer}}, \bibinfo {author} {\bibfnamefont {T.}~\bibnamefont {Yabuki}},
  \bibinfo {author} {\bibfnamefont {C.}~\bibnamefont {Monachon}}, \bibinfo
  {author} {\bibfnamefont {J.}~\bibnamefont {Wu}}, \ and\ \bibinfo {author}
  {\bibfnamefont {C.}~\bibnamefont {Dames}},\ }\href {\doibase
  10.1063/1.5001072} {\bibfield  {journal} {\bibinfo  {journal} {Appl. Phys.
  Rev.}\ }\textbf {\bibinfo {volume} {4}},\ \bibinfo {pages} {041304} (\bibinfo
  {year} {2017})}\BibitemShut {NoStop}%
\bibitem [{\citenamefont {Pereira}(2010)}]{pereira2010}%
  \BibitemOpen
  \bibfield  {author} {\bibinfo {author} {\bibfnamefont {E.}~\bibnamefont
  {Pereira}},\ }\href {\doibase 10.1103/PhysRevE.82.040101} {\bibfield
  {journal} {\bibinfo  {journal} {Phys. Rev. E}\ }\textbf {\bibinfo {volume}
  {82}},\ \bibinfo {pages} {040101} (\bibinfo {year} {2010})}\BibitemShut
  {NoStop}%
\bibitem [{\citenamefont {Pereira}(2011)}]{pereira2011}%
  \BibitemOpen
  \bibfield  {author} {\bibinfo {author} {\bibfnamefont {E.}~\bibnamefont
  {Pereira}},\ }\href {\doibase 10.1103/PhysRevE.83.031106} {\bibfield
  {journal} {\bibinfo  {journal} {Phys. Rev. E}\ }\textbf {\bibinfo {volume}
  {83}},\ \bibinfo {pages} {031106} (\bibinfo {year} {2011})}\BibitemShut
  {NoStop}%
\bibitem [{\citenamefont {Kubo~R.}\ and\ \citenamefont
  {N.}(1991)}]{Kubo1991statistical}%
  \BibitemOpen
  \bibfield  {author} {\bibinfo {author} {\bibfnamefont {T.~M.}\ \bibnamefont
  {Kubo~R.}}\ and\ \bibinfo {author} {\bibfnamefont {H.}~\bibnamefont {N.}},\
  }\href {https://doi.org/10.1007/978-3-642-58244-8} {\emph {\bibinfo {title}
  {Statistical Physics II Nonequilibrium Statistical Mechanics}}},\ Springer
  Series in Solid State Sciences\ (\bibinfo  {publisher} {Springer, Berlin,
  Heidelberg},\ \bibinfo {year} {1991})\BibitemShut {NoStop}%
\bibitem [{\citenamefont {Wang}\ \emph
  {et~al.}(2019{\natexlab{b}})\citenamefont {Wang}, \citenamefont {Liu},\ and\
  \citenamefont {Li}}]{Wang_2019LTE}%
  \BibitemOpen
  \bibfield  {author} {\bibinfo {author} {\bibfnamefont {L.}~\bibnamefont
  {Wang}}, \bibinfo {author} {\bibfnamefont {S.}~\bibnamefont {Liu}}, \ and\
  \bibinfo {author} {\bibfnamefont {B.}~\bibnamefont {Li}},\ }\href {\doibase
  10.1088/1367-2630/ab34a0} {\bibfield  {journal} {\bibinfo  {journal} {New J.
  Phys.}\ }\textbf {\bibinfo {volume} {21}},\ \bibinfo {pages} {083019}
  (\bibinfo {year} {2019}{\natexlab{b}})}\BibitemShut {NoStop}%
\bibitem [{\citenamefont {Sawaki}\ \emph {et~al.}(2011)\citenamefont {Sawaki},
  \citenamefont {Kobayashi}, \citenamefont {Moritomo},\ and\ \citenamefont
  {Terasaki}}]{sawaki2011}%
  \BibitemOpen
  \bibfield  {author} {\bibinfo {author} {\bibfnamefont {D.}~\bibnamefont
  {Sawaki}}, \bibinfo {author} {\bibfnamefont {W.}~\bibnamefont {Kobayashi}},
  \bibinfo {author} {\bibfnamefont {Y.}~\bibnamefont {Moritomo}}, \ and\
  \bibinfo {author} {\bibfnamefont {I.}~\bibnamefont {Terasaki}},\ }\href
  {\doibase 10.1063/1.3559615} {\bibfield  {journal} {\bibinfo  {journal}
  {Appl. Phys. Lett.}\ }\textbf {\bibinfo {volume} {98}},\ \bibinfo {pages}
  {081915} (\bibinfo {year} {2011})}\BibitemShut {NoStop}%
\bibitem [{\citenamefont {Rudin}\ \emph {et~al.}(1964)\citenamefont {Rudin}
  \emph {et~al.}}]{rudin1964principles}%
  \BibitemOpen
  \bibfield  {author} {\bibinfo {author} {\bibfnamefont {W.}~\bibnamefont
  {Rudin}} \emph {et~al.},\ }\href@noop {} {\emph {\bibinfo {title} {Principles
  of mathematical analysis}}},\ Vol.~\bibinfo {volume} {3}\ (\bibinfo
  {publisher} {McGraw-hill New York},\ \bibinfo {year} {1964})\BibitemShut
  {NoStop}%
\bibitem [{\citenamefont {Bender}\ and\ \citenamefont
  {Orszag}(2013)}]{bender2013advanced}%
  \BibitemOpen
  \bibfield  {author} {\bibinfo {author} {\bibfnamefont {C.~M.}\ \bibnamefont
  {Bender}}\ and\ \bibinfo {author} {\bibfnamefont {S.~A.}\ \bibnamefont
  {Orszag}},\ }\href@noop {} {\emph {\bibinfo {title} {Advanced mathematical
  methods for scientists and engineers I: Asymptotic methods and perturbation
  theory}}}\ (\bibinfo  {publisher} {Springer Science \& Business Media},\
  \bibinfo {year} {2013})\BibitemShut {NoStop}%
\bibitem [{\citenamefont {Yang}\ \emph {et~al.}(2020)\citenamefont {Yang},
  \citenamefont {Zheng}, \citenamefont {Liu}, \citenamefont {Zhang},
  \citenamefont {Bai}, \citenamefont {Yang}, \citenamefont {Chen},\ and\
  \citenamefont {Liu}}]{yang_experimental_2020}%
  \BibitemOpen
  \bibfield  {author} {\bibinfo {author} {\bibfnamefont {X.}~\bibnamefont
  {Yang}}, \bibinfo {author} {\bibfnamefont {X.}~\bibnamefont {Zheng}},
  \bibinfo {author} {\bibfnamefont {Q.}~\bibnamefont {Liu}}, \bibinfo {author}
  {\bibfnamefont {T.}~\bibnamefont {Zhang}}, \bibinfo {author} {\bibfnamefont
  {Y.}~\bibnamefont {Bai}}, \bibinfo {author} {\bibfnamefont {Z.}~\bibnamefont
  {Yang}}, \bibinfo {author} {\bibfnamefont {H.}~\bibnamefont {Chen}}, \ and\
  \bibinfo {author} {\bibfnamefont {M.}~\bibnamefont {Liu}},\ }\href {\doibase
  10.1021/acsami.0c07544} {\bibfield  {journal} {\bibinfo  {journal} {{ACS}
  Appl. Mater. Interfaces}\ }\textbf {\bibinfo {volume} {12}},\ \bibinfo
  {pages} {28306} (\bibinfo {year} {2020})}\BibitemShut {NoStop}%
\bibitem [{\citenamefont {Bae}\ \emph {et~al.}(2013)\citenamefont {Bae},
  \citenamefont {Li}, \citenamefont {Aksamija}, \citenamefont {Martin},
  \citenamefont {Xiong}, \citenamefont {Ong}, \citenamefont {Knezevic},\ and\
  \citenamefont {Pop}}]{bae2013ballistic}%
  \BibitemOpen
  \bibfield  {author} {\bibinfo {author} {\bibfnamefont {M.-H.}\ \bibnamefont
  {Bae}}, \bibinfo {author} {\bibfnamefont {Z.}~\bibnamefont {Li}}, \bibinfo
  {author} {\bibfnamefont {Z.}~\bibnamefont {Aksamija}}, \bibinfo {author}
  {\bibfnamefont {P.~N.}\ \bibnamefont {Martin}}, \bibinfo {author}
  {\bibfnamefont {F.}~\bibnamefont {Xiong}}, \bibinfo {author} {\bibfnamefont
  {Z.-Y.}\ \bibnamefont {Ong}}, \bibinfo {author} {\bibfnamefont
  {I.}~\bibnamefont {Knezevic}}, \ and\ \bibinfo {author} {\bibfnamefont
  {E.}~\bibnamefont {Pop}},\ }\href {\doibase 10.1038/ncomms2755} {\bibfield
  {journal} {\bibinfo  {journal} {Nat. Commun.}\ }\textbf {\bibinfo {volume}
  {4}},\ \bibinfo {pages} {1} (\bibinfo {year} {2013})}\BibitemShut {NoStop}%
\bibitem [{\citenamefont {Xu}\ \emph {et~al.}(2014)\citenamefont {Xu},
  \citenamefont {Pereira}, \citenamefont {Wang}, \citenamefont {Wu},
  \citenamefont {Zhang}, \citenamefont {Zhao}, \citenamefont {Bae},
  \citenamefont {Bui}, \citenamefont {Xie}, \citenamefont {Thong},
  \citenamefont {Hong}, \citenamefont {Loh}, \citenamefont {Donadio},
  \citenamefont {Li},\ and\ \citenamefont
  {{\"O}zyilmaz}}]{xu_length-dependent_2014}%
  \BibitemOpen
  \bibfield  {author} {\bibinfo {author} {\bibfnamefont {X.}~\bibnamefont
  {Xu}}, \bibinfo {author} {\bibfnamefont {L.~F.~C.}\ \bibnamefont {Pereira}},
  \bibinfo {author} {\bibfnamefont {Y.}~\bibnamefont {Wang}}, \bibinfo {author}
  {\bibfnamefont {J.}~\bibnamefont {Wu}}, \bibinfo {author} {\bibfnamefont
  {K.}~\bibnamefont {Zhang}}, \bibinfo {author} {\bibfnamefont
  {X.}~\bibnamefont {Zhao}}, \bibinfo {author} {\bibfnamefont {S.}~\bibnamefont
  {Bae}}, \bibinfo {author} {\bibfnamefont {C.~T.}\ \bibnamefont {Bui}},
  \bibinfo {author} {\bibfnamefont {R.}~\bibnamefont {Xie}}, \bibinfo {author}
  {\bibfnamefont {J.~T.~L.}\ \bibnamefont {Thong}}, \bibinfo {author}
  {\bibfnamefont {B.~H.}\ \bibnamefont {Hong}}, \bibinfo {author}
  {\bibfnamefont {K.~P.}\ \bibnamefont {Loh}}, \bibinfo {author} {\bibfnamefont
  {D.}~\bibnamefont {Donadio}}, \bibinfo {author} {\bibfnamefont
  {B.}~\bibnamefont {Li}}, \ and\ \bibinfo {author} {\bibfnamefont
  {B.}~\bibnamefont {{\"O}zyilmaz}},\ }\href {\doibase 10.1038/ncomms4689}
  {\bibfield  {journal} {\bibinfo  {journal} {Nat. Commun.}\ }\textbf {\bibinfo
  {volume} {5}},\ \bibinfo {pages} {3689} (\bibinfo {year} {2014})}\BibitemShut
  {NoStop}%
\bibitem [{\citenamefont {Alvarez}\ \emph {et~al.}(2010)\citenamefont
  {Alvarez}, \citenamefont {Jou},\ and\ \citenamefont
  {Sellitto}}]{hydrodynamicporeousAFX2010}%
  \BibitemOpen
  \bibfield  {author} {\bibinfo {author} {\bibfnamefont {F.~X.}\ \bibnamefont
  {Alvarez}}, \bibinfo {author} {\bibfnamefont {D.}~\bibnamefont {Jou}}, \ and\
  \bibinfo {author} {\bibfnamefont {A.}~\bibnamefont {Sellitto}},\ }\href
  {\doibase 10.1063/1.3462936} {\bibfield  {journal} {\bibinfo  {journal}
  {Appl. Phys. Lett.}\ }\textbf {\bibinfo {volume} {97}},\ \bibinfo {pages}
  {033103} (\bibinfo {year} {2010})}\BibitemShut {NoStop}%
\bibitem [{\citenamefont {Pandey}\ and\ \citenamefont
  {Leitner}(2017)}]{Pandey17PEG}%
  \BibitemOpen
  \bibfield  {author} {\bibinfo {author} {\bibfnamefont {H.~D.}\ \bibnamefont
  {Pandey}}\ and\ \bibinfo {author} {\bibfnamefont {D.~M.}\ \bibnamefont
  {Leitner}},\ }\href {\doibase 10.1063/1.4999411} {\bibfield  {journal}
  {\bibinfo  {journal} {J. Chem. Phys.}\ }\textbf {\bibinfo {volume} {147}},\
  \bibinfo {pages} {084701} (\bibinfo {year} {2017})}\BibitemShut {NoStop}%
\bibitem [{\citenamefont {Pandey}\ and\ \citenamefont
  {Leitner}(2016)}]{pandey_thermalization_2016}%
  \BibitemOpen
  \bibfield  {author} {\bibinfo {author} {\bibfnamefont {H.~D.}\ \bibnamefont
  {Pandey}}\ and\ \bibinfo {author} {\bibfnamefont {D.~M.}\ \bibnamefont
  {Leitner}},\ }\href {\doibase 10.1021/acs.jpclett.6b02539} {\bibfield
  {journal} {\bibinfo  {journal} {J. Phys. Chem. Lett.}\ }\textbf {\bibinfo
  {volume} {7}},\ \bibinfo {pages} {5062} (\bibinfo {year} {2016})}\BibitemShut
  {NoStop}%
\bibitem [{\citenamefont {Leitner}(2013)}]{leitner_thermal_2013}%
  \BibitemOpen
  \bibfield  {author} {\bibinfo {author} {\bibfnamefont {D.~M.}\ \bibnamefont
  {Leitner}},\ }\href {\doibase 10.1021/jp402012z} {\bibfield  {journal}
  {\bibinfo  {journal} {J. Phys. Chem. B}\ }\textbf {\bibinfo {volume} {117}},\
  \bibinfo {pages} {12820} (\bibinfo {year} {2013})}\BibitemShut {NoStop}%
\bibitem [{\citenamefont {Kobayashi}\ \emph {et~al.}(2012)\citenamefont
  {Kobayashi}, \citenamefont {Sawaki}, \citenamefont {Omura}, \citenamefont
  {Katsufuji}, \citenamefont {Moritomo},\ and\ \citenamefont
  {Terasaki}}]{kobayashi2012}%
  \BibitemOpen
  \bibfield  {author} {\bibinfo {author} {\bibfnamefont {W.}~\bibnamefont
  {Kobayashi}}, \bibinfo {author} {\bibfnamefont {D.}~\bibnamefont {Sawaki}},
  \bibinfo {author} {\bibfnamefont {T.}~\bibnamefont {Omura}}, \bibinfo
  {author} {\bibfnamefont {T.}~\bibnamefont {Katsufuji}}, \bibinfo {author}
  {\bibfnamefont {Y.}~\bibnamefont {Moritomo}}, \ and\ \bibinfo {author}
  {\bibfnamefont {I.}~\bibnamefont {Terasaki}},\ }\href {\doibase
  10.1143/APEX.5.027302} {\bibfield  {journal} {\bibinfo  {journal} {Appl.
  Phys. Express}\ }\textbf {\bibinfo {volume} {5}},\ \bibinfo {pages} {027302}
  (\bibinfo {year} {2012})}\BibitemShut {NoStop}%
\bibitem [{\citenamefont {Kang}\ \emph {et~al.}(2018)\citenamefont {Kang},
  \citenamefont {Yang},\ and\ \citenamefont {Urban}}]{kang2018a}%
  \BibitemOpen
  \bibfield  {author} {\bibinfo {author} {\bibfnamefont {H.}~\bibnamefont
  {Kang}}, \bibinfo {author} {\bibfnamefont {F.}~\bibnamefont {Yang}}, \ and\
  \bibinfo {author} {\bibfnamefont {J.~J.}\ \bibnamefont {Urban}},\ }\href
  {\doibase 10.1103/PhysRevApplied.10.024034} {\bibfield  {journal} {\bibinfo
  {journal} {Phys. Rev. Applied}\ }\textbf {\bibinfo {volume} {10}},\ \bibinfo
  {pages} {024034} (\bibinfo {year} {2018})}\BibitemShut {NoStop}%
\bibitem [{\citenamefont {Jiang}\ \emph {et~al.}(2020)\citenamefont {Jiang},
  \citenamefont {Hu}, \citenamefont {Ouyang}, \citenamefont {Ren},
  \citenamefont {Yu}, \citenamefont {Zhang},\ and\ \citenamefont
  {Chen}}]{rectificationcontacts2020JAP}%
  \BibitemOpen
  \bibfield  {author} {\bibinfo {author} {\bibfnamefont {P.}~\bibnamefont
  {Jiang}}, \bibinfo {author} {\bibfnamefont {S.}~\bibnamefont {Hu}}, \bibinfo
  {author} {\bibfnamefont {Y.}~\bibnamefont {Ouyang}}, \bibinfo {author}
  {\bibfnamefont {W.}~\bibnamefont {Ren}}, \bibinfo {author} {\bibfnamefont
  {C.}~\bibnamefont {Yu}}, \bibinfo {author} {\bibfnamefont {Z.}~\bibnamefont
  {Zhang}}, \ and\ \bibinfo {author} {\bibfnamefont {J.}~\bibnamefont {Chen}},\
  }\href {\doibase 10.1063/5.0004484} {\bibfield  {journal} {\bibinfo
  {journal} {J. Appl. Phys.}\ }\textbf {\bibinfo {volume} {127}},\ \bibinfo
  {pages} {235101} (\bibinfo {year} {2020})}\BibitemShut {NoStop}%
\end{thebibliography}%


%merlin.mbs apsrev4-1.bst 2010-07-25 4.21a (PWD, AO, DPC) hacked
%Control: key (0)
%Control: author (8) initials jnrlst
%Control: editor formatted (1) identically to author
%Control: production of article title (-1) disabled
%Control: page (0) single
%Control: year (1) truncated
%Control: production of eprint (0) enabled
\begin{thebibliography}{29}%
\makeatletter
\providecommand \@ifxundefined [1]{%
 \@ifx{#1\undefined}
}%
\providecommand \@ifnum [1]{%
 \ifnum #1\expandafter \@firstoftwo
 \else \expandafter \@secondoftwo
 \fi
}%
\providecommand \@ifx [1]{%
 \ifx #1\expandafter \@firstoftwo
 \else \expandafter \@secondoftwo
 \fi
}%
\providecommand \natexlab [1]{#1}%
\providecommand \enquote  [1]{``#1''}%
\providecommand \bibnamefont  [1]{#1}%
\providecommand \bibfnamefont [1]{#1}%
\providecommand \citenamefont [1]{#1}%
\providecommand \href@noop [0]{\@secondoftwo}%
\providecommand \href [0]{\begingroup \@sanitize@url \@href}%
\providecommand \@href[1]{\@@startlink{#1}\@@href}%
\providecommand \@@href[1]{\endgroup#1\@@endlink}%
\providecommand \@sanitize@url [0]{\catcode `\\12\catcode `\$12\catcode
  `\&12\catcode `\#12\catcode `\^12\catcode `\_12\catcode `\%12\relax}%
\providecommand \@@startlink[1]{}%
\providecommand \@@endlink[0]{}%
\providecommand \url  [0]{\begingroup\@sanitize@url \@url }%
\providecommand \@url [1]{\endgroup\@href {#1}{\urlprefix }}%
\providecommand \urlprefix  [0]{URL }%
\providecommand \Eprint [0]{\href }%
\providecommand \doibase [0]{http://dx.doi.org/}%
\providecommand \selectlanguage [0]{\@gobble}%
\providecommand \bibinfo  [0]{\@secondoftwo}%
\providecommand \bibfield  [0]{\@secondoftwo}%
\providecommand \translation [1]{[#1]}%
\providecommand \BibitemOpen [0]{}%
\providecommand \bibitemStop [0]{}%
\providecommand \bibitemNoStop [0]{.\EOS\space}%
\providecommand \EOS [0]{\spacefactor3000\relax}%
\providecommand \BibitemShut  [1]{\csname bibitem#1\endcsname}%
\let\auto@bib@innerbib\@empty
%</preamble>
\bibitem [{\citenamefont {Kubo~R.}\ and\ \citenamefont
  {N.}(1991)}]{Kubo1991statistical}%
  \BibitemOpen
  \bibfield  {author} {\bibinfo {author} {\bibfnamefont {T.~M.}\ \bibnamefont
  {Kubo~R.}}\ and\ \bibinfo {author} {\bibfnamefont {H.}~\bibnamefont {N.}},\
  }\href {https://doi.org/10.1007/978-3-642-58244-8} {\emph {\bibinfo {title}
  {Statistical Physics II Nonequilibrium Statistical Mechanics}}},\ Springer
  Series in Solid State Sciences\ (\bibinfo  {publisher} {Springer, Berlin,
  Heidelberg},\ \bibinfo {year} {1991})\BibitemShut {NoStop}%
\bibitem [{\citenamefont {Wang}\ \emph
  {et~al.}(2019{\natexlab{a}})\citenamefont {Wang}, \citenamefont {Liu},\ and\
  \citenamefont {Li}}]{Wang_2019LTE}%
  \BibitemOpen
  \bibfield  {author} {\bibinfo {author} {\bibfnamefont {L.}~\bibnamefont
  {Wang}}, \bibinfo {author} {\bibfnamefont {S.}~\bibnamefont {Liu}}, \ and\
  \bibinfo {author} {\bibfnamefont {B.}~\bibnamefont {Li}},\ }\href {\doibase
  10.1088/1367-2630/ab34a0} {\bibfield  {journal} {\bibinfo  {journal} {New J.
  Phys.}\ }\textbf {\bibinfo {volume} {21}},\ \bibinfo {pages} {083019}
  (\bibinfo {year} {2019}{\natexlab{a}})}\BibitemShut {NoStop}%
\bibitem [{\citenamefont {Zhang}\ \emph {et~al.}(2020)\citenamefont {Zhang},
  \citenamefont {Ouyang}, \citenamefont {Cheng}, \citenamefont {Chen},
  \citenamefont {Li},\ and\ \citenamefont {Zhang}}]{ZHANG2020}%
  \BibitemOpen
  \bibfield  {author} {\bibinfo {author} {\bibfnamefont {Z.}~\bibnamefont
  {Zhang}}, \bibinfo {author} {\bibfnamefont {Y.}~\bibnamefont {Ouyang}},
  \bibinfo {author} {\bibfnamefont {Y.}~\bibnamefont {Cheng}}, \bibinfo
  {author} {\bibfnamefont {J.}~\bibnamefont {Chen}}, \bibinfo {author}
  {\bibfnamefont {N.}~\bibnamefont {Li}}, \ and\ \bibinfo {author}
  {\bibfnamefont {G.}~\bibnamefont {Zhang}},\ }\href {\doibase
  https://doi.org/10.1016/j.physrep.2020.03.001} {\bibfield  {journal}
  {\bibinfo  {journal} {Phys. Rep.}\ }\textbf {\bibinfo {volume} {860}},\
  \bibinfo {pages} {1 } (\bibinfo {year} {2020})}\BibitemShut {NoStop}%
\bibitem [{\citenamefont {Bae}\ \emph {et~al.}(2013)\citenamefont {Bae},
  \citenamefont {Li}, \citenamefont {Aksamija}, \citenamefont {Martin},
  \citenamefont {Xiong}, \citenamefont {Ong}, \citenamefont {Knezevic},\ and\
  \citenamefont {Pop}}]{bae2013ballistic}%
  \BibitemOpen
  \bibfield  {author} {\bibinfo {author} {\bibfnamefont {M.-H.}\ \bibnamefont
  {Bae}}, \bibinfo {author} {\bibfnamefont {Z.}~\bibnamefont {Li}}, \bibinfo
  {author} {\bibfnamefont {Z.}~\bibnamefont {Aksamija}}, \bibinfo {author}
  {\bibfnamefont {P.~N.}\ \bibnamefont {Martin}}, \bibinfo {author}
  {\bibfnamefont {F.}~\bibnamefont {Xiong}}, \bibinfo {author} {\bibfnamefont
  {Z.-Y.}\ \bibnamefont {Ong}}, \bibinfo {author} {\bibfnamefont
  {I.}~\bibnamefont {Knezevic}}, \ and\ \bibinfo {author} {\bibfnamefont
  {E.}~\bibnamefont {Pop}},\ }\href {\doibase 10.1038/ncomms2755} {\bibfield
  {journal} {\bibinfo  {journal} {Nat. Commun.}\ }\textbf {\bibinfo {volume}
  {4}},\ \bibinfo {pages} {1} (\bibinfo {year} {2013})}\BibitemShut {NoStop}%
\bibitem [{\citenamefont {Xu}\ \emph {et~al.}(2014)\citenamefont {Xu},
  \citenamefont {Pereira}, \citenamefont {Wang}, \citenamefont {Wu},
  \citenamefont {Zhang}, \citenamefont {Zhao}, \citenamefont {Bae},
  \citenamefont {Bui}, \citenamefont {Xie}, \citenamefont {Thong},
  \citenamefont {Hong}, \citenamefont {Loh}, \citenamefont {Donadio},
  \citenamefont {Li},\ and\ \citenamefont
  {{\"O}zyilmaz}}]{xu_length-dependent_2014}%
  \BibitemOpen
  \bibfield  {author} {\bibinfo {author} {\bibfnamefont {X.}~\bibnamefont
  {Xu}}, \bibinfo {author} {\bibfnamefont {L.~F.~C.}\ \bibnamefont {Pereira}},
  \bibinfo {author} {\bibfnamefont {Y.}~\bibnamefont {Wang}}, \bibinfo {author}
  {\bibfnamefont {J.}~\bibnamefont {Wu}}, \bibinfo {author} {\bibfnamefont
  {K.}~\bibnamefont {Zhang}}, \bibinfo {author} {\bibfnamefont
  {X.}~\bibnamefont {Zhao}}, \bibinfo {author} {\bibfnamefont {S.}~\bibnamefont
  {Bae}}, \bibinfo {author} {\bibfnamefont {C.~T.}\ \bibnamefont {Bui}},
  \bibinfo {author} {\bibfnamefont {R.}~\bibnamefont {Xie}}, \bibinfo {author}
  {\bibfnamefont {J.~T.~L.}\ \bibnamefont {Thong}}, \bibinfo {author}
  {\bibfnamefont {B.~H.}\ \bibnamefont {Hong}}, \bibinfo {author}
  {\bibfnamefont {K.~P.}\ \bibnamefont {Loh}}, \bibinfo {author} {\bibfnamefont
  {D.}~\bibnamefont {Donadio}}, \bibinfo {author} {\bibfnamefont
  {B.}~\bibnamefont {Li}}, \ and\ \bibinfo {author} {\bibfnamefont
  {B.}~\bibnamefont {{\"O}zyilmaz}},\ }\href {\doibase 10.1038/ncomms4689}
  {\bibfield  {journal} {\bibinfo  {journal} {Nat. Commun.}\ }\textbf {\bibinfo
  {volume} {5}},\ \bibinfo {pages} {3689} (\bibinfo {year} {2014})}\BibitemShut
  {NoStop}%
\bibitem [{\citenamefont {Li}\ \emph {et~al.}(2012)\citenamefont {Li},
  \citenamefont {Ren}, \citenamefont {Wang}, \citenamefont {Zhang},
  \citenamefont {H\"anggi},\ and\ \citenamefont {Li}}]{RevModPhysLibaowen}%
  \BibitemOpen
  \bibfield  {author} {\bibinfo {author} {\bibfnamefont {N.}~\bibnamefont
  {Li}}, \bibinfo {author} {\bibfnamefont {J.}~\bibnamefont {Ren}}, \bibinfo
  {author} {\bibfnamefont {L.}~\bibnamefont {Wang}}, \bibinfo {author}
  {\bibfnamefont {G.}~\bibnamefont {Zhang}}, \bibinfo {author} {\bibfnamefont
  {P.}~\bibnamefont {H\"anggi}}, \ and\ \bibinfo {author} {\bibfnamefont
  {B.}~\bibnamefont {Li}},\ }\href {\doibase 10.1103/RevModPhys.84.1045}
  {\bibfield  {journal} {\bibinfo  {journal} {Rev. Mod. Phys.}\ }\textbf
  {\bibinfo {volume} {84}},\ \bibinfo {pages} {1045} (\bibinfo {year}
  {2012})}\BibitemShut {NoStop}%
\bibitem [{\citenamefont {Yang}\ \emph {et~al.}(2012)\citenamefont {Yang},
  \citenamefont {Xu}, \citenamefont {Zhang},\ and\ \citenamefont
  {Li}}]{yang2012a}%
  \BibitemOpen
  \bibfield  {author} {\bibinfo {author} {\bibfnamefont {N.}~\bibnamefont
  {Yang}}, \bibinfo {author} {\bibfnamefont {X.}~\bibnamefont {Xu}}, \bibinfo
  {author} {\bibfnamefont {G.}~\bibnamefont {Zhang}}, \ and\ \bibinfo {author}
  {\bibfnamefont {B.}~\bibnamefont {Li}},\ }\href {\doibase 10.1063/1.4773462}
  {\bibfield  {journal} {\bibinfo  {journal} {AIP Adv.}\ }\textbf {\bibinfo
  {volume} {2}},\ \bibinfo {pages} {041410} (\bibinfo {year}
  {2012})}\BibitemShut {NoStop}%
\bibitem [{\citenamefont {Wang}\ \emph {et~al.}(2017)\citenamefont {Wang},
  \citenamefont {Hu}, \citenamefont {Takahashi}, \citenamefont {Zhang},
  \citenamefont {Takamatsu},\ and\ \citenamefont {Chen}}]{wang2017}%
  \BibitemOpen
  \bibfield  {author} {\bibinfo {author} {\bibfnamefont {H.}~\bibnamefont
  {Wang}}, \bibinfo {author} {\bibfnamefont {S.}~\bibnamefont {Hu}}, \bibinfo
  {author} {\bibfnamefont {K.}~\bibnamefont {Takahashi}}, \bibinfo {author}
  {\bibfnamefont {X.}~\bibnamefont {Zhang}}, \bibinfo {author} {\bibfnamefont
  {H.}~\bibnamefont {Takamatsu}}, \ and\ \bibinfo {author} {\bibfnamefont
  {J.}~\bibnamefont {Chen}},\ }\href {\doibase 10.1038/ncomms15843} {\bibfield
  {journal} {\bibinfo  {journal} {Nat. Commun.}\ }\textbf {\bibinfo {volume}
  {8}},\ \bibinfo {pages} {15843} (\bibinfo {year} {2017})}\BibitemShut
  {NoStop}%
\bibitem [{\citenamefont {Wang}\ \emph {et~al.}(2014)\citenamefont {Wang},
  \citenamefont {Vallabhaneni}, \citenamefont {Hu}, \citenamefont {Qiu},
  \citenamefont {Chen},\ and\ \citenamefont {Ruan}}]{wang2014}%
  \BibitemOpen
  \bibfield  {author} {\bibinfo {author} {\bibfnamefont {Y.}~\bibnamefont
  {Wang}}, \bibinfo {author} {\bibfnamefont {A.}~\bibnamefont {Vallabhaneni}},
  \bibinfo {author} {\bibfnamefont {J.}~\bibnamefont {Hu}}, \bibinfo {author}
  {\bibfnamefont {B.}~\bibnamefont {Qiu}}, \bibinfo {author} {\bibfnamefont
  {Y.~P.}\ \bibnamefont {Chen}}, \ and\ \bibinfo {author} {\bibfnamefont
  {X.}~\bibnamefont {Ruan}},\ }\href {\doibase 10.1021/nl403773f} {\bibfield
  {journal} {\bibinfo  {journal} {Nano Lett.}\ }\textbf {\bibinfo {volume}
  {14}},\ \bibinfo {pages} {592} (\bibinfo {year} {2014})}\BibitemShut
  {NoStop}%
\bibitem [{\citenamefont {Rudin}\ \emph {et~al.}(1964)\citenamefont {Rudin}
  \emph {et~al.}}]{rudin1964principles}%
  \BibitemOpen
  \bibfield  {author} {\bibinfo {author} {\bibfnamefont {W.}~\bibnamefont
  {Rudin}} \emph {et~al.},\ }\href@noop {} {\emph {\bibinfo {title} {Principles
  of mathematical analysis}}},\ Vol.~\bibinfo {volume} {3}\ (\bibinfo
  {publisher} {McGraw-hill New York},\ \bibinfo {year} {1964})\BibitemShut
  {NoStop}%
\bibitem [{\citenamefont {Coddington}\ and\ \citenamefont
  {Levinson}(1955)}]{coddington1955theory}%
  \BibitemOpen
  \bibfield  {author} {\bibinfo {author} {\bibfnamefont {E.~A.}\ \bibnamefont
  {Coddington}}\ and\ \bibinfo {author} {\bibfnamefont {N.}~\bibnamefont
  {Levinson}},\ }\href@noop {} {\emph {\bibinfo {title} {Theory of ordinary
  differential equations}}}\ (\bibinfo  {publisher} {Tata McGraw-Hill
  Education},\ \bibinfo {year} {1955})\BibitemShut {NoStop}%
\bibitem [{\citenamefont {Bender}\ and\ \citenamefont
  {Orszag}(2013)}]{bender2013advanced}%
  \BibitemOpen
  \bibfield  {author} {\bibinfo {author} {\bibfnamefont {C.~M.}\ \bibnamefont
  {Bender}}\ and\ \bibinfo {author} {\bibfnamefont {S.~A.}\ \bibnamefont
  {Orszag}},\ }\href@noop {} {\emph {\bibinfo {title} {Advanced mathematical
  methods for scientists and engineers I: Asymptotic methods and perturbation
  theory}}}\ (\bibinfo  {publisher} {Springer Science \& Business Media},\
  \bibinfo {year} {2013})\BibitemShut {NoStop}%
\bibitem [{\citenamefont {S{\"u}li}\ and\ \citenamefont
  {Mayers}(2003)}]{Numericalanalysis}%
  \BibitemOpen
  \bibfield  {author} {\bibinfo {author} {\bibfnamefont {E.}~\bibnamefont
  {S{\"u}li}}\ and\ \bibinfo {author} {\bibfnamefont {D.}~\bibnamefont
  {Mayers}},\ }\href {https://books.google.co.jp/books?id=hj9weaqJTbQC} {\emph
  {\bibinfo {title} {An introduction to numerical analysis}}}\ (\bibinfo
  {publisher} {Cambridge University Press},\ \bibinfo {year}
  {2003})\BibitemShut {NoStop}%
\bibitem [{\citenamefont {Wang}\ \emph
  {et~al.}(2019{\natexlab{b}})\citenamefont {Wang}, \citenamefont {Yang},
  \citenamefont {Chen}, \citenamefont {Li},\ and\ \citenamefont
  {Zhang}}]{wang2019}%
  \BibitemOpen
  \bibfield  {author} {\bibinfo {author} {\bibfnamefont {H.}~\bibnamefont
  {Wang}}, \bibinfo {author} {\bibfnamefont {Y.}~\bibnamefont {Yang}}, \bibinfo
  {author} {\bibfnamefont {H.}~\bibnamefont {Chen}}, \bibinfo {author}
  {\bibfnamefont {N.}~\bibnamefont {Li}}, \ and\ \bibinfo {author}
  {\bibfnamefont {L.}~\bibnamefont {Zhang}},\ }\href {\doibase
  10.1103/PhysRevE.99.062111} {\bibfield  {journal} {\bibinfo  {journal} {Phys.
  Rev. E}\ }\textbf {\bibinfo {volume} {99}},\ \bibinfo {pages} {062111}
  (\bibinfo {year} {2019}{\natexlab{b}})}\BibitemShut {NoStop}%
\bibitem [{\citenamefont {Chen}\ \emph {et~al.}(2018)\citenamefont {Chen},
  \citenamefont {Wang}, \citenamefont {Yang}, \citenamefont {Li},\ and\
  \citenamefont {Zhang}}]{chen2018b}%
  \BibitemOpen
  \bibfield  {author} {\bibinfo {author} {\bibfnamefont {H.}~\bibnamefont
  {Chen}}, \bibinfo {author} {\bibfnamefont {H.}~\bibnamefont {Wang}}, \bibinfo
  {author} {\bibfnamefont {Y.}~\bibnamefont {Yang}}, \bibinfo {author}
  {\bibfnamefont {N.}~\bibnamefont {Li}}, \ and\ \bibinfo {author}
  {\bibfnamefont {L.}~\bibnamefont {Zhang}},\ }\href {\doibase
  10.1103/PhysRevE.98.032131} {\bibfield  {journal} {\bibinfo  {journal} {Phys.
  Rev. E}\ }\textbf {\bibinfo {volume} {98}},\ \bibinfo {pages} {032131}
  (\bibinfo {year} {2018})}\BibitemShut {NoStop}%
\bibitem [{\citenamefont {Mao}\ \emph {et~al.}(2008)\citenamefont {Mao},
  \citenamefont {Li},\ and\ \citenamefont {Deng}}]{mao2008heat}%
  \BibitemOpen
  \bibfield  {author} {\bibinfo {author} {\bibfnamefont {J.-W.}\ \bibnamefont
  {Mao}}, \bibinfo {author} {\bibfnamefont {Y.-Q.}\ \bibnamefont {Li}}, \ and\
  \bibinfo {author} {\bibfnamefont {L.-Y.}\ \bibnamefont {Deng}},\ }\href
  {\doibase 10.1142/S0217979208048693} {\bibfield  {journal} {\bibinfo
  {journal} {Int. J. Mor. Phys. B}\ }\textbf {\bibinfo {volume} {22}},\
  \bibinfo {pages} {3901} (\bibinfo {year} {2008})}\BibitemShut {NoStop}%
\bibitem [{\citenamefont {Go}\ and\ \citenamefont {Sen}(2010)}]{go2010}%
  \BibitemOpen
  \bibfield  {author} {\bibinfo {author} {\bibfnamefont {D.~B.}\ \bibnamefont
  {Go}}\ and\ \bibinfo {author} {\bibfnamefont {M.}~\bibnamefont {Sen}},\
  }\href {\doibase 10.1115/1.4002286} {\bibfield  {journal} {\bibinfo
  {journal} {J. Heat Transfer}\ }\textbf {\bibinfo {volume} {132}},\ \bibinfo
  {pages} {124502} (\bibinfo {year} {2010})}\BibitemShut {NoStop}%
\bibitem [{\citenamefont {{Criado-Sancho}}\ \emph {et~al.}(2013)\citenamefont
  {{Criado-Sancho}}, \citenamefont {Alvarez},\ and\ \citenamefont
  {Jou}}]{criado-sancho2013}%
  \BibitemOpen
  \bibfield  {author} {\bibinfo {author} {\bibfnamefont {M.}~\bibnamefont
  {{Criado-Sancho}}}, \bibinfo {author} {\bibfnamefont {F.~X.}\ \bibnamefont
  {Alvarez}}, \ and\ \bibinfo {author} {\bibfnamefont {D.}~\bibnamefont
  {Jou}},\ }\href {\doibase 10.1063/1.4816685} {\bibfield  {journal} {\bibinfo
  {journal} {J. Appl. Phys.}\ }\textbf {\bibinfo {volume} {114}},\ \bibinfo
  {pages} {053512} (\bibinfo {year} {2013})}\BibitemShut {NoStop}%
\bibitem [{\citenamefont {Naso}\ \emph {et~al.}(2019)\citenamefont {Naso},
  \citenamefont {Vuk},\ and\ \citenamefont {Zullo}}]{naso2019}%
  \BibitemOpen
  \bibfield  {author} {\bibinfo {author} {\bibfnamefont {M.~G.}\ \bibnamefont
  {Naso}}, \bibinfo {author} {\bibfnamefont {E.}~\bibnamefont {Vuk}}, \ and\
  \bibinfo {author} {\bibfnamefont {F.}~\bibnamefont {Zullo}},\ }\href
  {\doibase 10.1016/j.ijheatmasstransfer.2019.118520} {\bibfield  {journal}
  {\bibinfo  {journal} {Int. J.Heat Mass Transfer}\ }\textbf {\bibinfo {volume}
  {143}},\ \bibinfo {pages} {118520} (\bibinfo {year} {2019})}\BibitemShut
  {NoStop}%
\bibitem [{\citenamefont {Alvarez}\ \emph {et~al.}(2010)\citenamefont
  {Alvarez}, \citenamefont {Jou},\ and\ \citenamefont
  {Sellitto}}]{hydrodynamicporeousAFX2010}%
  \BibitemOpen
  \bibfield  {author} {\bibinfo {author} {\bibfnamefont {F.~X.}\ \bibnamefont
  {Alvarez}}, \bibinfo {author} {\bibfnamefont {D.}~\bibnamefont {Jou}}, \ and\
  \bibinfo {author} {\bibfnamefont {A.}~\bibnamefont {Sellitto}},\ }\href
  {\doibase 10.1063/1.3462936} {\bibfield  {journal} {\bibinfo  {journal}
  {Appl. Phys. Lett.}\ }\textbf {\bibinfo {volume} {97}},\ \bibinfo {pages}
  {033103} (\bibinfo {year} {2010})}\BibitemShut {NoStop}%
\bibitem [{\citenamefont {Holland}(1963)}]{holland1963analysis}%
  \BibitemOpen
  \bibfield  {author} {\bibinfo {author} {\bibfnamefont {M.~G.}\ \bibnamefont
  {Holland}},\ }\href {\doibase 10.1103/PhysRev.132.2461} {\bibfield  {journal}
  {\bibinfo  {journal} {Phys. Rev.}\ }\textbf {\bibinfo {volume} {132}},\
  \bibinfo {pages} {2461} (\bibinfo {year} {1963})}\BibitemShut {NoStop}%
\bibitem [{\citenamefont {Brockhouse}(1959)}]{brockhouse1959lattice}%
  \BibitemOpen
  \bibfield  {author} {\bibinfo {author} {\bibfnamefont {B.~N.}\ \bibnamefont
  {Brockhouse}},\ }\href {\doibase 10.1103/PhysRevLett.2.256} {\bibfield
  {journal} {\bibinfo  {journal} {Phys. Rev. Lett.}\ }\textbf {\bibinfo
  {volume} {2}},\ \bibinfo {pages} {256} (\bibinfo {year} {1959})}\BibitemShut
  {NoStop}%
\bibitem [{\citenamefont {Glassbrenner}\ and\ \citenamefont
  {Slack}(1964)}]{Glassbrenner64conductivity}%
  \BibitemOpen
  \bibfield  {author} {\bibinfo {author} {\bibfnamefont {C.~J.}\ \bibnamefont
  {Glassbrenner}}\ and\ \bibinfo {author} {\bibfnamefont {G.~A.}\ \bibnamefont
  {Slack}},\ }\href {\doibase 10.1103/PhysRev.134.A1058} {\bibfield  {journal}
  {\bibinfo  {journal} {Phys. Rev.}\ }\textbf {\bibinfo {volume} {134}},\
  \bibinfo {pages} {A1058} (\bibinfo {year} {1964})}\BibitemShut {NoStop}%
\bibitem [{\citenamefont {Flubacher}\ \emph {et~al.}(1959)\citenamefont
  {Flubacher}, \citenamefont {Leadbetter},\ and\ \citenamefont
  {Morrison}}]{CpSiGeExpriment1959}%
  \BibitemOpen
  \bibfield  {author} {\bibinfo {author} {\bibfnamefont {P.}~\bibnamefont
  {Flubacher}}, \bibinfo {author} {\bibfnamefont {A.~J.}\ \bibnamefont
  {Leadbetter}}, \ and\ \bibinfo {author} {\bibfnamefont {J.~A.}\ \bibnamefont
  {Morrison}},\ }\href {\doibase 10.1080/14786435908233340} {\bibfield
  {journal} {\bibinfo  {journal} {Philos. Mag.}\ }\textbf {\bibinfo {volume}
  {4}},\ \bibinfo {pages} {273} (\bibinfo {year} {1959})}\BibitemShut {NoStop}%
\bibitem [{\citenamefont {Li}\ \emph {et~al.}(2015)\citenamefont {Li},
  \citenamefont {Takahashi}, \citenamefont {Ago}, \citenamefont {Zhang},
  \citenamefont {Ikuta}, \citenamefont {Nishiyama},\ and\ \citenamefont
  {Kawahara}}]{japgraphene2015}%
  \BibitemOpen
  \bibfield  {author} {\bibinfo {author} {\bibfnamefont {Q.-Y.}\ \bibnamefont
  {Li}}, \bibinfo {author} {\bibfnamefont {K.}~\bibnamefont {Takahashi}},
  \bibinfo {author} {\bibfnamefont {H.}~\bibnamefont {Ago}}, \bibinfo {author}
  {\bibfnamefont {X.}~\bibnamefont {Zhang}}, \bibinfo {author} {\bibfnamefont
  {T.}~\bibnamefont {Ikuta}}, \bibinfo {author} {\bibfnamefont
  {T.}~\bibnamefont {Nishiyama}}, \ and\ \bibinfo {author} {\bibfnamefont
  {K.}~\bibnamefont {Kawahara}},\ }\href {\doibase 10.1063/1.4907699}
  {\bibfield  {journal} {\bibinfo  {journal} {J. Appl. Phys.}\ }\textbf
  {\bibinfo {volume} {117}},\ \bibinfo {pages} {065102} (\bibinfo {year}
  {2015})}\BibitemShut {NoStop}%
\bibitem [{\citenamefont {Dames}(2009)}]{dames2009}%
  \BibitemOpen
  \bibfield  {author} {\bibinfo {author} {\bibfnamefont {C.}~\bibnamefont
  {Dames}},\ }\href {\doibase 10.1115/1.3089552} {\bibfield  {journal}
  {\bibinfo  {journal} {J. Heat Transfer}\ }\textbf {\bibinfo {volume} {131}},\
  \bibinfo {pages} {061301} (\bibinfo {year} {2009})}\BibitemShut {NoStop}%
\bibitem [{\citenamefont {Kobayashi}\ \emph {et~al.}(2009)\citenamefont
  {Kobayashi}, \citenamefont {Teraoka},\ and\ \citenamefont
  {Terasaki}}]{kobayashi2009}%
  \BibitemOpen
  \bibfield  {author} {\bibinfo {author} {\bibfnamefont {W.}~\bibnamefont
  {Kobayashi}}, \bibinfo {author} {\bibfnamefont {Y.}~\bibnamefont {Teraoka}},
  \ and\ \bibinfo {author} {\bibfnamefont {I.}~\bibnamefont {Terasaki}},\
  }\href {\doibase 10.1063/1.3253712} {\bibfield  {journal} {\bibinfo
  {journal} {Appl. Phys. Lett.}\ }\textbf {\bibinfo {volume} {95}},\ \bibinfo
  {pages} {171905} (\bibinfo {year} {2009})}\BibitemShut {NoStop}%
\bibitem [{\citenamefont {Peyrard}(2006)}]{peyrard2006}%
  \BibitemOpen
  \bibfield  {author} {\bibinfo {author} {\bibfnamefont {M.}~\bibnamefont
  {Peyrard}},\ }\href {\doibase 10.1209/epl/i2006-10223-5} {\bibfield
  {journal} {\bibinfo  {journal} {EPL}\ }\textbf {\bibinfo {volume} {76}},\
  \bibinfo {pages} {49} (\bibinfo {year} {2006})}\BibitemShut {NoStop}%
\bibitem [{\citenamefont {Yang}\ \emph {et~al.}(2018)\citenamefont {Yang},
  \citenamefont {Chen}, \citenamefont {Wang}, \citenamefont {Li},\ and\
  \citenamefont {Zhang}}]{PhysRevE.98.042131}%
  \BibitemOpen
  \bibfield  {author} {\bibinfo {author} {\bibfnamefont {Y.}~\bibnamefont
  {Yang}}, \bibinfo {author} {\bibfnamefont {H.}~\bibnamefont {Chen}}, \bibinfo
  {author} {\bibfnamefont {H.}~\bibnamefont {Wang}}, \bibinfo {author}
  {\bibfnamefont {N.}~\bibnamefont {Li}}, \ and\ \bibinfo {author}
  {\bibfnamefont {L.}~\bibnamefont {Zhang}},\ }\href {\doibase
  10.1103/PhysRevE.98.042131} {\bibfield  {journal} {\bibinfo  {journal} {Phys.
  Rev. E}\ }\textbf {\bibinfo {volume} {98}},\ \bibinfo {pages} {042131}
  (\bibinfo {year} {2018})}\BibitemShut {NoStop}%
\end{thebibliography}%

\end{document}

\title{Perturbation theory of thermal rectification: Supplemental Material}

\author{Chuang Zhang}
\email{zhangcmzt@hust.edu.cn}
\affiliation{%
 State Key Laboratory of Coal Combustion, School of Energy and Power Engineering, Huazhong University of Science and Technology, Wuhan 430074, China}%
\author{Meng An}
\email{anmeng@sust.edu.cn}
\affiliation{%
College of Mechanical and Electrical Engineering, Shaanxi University of Science and Technology, 6 Xuefuzhong Road, Weiyangdaxueyuan, Xi’an 710021, China}
\author{Zhaoli Guo}%
\email{zlguo@hust.edu.cn}
\affiliation{%
 State Key Laboratory of Coal Combustion, School of Energy and Power Engineering, Huazhong University of Science and Technology, Wuhan 430074, China}%
\author{Songze Chen}
\email{Corresponding author: jacksongze@hust.edu.cn}
\affiliation{%
 State Key Laboratory of Coal Combustion, School of Energy and Power Engineering, Huazhong University of Science and Technology, Wuhan 430074, China}%
\date{\today}
\maketitle

\section{Physical descriptions}

Suppose that an effective thermal conductivity $\kappa_e$ can be identified in a thermal system where the local thermal equilibrium~\cite{Kubo1991statistical,Wang_2019LTE} is satisfied, so that the Fourier law is satisfied formally,
\begin{align}
q = -\kappa_e (W, T, L) \frac{dT}{dx}, \label{eq:siconductivity}
\end{align}
where $q$ is heat flux, $T$ is the temperature, $x$ is the spatial position, $W$ is the local physical quantity.
The physical quantities except for the temperature that influence the effective thermal conductivity $\kappa_e (W, T, L)$ are grouped into two categories.
The first category stands for the local physical quantity varying with position, i.e., $W=W(x)$, for instance, the characteristic length in other directions, porosity.
The other one represents a kind of global physical quantity like the system length $L$.
Consider a one-dimensional thermal conduction system with spatial position $x \in [x_0-L/2, x_0+L/2]$, the heat conduction at steady state satisfies,
\begin{align}
\frac{\partial q}{\partial x} = 0, \label{eq:silaplacian}
\end{align}
where $x_0$ is the central position of this system, $L$ is the system length.
Two temperatures ($T_0-\Delta T/2, T_0+\Delta T/2$) are imposed at the two ends.
Then the thermal conductivity in Eq.~\eqref{eq:siconductivity} can be written as
\begin{align}
\kappa_e =\kappa_e (W(x), T, L)=\kappa_e (x, T, L).
\end{align}
Note that as the system length $L$ of thermal system is comparable to the phonon mean free path, $\kappa_e(L)$ may change as the system length $L$ changes~\cite{ZHANG2020,bae2013ballistic,xu_length-dependent_2014,RevModPhysLibaowen,yang2012a,wang2017,wang2014}.

Given that $\kappa_e(x,T, L)$ is differentiable in the whole thermal system (or changes smoothly and slightly), then it can be approximated by the Taylor expansion at $(x_0,T_0)$~\cite{rudin1964principles}, i.e.,
\begin{align}
\kappa_e (x,T) &= \kappa_0 + a (T-T_0) +b (x-x_0) + c(x-x_0)(T-T_0) +d(x-x_0)^2+ f(T-T_0)^2 , \label{eq:sitaylorsm}
\end{align}
where the high order Taylor expansion terms are assumed to be ignorable in Eq.~\eqref{eq:sitaylorsm} and
\begin{align}
\kappa_0 &=\kappa_e (x_0,T_0, L) \neq 0 , &\quad
a &= \left.\ \frac{\partial \kappa_e }{\partial T} \right|_{(x,T)=(x_0,T_0)}, &\quad
b &= \left.\  \left( \frac{\partial \kappa_e }{\partial x}  \right)  \right|_{(x,T)=(x_0,T_0)} ,  \\
c &= \left.\  \left( \frac{\partial^2 \kappa_e }{\partial x \partial T}  \right)  \right|_{(x,T)=(x_0,T_0)} , &\quad
d &=\frac{1}{2} \left.\  \left( \frac{\partial^2 \kappa_e }{\partial x^2}  \right)  \right|_{(x,T)=(x_0,T_0)} ,  &\quad
f &= \frac{1}{2} \left.\  \left( \frac{\partial^2 \kappa_e }{\partial T^2}  \right)  \right|_{(x,T)=(x_0,T_0)} .  \label{eq:sifsm}
\end{align}
Note that all the partial derivatives of the effective thermal conductivity in this work are calculated at $(x_0,T_0)$ according to Taylor expansion theory~\cite{rudin1964principles}.
In addition, above equations are solved with two sets of
boundary conditions respectively,
\begin{align}
\text{forward (`+')}:\quad T( x_0-L/2 ) &= T_0-\Delta T/2, &\quad
  T( x_0+L/2 ) &= T_0+\Delta T/2 ,\label{eq:forwardBc} \\
\text{backward (`-')}:\quad T( x_0-L/2 ) &= T_0+\Delta T/2,  &\quad
T( x_0+L/2 ) &= T_0-\Delta T/2 . \label{eq:backwardBc}
\end{align}

For the forward direction (Eq.~\eqref{eq:forwardBc}), the heat flux $q$ is denoted by $q_{+}$, and as the temperature gradient is reversed (Eq.~\eqref{eq:backwardBc}), we denote $q$ as $q_{-}$.
In what follows, all variables $V$ are labeled as `$V_{+}$' for the forward direction and `$V_{-}$' for the backward direction.
The thermal rectification ratio $\beta$ is
\begin{align}
\beta = \frac{q_{+} +q_{-} }{q_{+} - q_{-} }.
\label{eq:definitionbeta}
\end{align}

\section{Dimensionless treatments}

Before solving above equations analytically, the dimensionless treatment of Eq.~\eqref{eq:sitaylorsm} is implemented first, i.e.,
\begin{align}
\kappa^{*}_e= 1+\alpha_T T^{*}+\alpha_x x^{*}+\alpha_{xT} x^{*}T^{*} + \alpha_{x^2}  x^{*2}  + \alpha_{T^2} T^{*2}, \label{eq:sinormalizedtaylorsm}
\end{align}
where
\begin{align}
\kappa^*_e &=\frac{\kappa_e}{ \kappa_0 },  &\quad
T^*&=\frac{T-T_0}{\Delta T}, &\quad
x^*&=\frac{x-x_0}{L}, &\quad  \\
\alpha_x & = \frac{ L }{ \kappa_0} \left.\ \frac{\partial \kappa_e }{\partial x} \right|_{(x,T)=(x_0,T_0)},  &\quad
\alpha_T & = \frac{\Delta T }{ \kappa_0} \left.\ \frac{\partial \kappa_e }{\partial T} \right|_{(x,T)=(x_0,T_0)} , &\quad
\alpha_{xT} &= \frac{ L \Delta T}{ \kappa_0} \left.\ \frac{\partial^2 \kappa_e }{\partial T \partial x} \right|_{(x,T)=(x_0,T_0)}, \\
\alpha_{x^2} &= \frac{L ^2}{ 2\kappa_0} \left.\ \frac{\partial^2 \kappa_e }{\partial x^2} \right|_{(x,T)=(x_0,T_0)}, &\quad
\alpha_{T^2} &= \frac{ \Delta T ^2 }{ 2\kappa_0} \left.\ \frac{\partial^2 \kappa_e }{\partial T^2} \right|_{(x,T)=(x_0,T_0)} .
\end{align}
%Equation~\eqref{eq:sinormalizedtaylorsm} is valid as
Here we assume that
\begin{align}
\quad \alpha_x &\rightarrow 0, &\quad \alpha_T  &\rightarrow 0 , &\quad \alpha_{xT}  & \rightarrow 0, \notag \\
\alpha_{x^2}  & \rightarrow 0, &\quad  \alpha_{T^2} & \rightarrow 0. &
\label{eq:silimitation11}
\end{align}
Similar treatment is implemented on Eq.~\eqref{eq:siconductivity}, i.e.,
\begin{align}
q^*=- \kappa^*_e \frac{dT^*}{dx^*},
\label{eq:sinondimensionalconductivity}
\end{align}
where the normalized heat flux is $q^*=(qL)/ ( \kappa_0 \Delta T) $.
Combining Eqs.~\eqref{eq:sinondimensionalconductivity} and~\eqref{eq:sinormalizedtaylorsm} leads to
\begin{align}
q^* = - \left(  1+\alpha_T T^{*}+ \alpha_x  x^{*}+ \alpha_{xT} x^{*} T^{*}
 + \alpha_{x^2}  x^{*2}   + \alpha_{T^2} T^{*2} \right)  \frac{dT^*}{dx^*}, \label{eq:sipartialeqsm}
\end{align}
which is a high-order nonlinear differential equation of $x^*$ and $T^*$~\cite{coddington1955theory}.
Based on Eqs.~\eqref{eq:forwardBc} and~\eqref{eq:backwardBc}, we have dimensionless boundary conditions, i.e.,
\begin{align}
\text{forward (`+')}:\quad T( -1/2 ) &= -1/2, &\quad
  T( 1/2 ) &= 1/2 ,\label{eq:lessforwardBc} \\
\text{backward (`-')}:\quad T( -1/2 ) &= 1/2, &\quad
  T( 1/2 ) &= -1/2 . \label{eq:lessbackwardBc}
\end{align}
For the forward direction, the normalized heat flux $q^*$ is denoted by $q^*_{+}$, and as the temperature gradient is reversed, we denote $q^*$ as $q^*_{-}$.
Then Eq.~\eqref{eq:definitionbeta} becomes
\begin{align}
\beta = \frac{q_{+}^* +q_{-}^* }{q_{+}^* - q_{-}^* }.
\end{align}

First, we consider the forward direction so that $\left( \frac{1}{2}, \frac{1}{2} \right)$ and $\left( -\frac{1}{2}, -\frac{1}{2} \right)$ are both the solutions of Eqs.~\eqref{eq:sinormalizedtaylorsm} and~\eqref{eq:sinondimensionalconductivity}.
Then similar derivations can be implemented for the backward direction directly.

\section{Perturbation theory of the thermal rectification}

In order to solve the following equation,
\begin{align}
q^* = - \left(  1+\alpha_T T^{*}+ \alpha_x  x^{*}+ \alpha_{xT} x^{*} T^{*}
 + \alpha_{x^2}  x^{*2}  + \alpha_{T^2} T^{*2} \right)  \frac{dT^*}{dx^*}, \label{eq:sipartialeqsm}
\end{align}
the perturbation method~\cite{bender2013advanced} is used.
Firstly, we convert Eq.~\eqref{eq:sipartialeqsm} into a perturbation problem by introducing a parameter $\epsilon$ in the right side of the equation, i.e,
\begin{align}
q^* = - \left(  1+\alpha_T T^{*}+\epsilon \left( \alpha_x  x^{*}+ \alpha_{xT} x^{*} T^{*}
 + \alpha_{x^2}  x^{*2}  + \alpha_{T^2} T^{*2} \right) \right)  \frac{dT^*}{dx^*}. \label{eq:pertubation11}
\end{align}
It can be found that Eq.~\eqref{eq:pertubation11} returns to the original equation~\eqref{eq:sipartialeqsm} by assigning $\epsilon=1$.
We further assume a perturbation series in powers of $\epsilon$, i.e.,
\begin{align}
T^* &= T_0^* + \epsilon T_1^* +\epsilon^2 T_2^*+... =\sum_{i=0}^{\infty} \epsilon^i T_i^*,\label{eq:Tpertubation} \\
q^*&= q_0^* + \epsilon q_1^* +\epsilon^2 q_2^* +...=\sum_{i=0}^{\infty} \epsilon^i q_i^*.
\label{eq:qpertubation}
\end{align}

The zeroth-order problem is obtained by setting $\epsilon=0$, i.e.,
\begin{align}
q_0^* =-(1+ \alpha_T T_0^*) \frac{dT_0^*}{dx^*}, \label{eq:epsilonzero}
\end{align}
which is an ordinary differential equation of $x^*$ and $T_0^*$.
The constant variation method~\cite{coddington1955theory} is used to solve Eq.~\eqref{eq:epsilonzero} and the analytical solution is
\begin{align}
x^*= - \frac{T_0^* + \frac{\alpha_T}{2} T_0^{*2} +C_0 }{q_0^*}, \label{eq:xzero}
\end{align}
where $C_0$ is a constant.
Combining the boundary conditions (Eqs.~\eqref{eq:lessforwardBc} and~\eqref{eq:lessbackwardBc}),
%\begin{align}
%\text{forward ('+')}:\quad T_0^*( -1/2 ) &= -1/2, \quad T_0^*( 1/2 ) = 1/2 ,\label{eq:lessforwardBc} \\
%\text{backward ('-')}:\quad T_0^*( -1/2 ) &= 1/2, \quad T_0^*( 1/2 ) = -1/2, \label{eq:lessbackwardBc}
%\end{align}
we can determine the heat flux $q_0^*$ and integration constant $C_0$ for the zeroth-order solution,
\begin{align}
q_{0+}^* &= -1, \quad C_{0+} = -\frac{\alpha_T}{8} , \label{eq:qzero1}\\
q_{0-}^* &= 1, \quad C_{0-} = -\frac{\alpha_T}{8}. \label{eq:qzero2}
\end{align}

The first order problem is then obtained by equating the coefficient of $\epsilon$ on the left and right hand sides of Eq.~\eqref{eq:pertubation11}, i.e.,
\begin{align}
q_1^* &=-(1+\alpha_T T_0^*)\frac{dT_1^*}{dx^*}
-\left( \alpha_T T_1^* +\alpha_x x^*+\alpha_{xT} x^* T_0^* +\alpha_{x^2} x^{*2} +\alpha_{T^2} T_0^{*2}   \right) \frac{dT_0^*}{dx^*}.
\label{eq:firstpertubation}
\end{align}
Substitute Eqs.~\eqref{eq:epsilonzero} into Eq.~\eqref{eq:firstpertubation} to eliminate the derivative of $x^*$, and obtain an ordinary differential equation,
\begin{align}
\frac{dT_1^*}{dT_0^*}& =-\frac{\alpha_T  T_1^*}{1+\alpha_T T_0^*} + \frac{q_1^*}{q_0^*} - \frac{ \alpha_x x^*+\alpha_{xT} x^* T_0^* +\alpha_{x^2} x^{*2} +\alpha_{T^2} T_0^{*2}   }{1+\alpha_T T_0^* } ,
\label{eq:firsteq}
\end{align}
This equation is solved by the constant variation method~\cite{coddington1955theory}, and the analytical solution is
\begin{align}
T_1^* =\frac{1}{1+\alpha_T  T_0^*}  g(T_0^*), \label{eq:answerfirst}
\end{align}
where $g(T_0^*)$ is a function of $T_0^*$ and satisfies
\begin{align}
\frac{\partial{g(T_0^*)}  }{\partial{T_0^*} }  = \frac{q_1^*}{q_0^*} (1+\alpha_T T_0^*) -\left(  \alpha_x x^*+\alpha_{xT} x^* T_0^* +\alpha_{x^2} x^{*2} +\alpha_{T^2} T_0^{*2}   \right).
\label{eq:functiong}
\end{align}
Substitute \cref{eq:xzero} into \cref{eq:functiong} and integrate with respect to $T_0^*$, then $g(T_0^*)$ reads,
\begin{align}
g(T_0^*) &= A_0 T_0^* +\frac{A_1}{2}T_0^{*2} +\frac{A_2}{3} T_0^{*3}
+\frac{A_3}{4}T_0^{*4} +\frac{A_4}{5}T_0^{*5} + C_1,
\label{eq:gfunctionvalue}
\end{align}
where
\begin{align}
A_0 &= \frac{q_1^*}{q_0^*}  +\frac{\alpha_x \alpha_T}{8}  -\frac{\alpha_T ^2 \alpha_{x^2}   }{64}, \label{eq:A0} \\
A_1 &= \alpha_T \frac{q_1^*}{q_0^*} -\alpha_x +\frac{\alpha_{xT} \alpha_T}{8} +\frac{\alpha_T  \alpha_{x^2}   }{4} , \label{eq:A1} \\
A_2 &= -\frac{\alpha_x \alpha_T}{2} -\alpha_{xT} -\alpha_{x^2} -\alpha_{T^2} + \frac{\alpha_T ^2 \alpha_{x^2}   }{8},\label{eq:A2} \\
A_3 &= \frac{\alpha_{xT} \alpha_T}{2}  - \alpha_T  \alpha_{x^2}, \label{eq:A3} \\
A_4 &= -\frac{\alpha_T ^2 \alpha_{x^2}   }{4} ,\label{eq:A4}
\end{align}
$q_1^*$ and $C_1$ are integration constants.
And then the boundary conditions for $T_1^*$,
\begin{align}
g \left( -\frac{1}{2} \right)=0, \quad  g \left( \frac{1}{2} \right)=0, \label{eq:bcsp22}
\end{align}
are imposed to determine $q_1^*$ (and $C_1$),
\begin{align}
q_1^* &=-\frac{q_0^*}{12} \left(\alpha_x \alpha_T - \alpha_{xT} -\alpha_{x^2 }-\alpha_{T^2} -\frac{1}{10} \alpha_T ^2 \alpha_{x^2}  \right)
\label{eq:firstq}
\end{align}
According to Eq.~\eqref{eq:qpertubation}, the heat flux can be approximated as $q^* \approx q_0^*+ \epsilon q_1^*$ with $\epsilon = 1$. Then the forward and backward heat flux are expressed as follows,
\begin{align}
q^*_{+}=& -1 + \frac{1}{12} \left(\alpha_x \alpha_T - \alpha_{xT} -\alpha_{x^2 }-\alpha_{T^2} -\frac{1}{10} \alpha_T^2 \alpha_{x^2}  \right),
 \label{eq:heatflux22} \\
q^*_{-}=& 1 + \frac{1}{12} \left(\alpha_x \alpha_T - \alpha_{xT} + \alpha_{x^2 } + \alpha_{T^2}  + \frac{1}{10} \alpha_T^2 \alpha_{x^2}  \right).
 \label{eq:heatflux33}
\end{align}

Combining Eqs.~\eqref{eq:heatflux22} and~\eqref{eq:heatflux33}, the thermal rectification ratio is approximated as,
\begin{align}
\beta \approx  \frac{L\Delta T }{12} \left.\  \left( \frac{1}{\kappa_0}\frac{\partial^2 \kappa_e}{\partial x \partial T}-\frac{1}{\kappa_0^2}\frac{\partial \kappa_e}{\partial x}\frac{\partial \kappa_e}{\partial T} \right) \right|_{(x,T)=(x_0,T_0)} = \frac{1}{12}(\alpha_{xT} -\alpha_x \alpha_T).
\label{eq:siCVMsq15}
\end{align}

\begin{figure}
 \centering
 \includegraphics[scale=0.33,viewport=0 150 900 760,clip=true]{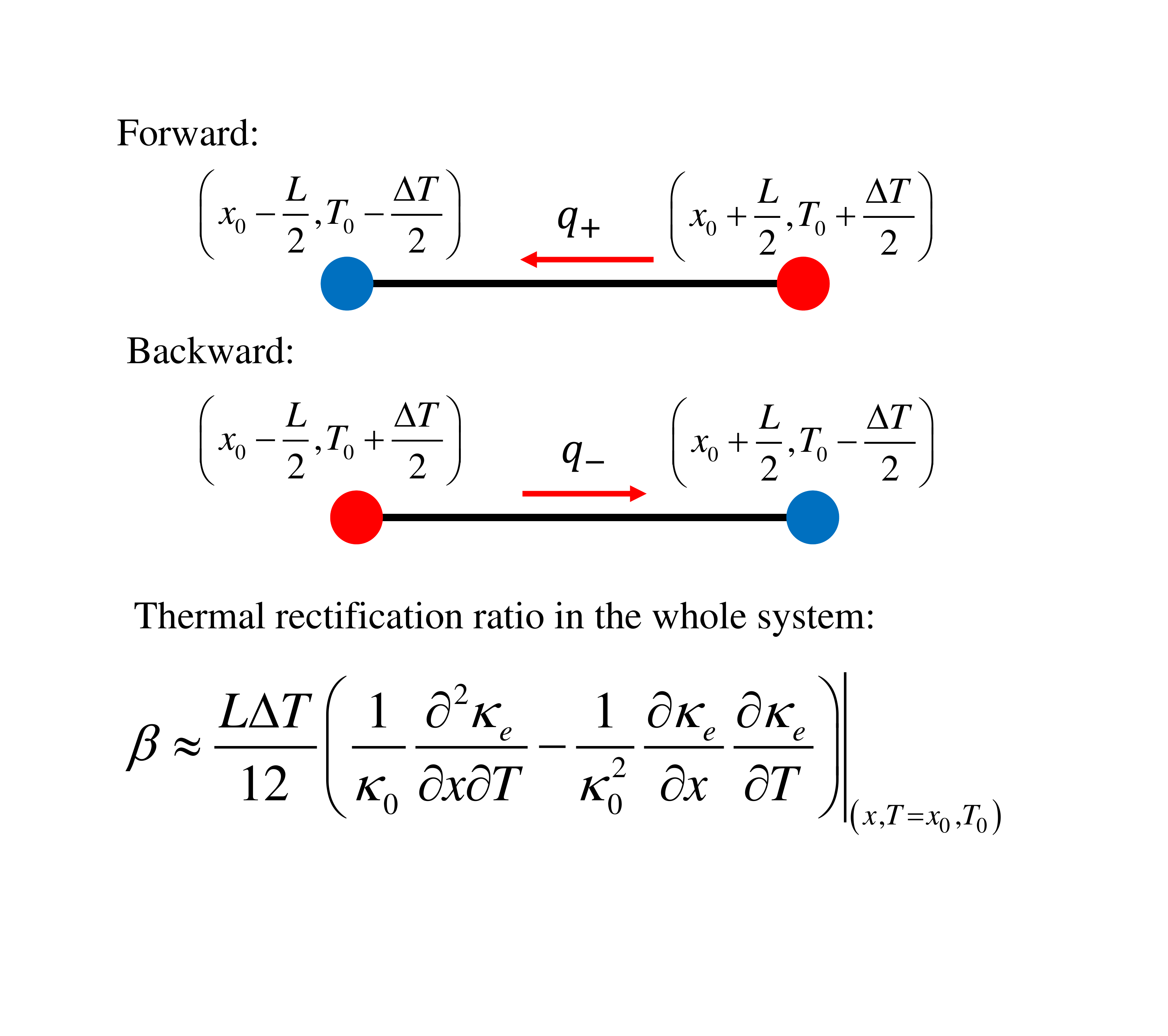}
\caption{In a thermal system with total system length $L$ coupled to two heat baths at $T_H=T_0 + \Delta T/2$ and $T_L=T_0-\Delta T/2$, given a differentiable effective thermal conductivity $\kappa_e (x,T, L)$ in the whole thermal system, the thermal rectification ratio $\beta$ in this system can be derived analytically (Eq.~\eqref{eq:siCVMsq15}) based on perturbation method~\cite{bender2013advanced} or Taylor expansion~\cite{rudin1964principles}, where $\kappa_0 =\kappa_e (x_0,T_0,L)$. The theoretical formula requires that the effective thermal conductivity changes smoothly and slightly in this system. }
 \label{keywork}
\end{figure}
Equation~\eqref{eq:siCVMsq15} is the central result of present study (see~\cref{keywork}).
Based on theoretical constraints and the numerical validations, Eq.~\eqref{eq:siCVMsq15} is valid as the effective thermal conductivity $\kappa_e(x,T,L)$ changes smoothly and slightly in the whole system or these dimensionless parameters should be small.
Note that small $\alpha_x$ ($\alpha_T$) are not equivalent to small $L$ ($\Delta T$).

Based on Eq.~\eqref{eq:siCVMsq15}, it is interesting to find that the leading order term of the thermal rectification is a cross term ($L \Delta T$).
It is stemmed from the linear term and the cross term of the expansion of $\kappa_e^*$ according to Eq.~\eqref{eq:sinormalizedtaylorsm}.
And the other second-order terms have no impact on the leading order term of the thermal rectification.
%These three dimensionless parameters denote the relative change of the effective thermal conductivity throughout the whole system due to the temperature change and the heterogeneity of the other physical properties.

\section{Solving the linear mode with the Taylor expansion}

It is very complex to obtain the thermal rectification ratio of \cref{eq:sipartialeqsm} through the Taylor expansion.
We only show how to solve the linear mode as an illustration. The dimensionless form of the linear mode reads,
\begin{align}
\frac{dq^*}{dx^*} = 0, \quad q^* = - \left( 1+\alpha_T T^{*}+ \alpha_x  x^{*}\right)  \frac{dT^*}{dx^*}. \label{eq:linearModeEqn}
\end{align}
Its analytical solution is derived by the constant variation method~\cite{coddington1955theory}, i.e.,
\begin{equation}
x^*= - \frac{\alpha_T}{\alpha_x} T^*  +\frac{\alpha_T}{\alpha_x ^2}  q^*- \frac{1}{\alpha_x}  + C_1 \exp{\left( -\alpha_x T^*/q^* \right)  }, \label{eq:taylorsiCVMs3}
\end{equation}
where $q^*$ and $C$ are integration constant.
%The boundary conditions are expressed as follows,
%\begin{align}
%\text{forward ('+')}:\quad T^*( -1/2 ) &= -1/2, \quad T^*( 1/2 ) = 1/2,\label{eq:forwardBc} \\
%\text{backward ('-')}:\quad T^*( -1/2 ) &= 1/2, \quad T^*( 1/2 ) = -1/2. \label{eq:backwardBc}
%\end{align}
For the forward direction, according to boundary conditions (Eqs.~\eqref{eq:lessforwardBc} and~\eqref{eq:lessbackwardBc}), we can get
\begin{align}
-\frac{1}{2} &=  \frac{\alpha_T}{\alpha_x} \frac{1}{2}  +\frac{\alpha_T}{\alpha_x ^2}  q^*_+ - \frac{1}{\alpha_x}  + C_+ \exp{\left( \alpha_x /(2q^*_+) \right)  },  \label{eq:taylorsiCVMbc1} \\
\frac{1}{2} &= - \frac{\alpha_T}{\alpha_x} \frac{1}{2}  +\frac{\alpha_T}{\alpha_x ^2}  q^*_+ - \frac{1}{\alpha_x}  + C_+ \exp{\left(- \alpha_x /(2q^*_+) \right)  } . \label{eq:taylorsiCVMbc2}
\end{align}
Combining above two equations, we have
\begin{align}
\frac{ \alpha_x +\alpha_T  }{ \alpha_T/\epsilon_+ - 2 }  &= \frac{ \exp{\left( \epsilon_+ \right) }- \exp{\left( -\epsilon_+ \right)  } }{ \exp{\left( \epsilon_+ \right)  } +\exp{\left( -\epsilon_+ \right) } } = \tanh \left( \epsilon_+  \right),  \label{eq:taylorsiCVMs6}
\end{align}
where $\epsilon_+ = \alpha_x/(2q^*_+)$. According to the solution of Eq.~\eqref{eq:epsilonzero}, $q^*_+ \approx -1$ as long as $\alpha_x$ is infinitesimal.
In another word, if $\alpha_x \rightarrow 0$ , $\epsilon_+ \rightarrow 0$.
Then $\tanh(\epsilon_+)$ is expanded under assumption $\alpha_x \rightarrow 0$,
\begin{align}
\tanh (\epsilon_+) =  \epsilon_+ - \epsilon_+ ^3 /3 + O( \epsilon_+^5 ). \label{eq:taylorsitanhepision}
\end{align}
{\color{black}{It is worth noting that no assumption is made on $\alpha_T$.}}
Then Eq.~\eqref{eq:taylorsiCVMs6} becomes
\begin{align}
\alpha_x &\approx - 2\epsilon_+ - \frac{1}{3}(\alpha_T \epsilon_+^2 - 2\epsilon_+^3) \\
\Longrightarrow q^*_+ &\approx -1 + \frac{\alpha_x  \alpha_T}{12 } + \frac{ \alpha_x  ^2 }{12  } .  \label{eq:taylorsiCVMs10}
\end{align}
Similarly, we can get
\begin{align}
q^*_{-} &\approx 1 + \frac{\alpha_x  \alpha_T}{12 } - \frac{ \alpha_x  ^2 }{12  } .  \label{eq:taylorsiCVMs11}
\end{align}
Therefore, the thermal rectification ratio of linear mode is,
\begin{align}
\beta_l= \frac{q_{+} +q_{-} }{q_{+} - q_{-} } &\approx -\frac{1}{12} \alpha_x \alpha_T.    \label{eq:taylorsiCVMsq15}
\end{align}

Based on above mathematical derivations, similarly we can get other solutions of the thermal rectification, i.e.,
\begin{align}
\kappa^{*}_e &= 1 +\alpha_{xT} x^{*}T^{*}, &\quad \beta &= \frac{1}{12} \alpha_{xT}.   \\
\kappa^{*}_e &= 1+\alpha_T T^{*}+\alpha_x x^{*}+\alpha_{xT} x^{*}T^{*}, &\quad  \beta &=\frac{1}{12} \left( \alpha_{xT}- \alpha_x \alpha_T \right). \label{eq:fullzc}
\end{align}

\section{Numerical validation}

In this section, numerical simulations are implemented to validate the perturbation theory.

\subsection{Numerical discretization and solutions}

Based on Eqs.~\eqref{eq:siconductivity} and~\eqref{eq:silaplacian}, we can get
\begin{align}
\frac{\partial }{\partial x} \left( \kappa_e (x,T(x))  \frac{\partial T}{\partial x}  \right) =0.
\end{align}
The iterative method is used to solve above equation, i.e.,
\begin{align}
\frac{\partial }{\partial x} \left( \kappa_e  \frac{\partial \delta T^n }{\partial x}  \right) = -\frac{\partial }{\partial x} \left( \kappa_e  \frac{\partial T^n }{\partial x}  \right),
\label{eq:iterative}
\end{align}
where $n$ is the iteration index, $\delta T^n =T^{n+1} -T^{n}$ is the temperature variance between two successive iteration steps.
When $n=0$, the initial temperature inside the system is $T^n=T_0$.
To solve it numerically, the finite difference method is used and we discretize the computational domain into $(M-1)$ uniform cells with $M$ grid points, i.e.,
\begin{align}
x_i = x_0 - \frac{L}{2} + \frac{(i-1)L}{M-1},
\end{align}
where $x_i$ is position of the grid point $i$, $i=1,2,...,M$. It can be found that $x_1$ and $x_M$ are two boundaries with fixed temperatures.
Then Eq.~\eqref{eq:iterative} becomes
\begin{align}
&\sum_{j \in N(i) } \kappa_{e,ij} \delta T_{j}^n - \left( \sum_{j \in N(i) } \kappa_{e,ij} \right) \delta T_{i}^n  \notag \\
=& - \sum_{j \in N(i) } \kappa_{e,ij}  T_{j}^n + \left( \sum_{j \in N(i) } \kappa_{e,ij} \right)  T_{i}^n ,  \quad i \in [2, M-1]
\label{eq:fourierDsolver}
\end{align}
where $N(i)$ denotes the sets of neighbor grid points of grid point $i$.
In addition,
\begin{align}
\delta T^n_1 &= \delta T^n_M=0, \\
2T_{ij} &= T_{i} +T_{j}, \\
2x_{ij} &= x_{i} +x_{j}, \\
\kappa_{e,ij} &=\kappa_e (x_{ij}, T_{ij}).
\end{align}
Combining the boundary conditions (Eqs.~\eqref{eq:forwardBc},~\eqref{eq:backwardBc}),
Eq.~\eqref{eq:fourierDsolver} can be solved iteratively.
Here, the Thomas algorithm~\cite{Numericalanalysis} is used to solve Eq.~\eqref{eq:fourierDsolver} and the iteration converges as
\begin{align}
\frac{ \sqrt{  \sum_{2}^{M-1} \left|  \delta T_i^n \right|^2 }   }{  \sqrt{ (M-2)\Delta T ^2 } } < 10^{-14}.
\end{align}

For the dimensionless equations (Eqs.~\eqref{eq:sinondimensionalconductivity} and~\eqref{eq:sinormalizedtaylorsm}), similar iterative method can also be implemented directly.
For all numerical simulations, the mesh independence has been tested.
Without special statements, we set $M=10001$.

\subsection{Linear mode}

Above derivations have given the thermal rectification of the linear mode, i.e.,
\begin{align}
\kappa_e^* = 1+\alpha_x  x^* + \alpha_T T^*,\quad \beta_l= -\frac{1}{12} \alpha_x \alpha_T.
\label{eq:linearmodelaw}
\end{align}
In order to validate it, numerical simulations are conducted.
{\color{black}{For simplicity, the dimensionless parameters satisfy $\alpha_{x}, \alpha_{T} \in (0,2)$,  and $\alpha_{x}+\alpha_{T} \leq 2$, so that inside the system $x^* \in (-1/2,1/2)$, the thermal conductivity is positive, i.e., $\kappa_e^* >0$.}}
We discrete $\alpha_x \in (0,2)$ into $200$ uniform pieces as well as $\alpha_T \in (0,2)$.
For each discretized piece $\alpha_x$ and $\alpha_T$, the thermal rectification can be predicted by numerical iterative solutions and the theoretical law (Eq.~\eqref{eq:linearmodelaw}).
Especially, for the line $\alpha_x + \alpha_T =2$, we discrete it into $2000$ uniform pieces.

\begin{figure}
 \centering
 \includegraphics[scale=0.35,viewport=30 120 950 1080,clip=true]{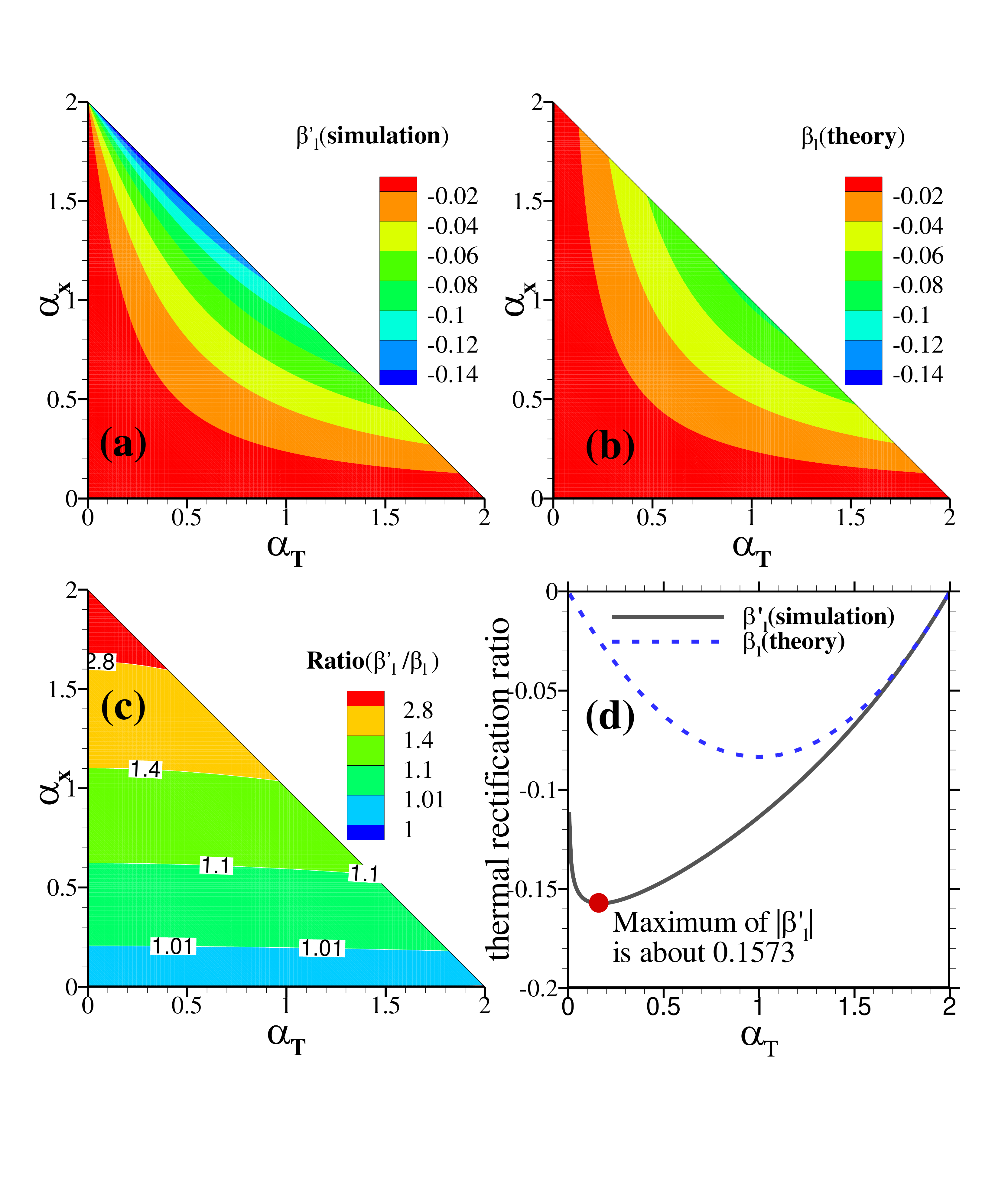}
 \caption{Thermal rectification ratio ($\beta_l$) of the linear mode. (a) $\beta_l'$ is solved by numerical method; (b) $\beta_l$ is predicted by perturbation theory (Eq.\ref{eq:linearmodelaw}); (c) the ratio between $\beta_l'$ and $\beta_l$ validates the perturbation theory for the linear mode when $\alpha_{x}$ approaches zero; (d) $\beta_l'$ along the line $(\alpha_{x}+\alpha_{T} = 2.0)$ reveals the maximum rectification ratio ($|\beta_l'|$) is about 0.1573 for the linear mode.}
 \label{fig:linearMode}
\end{figure}

Numerical results are shown in~\cref{fig:linearMode}.
%Figure~\ref{fig:linearMode} shows the thermal rectification ratio of the linear mode (Eq.~\eqref{eq:linearMode}).
%{\color{black}{The dimensionless parameters satisfy $\alpha_{x}, \alpha_{T} \in (0,2)$,  and $\alpha_{x}+\alpha_{T} \leq 2$, so that inside the system $x^* \in (-1/2,1/2)$, the thermal conductivity is positive, i.e., $\kappa_e^* >0$.}}
The numerical solution ($\beta'_l$) is accurate with at least 4 significant digits. It can be taken as the real thermal rectification ratio of the linear mode.
As shown in \cref{fig:linearMode}(c), the predicted $\beta_l$ (Eq.~\eqref{eq:linearmodelaw}) is accurate as long as $\alpha_x \rightarrow 0$. Even when $\alpha_x$ is finite value, say $0.5$, the relative error is less than $10\%$. When $\alpha_x$ increases beyond $0.5$, the theory deviates from the reference solution, becomes inaccurate.
It is obvious that $\beta_l$ is symmetric about the line ($\alpha_x = \alpha_T$), but $\beta_l'$ is not.
The thermal rectification ratio of the linear mode reaches its maximum ($|\beta_l'| = 0.1573$) near the point ($\alpha_{x} = 1.835, \alpha_{T} = 0.165$) (\cref{fig:linearMode}(a,d)). It suggests that the thermal rectification ratio is less than $0.1573$, if the effective thermal conductivity varies linearly in terms of position and temperature.
\begin{figure}
 \centering
 \includegraphics[width=0.5\textwidth]{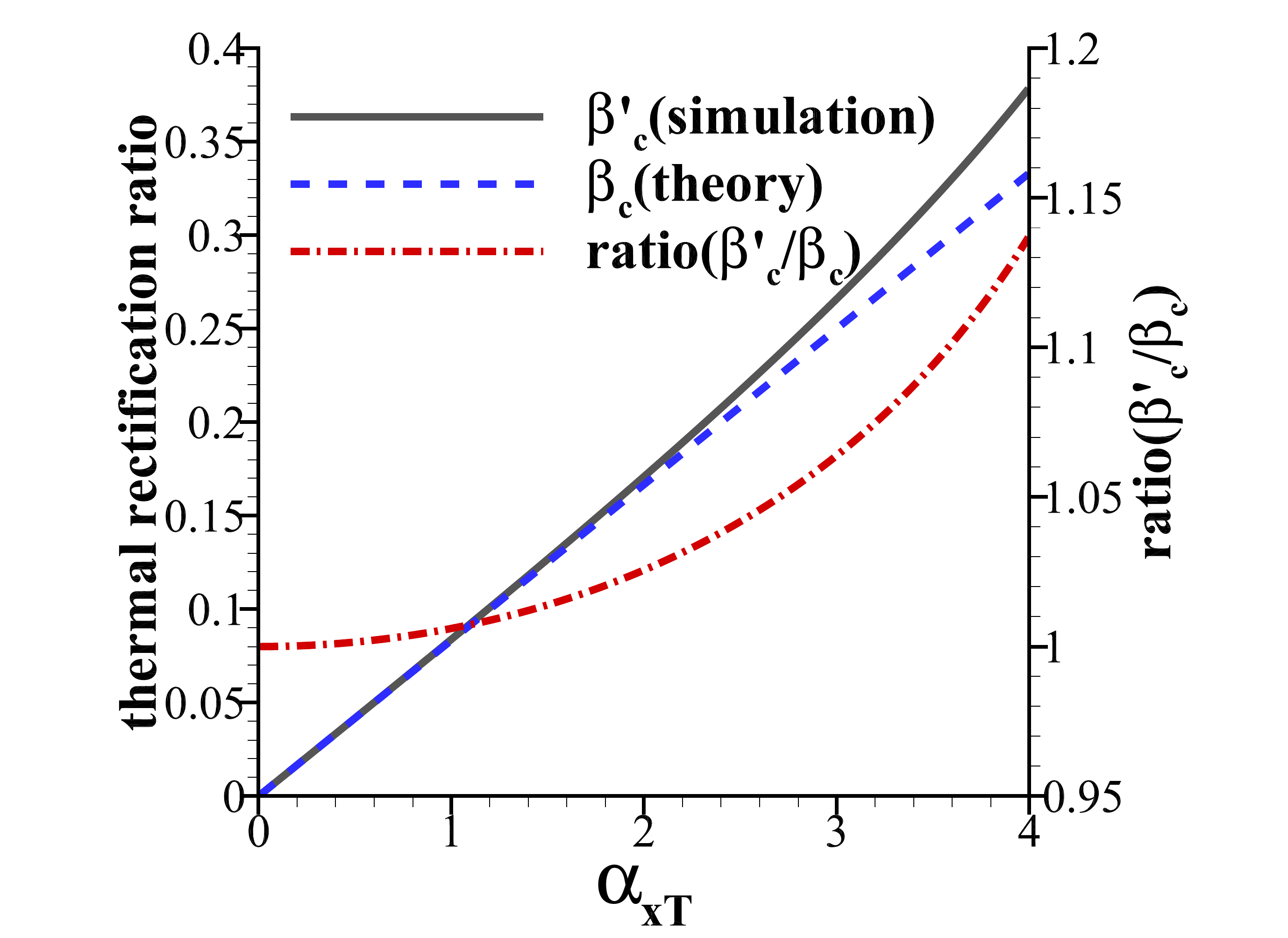}
 \caption{Thermal rectification ratio ($\beta_c$) of the cross mode. The dash line represents $\beta_c$ predicted by perturbation theory (Eq.~\eqref{eq:crossmodelaw}); the solid line represents $\beta_c'$ solved by numerical method; the dash dot line represents the ratio between $\beta_c'$ and $\beta_c$. The relative error is less than $1\%$ as $|\alpha_{xT}|$ is smaller than 1.2, and the maximum relative error is less than $15\%$. The maximum thermal rectification ratio ($|\beta_c'|$) occurs near $\alpha_{xT} = 4$, is about 0.3791 for the cross mode.}
 \label{fig:crossMode}
\end{figure}

\subsection{Cross mode}

Numerical simulations are conducted to validate the theoretical results of the cross mode, i.e.,
\begin{align}
\kappa_e^* = 1+\alpha_{xT}   x^*   T^* , \quad
\beta_c= -\frac{1}{12} \alpha_x \alpha_T.
\label{eq:crossmodelaw}
\end{align}
{\color{black}{The dimensionless parameter satisfies $\alpha_{xT} \in (0,4]$, so that inside the system $x^* \in (-1/2,1/2)$, the thermal conductivity is positive, i.e., $\kappa_e^* >0$.}}
We discrete $\alpha_{xT} \in (0,4]$ into $4000$ uniform pieces.
For each discretized $\alpha_{xT}$, the thermal rectification can be predicted by numerical iterative solutions and the theoretical law (Eq.~\eqref{eq:crossmodelaw}).

Numerical results are shown in~\cref{fig:crossMode}.
%Figure~\ref{fig:crossMode} shows the thermal rectification ratio of the cross mode, which is generated from the cross term ($x^*T^*$, Eq.~\eqref{eq:crossMode}). {\color{black}{As shown in \cref{fig:crossMode}, $\alpha_{xT} \in (0,4]$, so that inside the system $x^* \in (-1/2,1/2)$, the thermal conductivity is positive.}}
$\beta_c$ in Eq.~\eqref{eq:crossmodelaw} is a very accurate approximation for the cross mode. Its relative error to the reference numerical solution ($\beta_c'$) is less than $1\%$ as $|\alpha_{xT}|$ is smaller than 1.2, and the maximum relative error is less than $15\%$ in all cases. Moreover, the maximum thermal rectification ratio ($|\beta_c'|$) is considerably larger than that of the linear mode, which occurs at $\alpha_{xT} = 4$ and is about $0.3791$.

{\color{black}{Above theoretical derivations (Eqs.~\eqref{eq:siCVMsq15},~\eqref{eq:linearmodelaw} and~\eqref{eq:crossmodelaw}) and numerical results show that the linear mode and cross mode affect the thermal rectification ratio independently as $\alpha_x,\alpha_t,\alpha_{xT}$ are small.}}

%As shown in~\cref{fig:crossMode}, $\beta_c$ of the cross mode is very accurate in a wide range.
%However, the accuracy of $\beta_l$ of the linear mode behaves quite differently with respect to $\alpha_x$ and $\alpha_T$.
%{\color{black}{When $\alpha_x$ is less than $1$, the accuracy of $\beta_l$ is nearly independent of $\alpha_T$, which is reflected in \cref{fig:linearMode}(c) that the contour lines of $\beta_l’/\beta_l$ are almost parallel to the $\alpha_T$ axis.
%Therefore, the real thermal rectification ratio of linear mode $\beta_l'$ can be further approximated by parameterizing the factor in the expression of $\beta_l$ (Eq.~\eqref{eq:linearmodelaw}), i.e.,
%\begin{align}
%\beta_l' = -\frac{1+R(\alpha_x)}{12} \alpha_x\alpha_T,
%\end{align}
%where $1+R(\alpha_x)$ is actually the ratio between $\beta_l'$ and $\beta_l$, and $R \rightarrow 0$ as $\alpha_x \rightarrow 0$. Then, the real thermal rectification ratio $\beta$, which is the combination of the linear mode and the cross mode based on Eqs.~\eqref{eq:siCVMsq15},~\eqref{eq:linearmodelaw}, and~\eqref{eq:crossmodelaw}, becomes
%\begin{align}
%\beta \approx \beta_p - \frac{R(\alpha_x)}{12}\frac{1}{\kappa_0^2}\frac{\partial \kappa_e}{\partial x}\frac{\partial \kappa_e}{\partial T}L\Delta T . \label{eq:realThermalConductivity}
%\end{align}

\subsection{Arbitrary effective thermal conductivity}

\begin{figure}
 \centering
 \includegraphics[scale=0.5,viewport=0 600 1000 1100,clip=true]{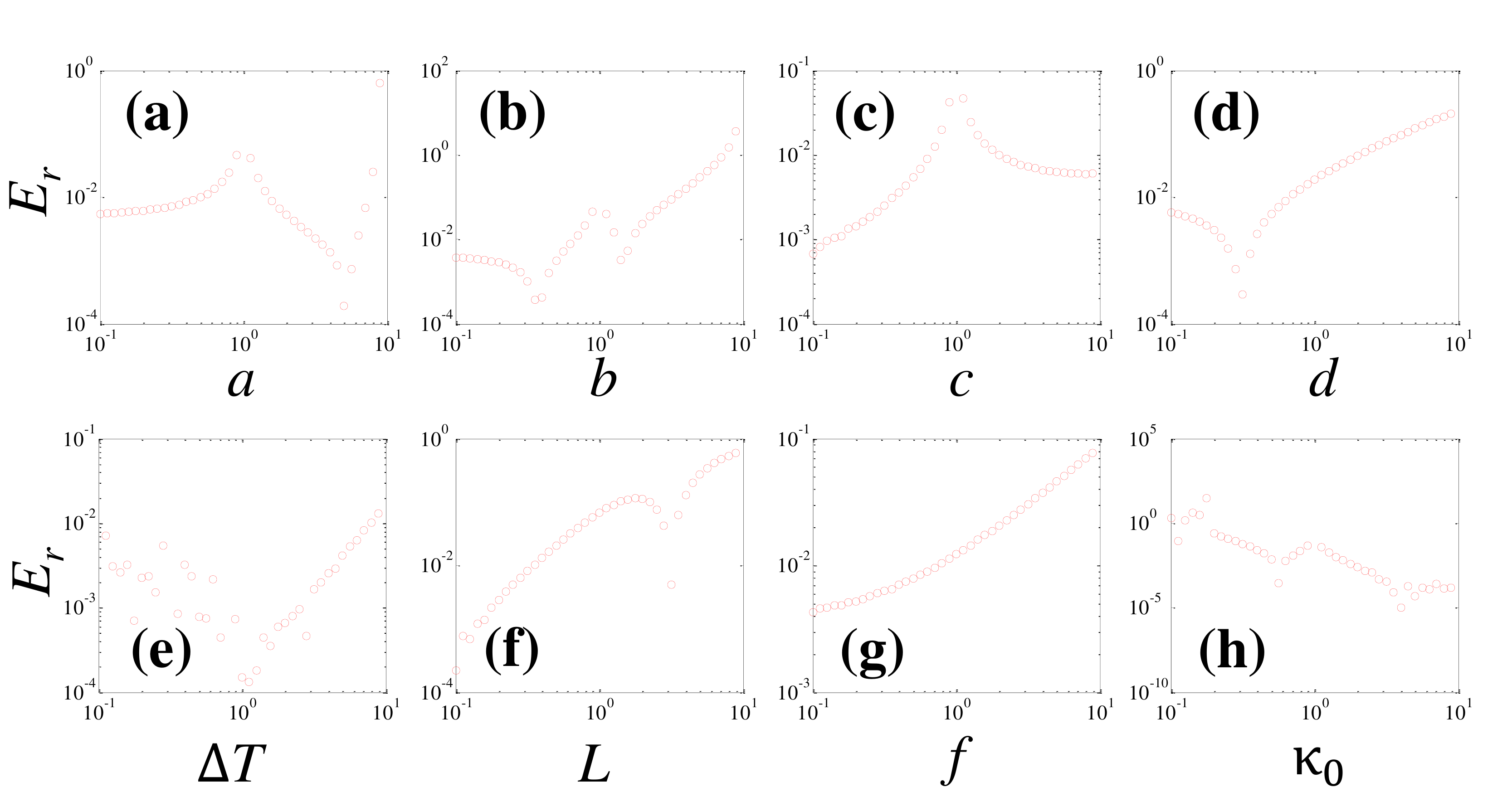}
 \caption{The relative errors between the thermal rectification predicted by the numerical and analytical solutions, i.e., $E_r = \left| ( \beta_{\text{numerical}} -\beta_{\text{theory}}   )/ \beta_{\text{theory}}  \right|$. In Eq.~\eqref{eq:sitaylorsm}, we fix $x_0=1,~T_0=1$. $Y$ axis is $E_r$ and $X$ axis are eight variables. (a) Change $a$ and fix $b=1,~c=1,~d=0.5,~f=0.2,~\kappa_0=1,~L=0.2,~\Delta T=0.2$. (b) Change $b$ and fix $a=1,~c=1,~d=0.5,~f=0.2,~\kappa_0=1,~L=0.2,~\Delta T=0.2$. (c) Change $c$ and fix $a=1,~b=1,~d=0.5,~f=0.2,~\kappa_0=1,~L=0.2,~\Delta T=0.2$. (d) Change $d$ and fix $a=1,~b=1,~c=0.5,~f=0.2,~\kappa_0=1,~L=0.2,~\Delta T=0.2$. (e) Change $\Delta T$ and fix $a=0.2,~b=1,~c=0.1,~d=0.1,~f=0.01,~\kappa_0=5,~L=0.1 $. (f) Change $L$ and fix $a=1,~b=0.5,~c=1,~d=0.5,~f=0.2,~\kappa_0=1,~\Delta T=0.2$. (g) Change $f$ and fix $a=1,~b=1,~c=0.5,~d=0.5,~\kappa_0=1,~L=0.2,~\Delta T=0.2$. (h) Change $\kappa_0$ and fix $a=1,~b=1,~c=1,~d=0.5,~f=0.2,~L=0.2,~\Delta T=0.2$. }
 \label{numericalvalidation}
\end{figure}
%\begin{figure}
% \centering
% \includegraphics[scale=0.4,viewport=50 400 1000 1100,clip=true]{arbitrary.pdf}
% \caption{Comparison of the predicted thermal rectification ratio $\beta$ between the numerical solution and our theory, where $\kappa_e =1+x^2 + 3T^4$, $x_0=L/2$, $T_0 =1$.  (a) Fix $L=5.0$, change $\Delta T\in [0.1, 1.9]$. (b) Fix $L=0.2$, change $\Delta T \in [0.1, 1.9]$. (c) Fix $\Delta T=0.2$, change $L \in [0.01, 10]$. (d) Fix $\Delta T=1.8$, change $L \in [0.01, 10]$.
% }
% \label{arbitrary}
%\end{figure}

We take Eq.~\eqref{eq:sitaylorsm} as an example.
We set $x_0=1,~T_0=1$.
There are eight independent variables in Eq.~\eqref{eq:sitaylorsm}, i.e., $a,~b,~c,~d,~f,~L,~\Delta T,~\kappa_0$.
Next, we change one of them and fix others.
The predicted thermal rectification ratio $\beta$ are compared with our derived analytical solutions, i.e., Eq.~\eqref{eq:siCVMsq15}.
A parameter $E_r$ is introduced to show the relative errors between the numerical ($\beta_{\text{numerical}}$) and theoretical ($\beta_{\text{theory}}$) results, i.e.,
\begin{align}
E_r = \left| \frac{\beta_{\text{numerical}} -\beta_{\text{theory}}  }{\beta_{\text{theory}}}  \right|.
\end{align}
Numerical results are shown in~\cref{numericalvalidation}, where $Y$ axis is $E_r$ and $X$ axis are eight variables, respectively.
It can be observed that the numerical results are in excellent agreement with our derived theoretical solutions within our assumptions (Eq.~\eqref{eq:silimitation11}).
However, as $\alpha_x,~\alpha_T,~\alpha_{xT}$ are large or $| \kappa_e/ \kappa_0 | \gg 1$, the theoretical results deviate the numerical results significantly.

\section{Usage of three dimensionless parameters in the thermal rectification in 2D Lorentz gas model}

\begin{figure}
 \centering
 \includegraphics[scale=0.35,viewport=50 150 1050 550,clip=true]{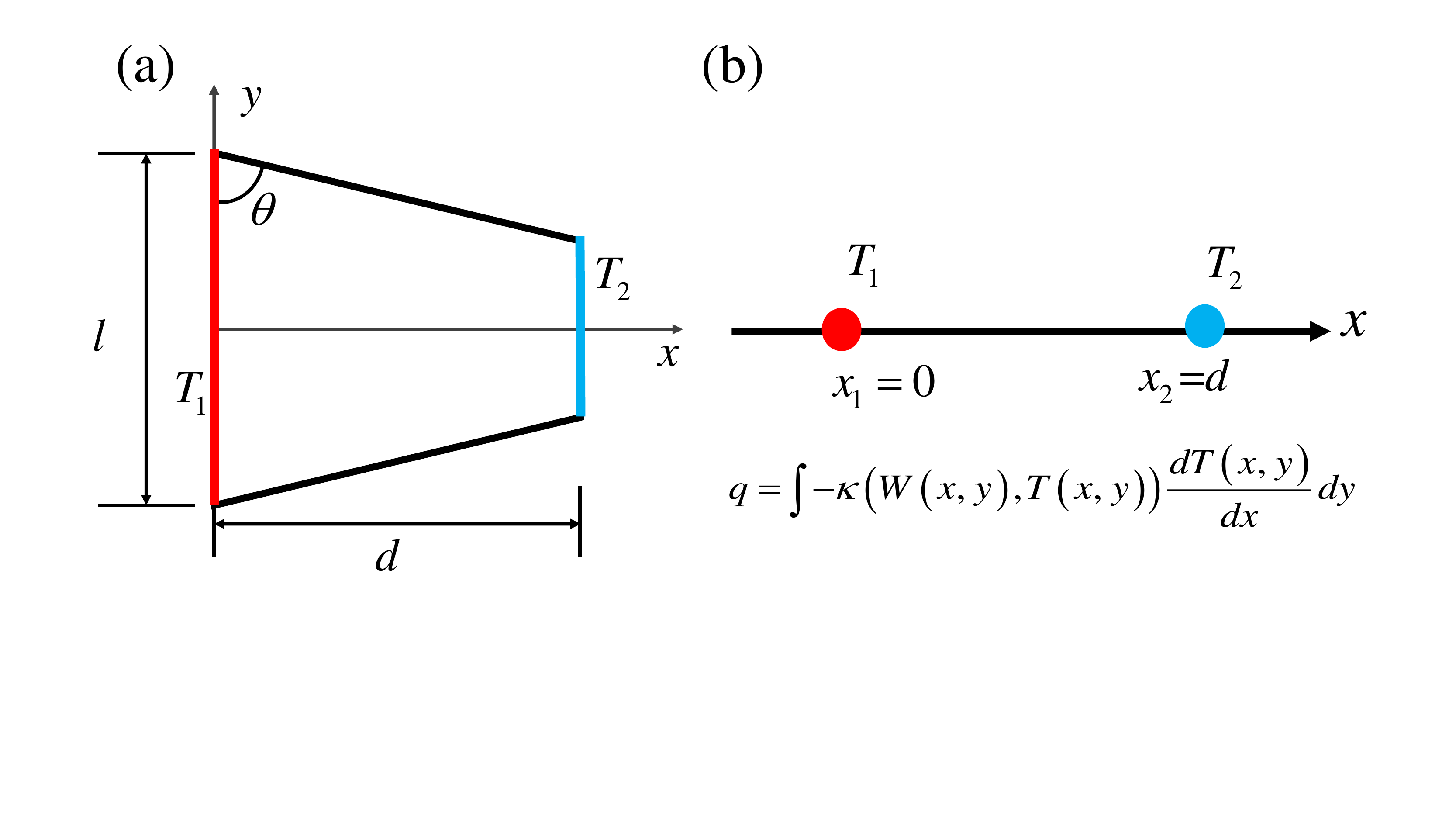}
 \caption{(a) A schematic of the 2D Lorentz gas model with an asymmetric trapezoidal shape (Fig.1(b) in Ref~\cite{wang2019}). (b) A reduced (quasi) one-dimensional thermal system and its heat flux, where $(x_1,T_1)$ and $(x_2,T_2)$ are boundary conditions, $L=d,~x_0=d/2$.}
 \label{PREproof}
\end{figure}

To show the utility of these dimensionless parameters $\alpha_x,~\alpha_T,~\alpha_{xT}$, we analyse a universal relation between thermal rectification ratio and the geometric parameters and source temperatures in a two-dimensional multiparticle Lorentz gas model~\cite{wang2019}, i.e.,
\begin{align}
\left| \frac{q_{+} +q_{-} }{q_{+} - q_{-} } \right| \propto \frac{d}{l \tan \theta } \frac{\Delta T}{T_0},
\label{eq:PREresults}
\end{align}
where $d$ and $l$ are the system length and width respectively in the 2D homogeneous and asymmetric trapezoidal domain as shown in \cref{PREproof}(a).
$\theta$ is the inclined angle, $T_0=(T_1 +T_2)/2$ and $\Delta T = |T_1 -T_2|$ are the average temperature and temperature difference between the left and right boundaries, respectively.
In this example, notice that there are obvious temperature slips near the boundaries~\cite{wang2019}, then the theoretical formula can not be used directly.
Therefore, here we just try to find out the dimensionless parameters for this system.

At first, the two-dimension heat conduction problem is reduced into one-dimensional problem (\cref{PREproof}(b)) and its heat flux along $x$ direction can be described by
\begin{align}
q  =  - \int \kappa \frac{dT}{dx} dy ,
\label{eq:fourier2D}
\end{align}
where $\kappa$ is the thermal conductivity.
Fortunately, previous studies have proven that the thermal conductivity almost keeps a constant as the length of the rectangular space changes~\cite{wang2019,chen2018b}.
Hence we can assume that the leading order term of the local thermal conductivity is only dependent on the local temperature,
\begin{align}
\kappa = \kappa(T) + \epsilon_3,
\end{align}
where $\kappa(T)= C T^{1/2}$~\cite{mao2008heat}, $C$ is a constant, $\epsilon_3$ represents small deviation from the leading order term.
Given that the heat flux is mainly parallel to the $x$ coordinate if $\theta \rightarrow \pi/2$, thereby, the variation in $y$ direction can be ignored, so that,
$T(x,y) \approx T(x)$.
Under these assumptions, an effective thermal conductivity can be defined,
\begin{align}
\kappa_e(x,T,L) &= \int \kappa dy = \overline{W}(x) \kappa(T)+\epsilon', \label{eq:ke-lorentz} \\
\overline{W}(x) &= \int dy = l - 2\frac{x}{\tan \theta}, \\
\epsilon' &= \int \epsilon_3 dy, \quad x_0 = d/2,
\end{align}
where $\overline{W}(x)$ is actually the transverse length in $y$ direction.
Note that $\epsilon_3$ and $\epsilon'$ are essential to keep $\kappa_e$ nonseparable so that the thermal rectification occurs~\cite{go2010}.

After a simple derivation, the dimensionless parameters are obtained by ignoring the small deviation ($\epsilon,\epsilon'$),
\begin{align}
\alpha_x &=\frac{L}{\kappa_e (x_0,T_0, L) } \left.\  \left( \frac{\partial \kappa_e }{\partial x}  \right) \right|_{(x,T)=(x_0,T_0)}  \approx -\frac{2d}{l\tan\theta-d}, \\
\alpha_T &= \frac{\Delta T}{\kappa_e (x_0,T_0, L) } \left.\ \left( \frac{\partial \kappa_e }{\partial T}  \right) \right|_{(x,T)=(x_0,T_0)} \approx \frac{\Delta T}{2T_0}, \\
\alpha_{xT} &=\frac{L \Delta T }{\kappa_e (x_0,T_0, L) }  \left.\ \left( \frac{\partial^2 \kappa_e }{\partial x \partial T}  \right) \right|_{(x,T)=(x_0,T_0)} \approx -\frac{d}{l\tan\theta-d}\frac{\Delta T}{T_0},
\end{align}
where $x_0=d/2$ is the central position of the whole thermal system, $L = d$ is the system length, spatial position $x \in [x_0 -L/2, x_0 +L/2]$.
As $\theta \rightarrow \pi/2$, we have $l\tan\theta \gg d$.
Then \cref{eq:PREresults} becomes,
\begin{align}
\left| \frac{q_{+} +q_{-} }{q_{+} - q_{-} } \right| \propto \alpha_{xT} \propto \alpha_x\alpha_T,
\end{align}
which indicates that the thermal rectification in 2D Lorentz gas model is related to these dimensionless parameters.

This example demonstrates the significance of $\alpha_x,\alpha_T,\alpha_{xT}$ to general thermal rectification.
Actually, these three dimensionless parameters denote the relative change of the effective thermal conductivity throughout the whole system due to the temperature change and the heterogeneity of the other physical properties.
In other words, this example indicates that the size-dependent thermal rectification is related to the thermal conductivity in 2D Lorentz gas model, which almost keeps a constant as the length of the rectangular space changes~\cite{wang2019,chen2018b}.

\section{Thermal rectification in inhomogeneous nanoporous silicon systems}

\begin{figure}
 \centering
 \includegraphics[scale=0.3,viewport=100 50 1500 1650,clip=true]{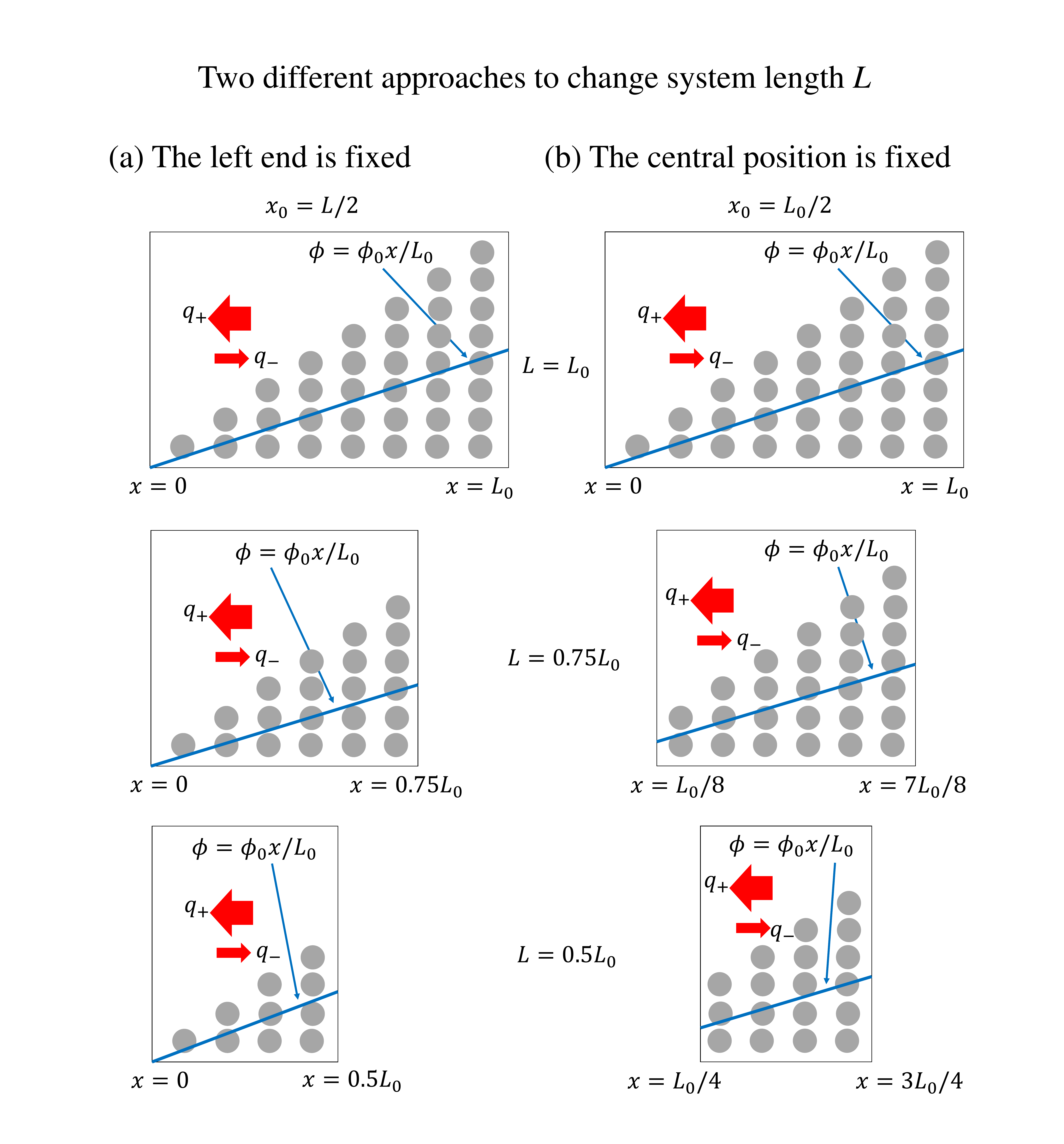}
 \caption{Thermal rectification of inhomogeneous porous silicon device~\cite{criado-sancho2013}. A sketch for graded inhomogeneous porous silicon device. The spatial porosity distributions satisfy $\phi (x)=\phi_0 x/L_0$, where $\phi_0 =0.1$, $L_0 =100$ mm is the reference length, $x$ is the spatial position. $L$ is the system length, $x_0$ is the central position. $x=x_0 -L/2 $ and $x=x_0 +L/2$ are the position of left and right boundaries, respectively. There are two different approaches to change the system length $L$. (a) First: The left end of the system is fixed ($x=0$), and the other end changes with system length, so that $x_0=L/2$ changes with system length $L$. Schematics of three geometries with different system length $L=L_0,~0.75L_0,~0.5L_0$. (b) Second: The central position $x_0=L_0 /2$ is anchored to a fixed point and the system is shortened towards two ends symmetrically, so that $x_0$ is independent of system length $L$. Schematics of three geometries with different system length $L=L_0,~0.75L_0,~0.5L_0$.}
 \label{porousNDSM}
\end{figure}
The thermal rectification coefficient of a quasi-one dimensional graded inhomogeneous porous silicon device is studied~\cite{criado-sancho2013,naso2019}.
According to previous studies~\cite{hydrodynamicporeousAFX2010,criado-sancho2013}, an effective thermal conductivity can be identified as
\begin{align}
\kappa_e( x,T )= \kappa_e(\phi,\lambda/r ,T )= \frac{ \kappa_{\text{bulk}} (1- \phi)^3    }{ 1+ \frac{9}{2} \phi (1- \phi)^3  (1+ \frac{3 \sqrt{\phi}}{\sqrt{2}} ) \frac{ (\lambda /r)^2 }{ 1+A'( \lambda/r)}},
\label{eq:kappaporoussi}
\end{align}
where $r$ is pore radius, $\phi (x)=\phi_0 x/L_0$ is the porosity (ratio of the volume of the pores divided by the total volume), $\phi_0= 0.10$, $L_0=100$ mm is the reference length, $A'(\lambda/r)= 0.864+ 0.290 \times \exp( -1.25 r/ \lambda )$.
$\kappa_{\text{bulk}}$ and $\lambda$ are the temperature dependent thermal conductivity and phonon mean free path of bulk silicon.

Here the Debye approximation and gray model are used~\cite{holland1963analysis}, where phonon dispersion and polarization are not considered.
The average phonon group velocity is $v_s =6400$ m/s~\cite{holland1963analysis,brockhouse1959lattice}.
The thermal conductivity $\kappa_{\text{bulk}}$ and specific heat $C_V$ in bulk silicon in different temperatures are obtained by experiments~\cite{Glassbrenner64conductivity,CpSiGeExpriment1959}.
The experimental data of thermal conductivity~\cite{Glassbrenner64conductivity} and specific heat~\cite{CpSiGeExpriment1959} as $60 ~\text{K} \leq T \leq 130 ~\text{K}$ is fitted by polynomial function, i.e.,
\begin{align}
\kappa_{\text{bulk}}(T) &= -0.00171 T^3 +0.741 T^2 -114.0 T+6677.0, \\
C_V (T) &= -0.0155 T^3 - 1.01  T^2 + 8922.0  T -2.58 \times 10^5.
\end{align}
The phonon mean free path $\lambda$ used in Eq.~\eqref{eq:kappaporoussi} is obtained from the expression for the thermal conductivity in terms of the mean-free path $\kappa_{\text{bulk}} = C_V v_s \lambda /3$~\cite{criado-sancho2013}.

There are two different ways to change the system length $L$.
The first way is to fix the left end of system and change the right end (\cref{porousNDSM}(a)). In this case, $x_0 = L/2$, moves as the system length changes.
The second way is to anchor the central position $x_0$ to a fixed point and move left and right ends symmetrically (\cref{porousNDSM}(b)), so that $x_0=L_0/2$ is independent of system length.

\section{Calculations of these dimensionless parameters in trapezoid suspended graphene}

Actually, as long as a differentiable effective thermal conductivity $\kappa_e(x,T,L)$ is given in the thermal system and system length $L$, temperature difference $\Delta T$, $(x_0,T_0)$ are known, these dimensionless parameters can be obtained mathematically.
There is no need to calculate the temperature field inside the thermal system.
Let's take the thermal rectification in trapezoid suspended graphene~\cite{wang2014} as an example.

The mathematical formulas of the effective thermal conductivity $\kappa_e (x , T , L )$ is
\begin{align}
%\frac{G_{\text{ball}} } {A} &= \left( \frac{1}{4.4 \times 10^5 T^{1.68}  } +\frac{1}{ 1.2 \times  10^{10} }   \right)^{-1},  \\
%\kappa_{1} (L,T) &=\frac{G_{\text{ball}} }{A} \left(\frac{1}{L} +\frac{1}{\pi \lambda /2 }     \right)^{-1}, \\
W_3 (x) & = \frac{ W_2 -W_1 }{L} x +W_1, \\
\kappa_{\text{GNR} } (L,T,W_3) & = \left( \frac{1}{c_1}  \left(\frac{\Delta}{W_3 } \right)^{m} +\left(\frac{1}{L} +\frac{1}{\pi \lambda /2 }     \right)  \left( \frac{1}{4.4 \times 10^5 T^{1.68}  } +\frac{1}{ 1.2 \times  10^{10} }   \right)    \right)^{-1}, \\
\kappa_e (x , T , L ) &= \kappa_e (W_3 , T , L ) \approx W_3 (x) \times \kappa_{\text{GNR} } (L, T ,W_3), \notag \\
&= \frac{ \left( w_1 x +w_2 \right) }{ \left( w_3  \left( w_1 x +w_2 \right)^{-m} + w_4  \left( w_5 T^{-1.68} +w_6 \right)   \right) } =  \frac{D_1}{D_2}, \label{eq:kappaGNR}
\end{align}
where $x_0=L/2$ and
\begin{align}
w_1 &= \frac{ W_2 -W_1 }{L}, &\quad w_2 &= W_1 , &\quad  w_3 &=  \frac{ \Delta ^m }{c_1} ,\\
w_4 &=\left(\frac{1}{L} +\frac{1}{\pi \lambda /2 }     \right) , &\quad w_5 &=\frac{1}{4.4 \times 10^5   }, &\quad w_6 &=  \frac{1}{ 1.2 \times  10^{10} }  ,
\end{align}
\begin{align}
D_1 = \left( w_1 x +w_2 \right), \quad \quad D_2 = \left( w_3  \left( w_1 x +w_2 \right)^{-m} + w_4  \left( w_5 T^{-1.68} +w_6 \right)   \right).
\end{align}
Note that $w_1,~w_2,~w_3,~w_4,~w_5,~w_6$ are related to the geometry of thermal system and $(x_0,T_0)$.
For a given system, we assumed that they are constants based on previous experiments~\cite{bae2013ballistic,japgraphene2015,xu_length-dependent_2014}.
Here we set $\Delta =0.6$ nm, $c_1=0.04$ W/(mK), $T_0=300~\text{K}$, $m=1.8$, $ | \Delta T |=30~\text{K}$~\cite{wang2014}, $\lambda=240$ nm~\cite{xu_length-dependent_2014}.
Then $D_1$ and $D_2$ are both differentiable functions of $x$ and $T$.
So that based on Eq.~\eqref{eq:kappaGNR}, we can derive
\begin{align}
\frac{\partial \kappa_e }{\partial x}  &= \frac{ w_1 D_2 + m w_3 w_1 (w_1 x +w_2)^{-m}  }{  D_2 ^2  }, \\
\frac{\partial \kappa_e }{\partial T}  &= \frac{ (w_1 x+ w_2 ) w_4 (1.68 w_5 T^{-2.68}) }{  D_2 ^2  },  \\
\frac{\partial ^2 \kappa_e }{\partial T \partial x} &=  w_4 (1.68 w_5 T^{-2.68}) \frac{ w_1 D_2^2 + 2 D_2 (w_1 x+ w_2)^{-m}   w_3 w_1 m       }{D_2^4}.
\end{align}
Then even if the temperature field inside the system is unknown, these three dimensionless parameters ($\alpha_x,~\alpha_T,~\alpha_{xT}$) can still be calculated directly at $(x_0,T_0)$ based on above equations in this section, i.e.,
\begin{align}
\alpha_x &=\frac{L}{\kappa_e (x_0,T_0, L) } \left.\  \left( \frac{\partial \kappa_e }{\partial x}  \right) \right|_{(x,T)=(x_0,T_0)}, \\
\alpha_T &= \frac{\Delta T}{\kappa_e (x_0,T_0, L) } \left.\ \left( \frac{\partial \kappa_e }{\partial T}  \right) \right|_{(x,T)=(x_0,T_0)},\\
\alpha_{xT} &=\frac{L \Delta T }{\kappa_e (x_0,T_0, L) }  \left.\ \left( \frac{\partial^2 \kappa_e }{\partial x \partial T}  \right) \right|_{(x,T)=(x_0,T_0)}.
\end{align}

\section{An extension of the perturbation theory in the thermal rectification in two-segment bulk materials}

Although our present theoretical formula can not directly be used in two-segment bulk materials with different temperature-dependent thermal conductivity~\cite{dames2009}, the perturbation method and Taylor expansion used in our theoretical derivations can be extended to two-segment bulk materials directly.
Here a simple introduction is made about this extension and results are shown in~\cref{figapl}.

First, let's see what we have done.
Given a (quasi) one-dimensional thermal conduction system with spatial position $x \in [x_0 -L/2, x_0 +L/2 ]$, two temperatures ($T_0-\Delta T/2, T_0+\Delta T/2$) are imposed at the two boundaries, where $T_0$ and $\Delta T$ are the average temperature and temperature difference between the temperatures of two boundaries.
As the effective thermal conductivity $\kappa_e (x,T,L)$ is known and differentiable, the dimensionless heat flux $q^*_{+}$ (or $q^*_{-}$) can be derived based on perturbation theory, i.e. (see Eqs.(S40)(S41) in Supplemental Material or Eqs.(12)(13) in the revised manuscript)
\begin{align}
q^*_{+}&=q^*_{+}(x_0,L,T_0,\Delta T,\kappa_e) = -1 + \frac{1}{12} \left(\alpha_x \alpha_T - \alpha_{xT} -\alpha_{x^2 }-\alpha_{T^2} -\frac{1}{10} \alpha_T^2 \alpha_{x^2}  \right), \label{eq:qzheng} \\
q^*_{-}&=q^*_{-}(x_0,L,T_0,\Delta T,\kappa_e) = 1 + \frac{1}{12} \left(\alpha_x \alpha_T - \alpha_{xT} + \alpha_{x^2 } + \alpha_{T^2}  + \frac{1}{10} \alpha_T^2 \alpha_{x^2}  \right). \label{eq:qfu}
\end{align}
For (quasi) one-dimensional homogeneous single bulk materials, the thermal conductivity is only a function of temperature, then Eqs.~\eqref{eq:qzheng} and~\eqref{eq:qfu} becomes
\begin{align}
q_{+}^{*} &=\frac{q_{+} L}{\kappa_e(x_0,T_0) \Delta T } = -1 -\frac{\alpha_{T^2}}{12} , \label{eq:qzhengfirstapprox} \\
q_{-}^{*} &=\frac{q_{-} L}{\kappa_e(x_0,T_0) \Delta T } = 1 + \frac{\alpha_{T^2}}{12} ,  \label{eq:qfufirstapprox}
\end{align}
so that we have
\begin{align}
q_{+} &=\left( \kappa_e(x_0,T_0) \frac{\Delta T}{L} \right) \left(-1 -\frac{\alpha_{T^2}}{12} \right), \label{eq:qzhenglinear} \\
q_{-} &=\left( \kappa_e(x_0,T_0) \frac{\Delta T}{L} \right) \left( 1 + \frac{\alpha_{T^2}}{12}  \right). \label{eq:qfulinear}
\end{align}
%Then there is no thermal rectification for single homogeneous bulk materials. However, for two-segment bulk materials, $(x_0,T_0)$ in above two equations of each segment are different and the interface between the two segments causes the nonseparable thermal conductivity in the whole domain.

\begin{figure}
 \centering
 \includegraphics[scale=0.35,viewport=0 380 1000 800,clip=true]{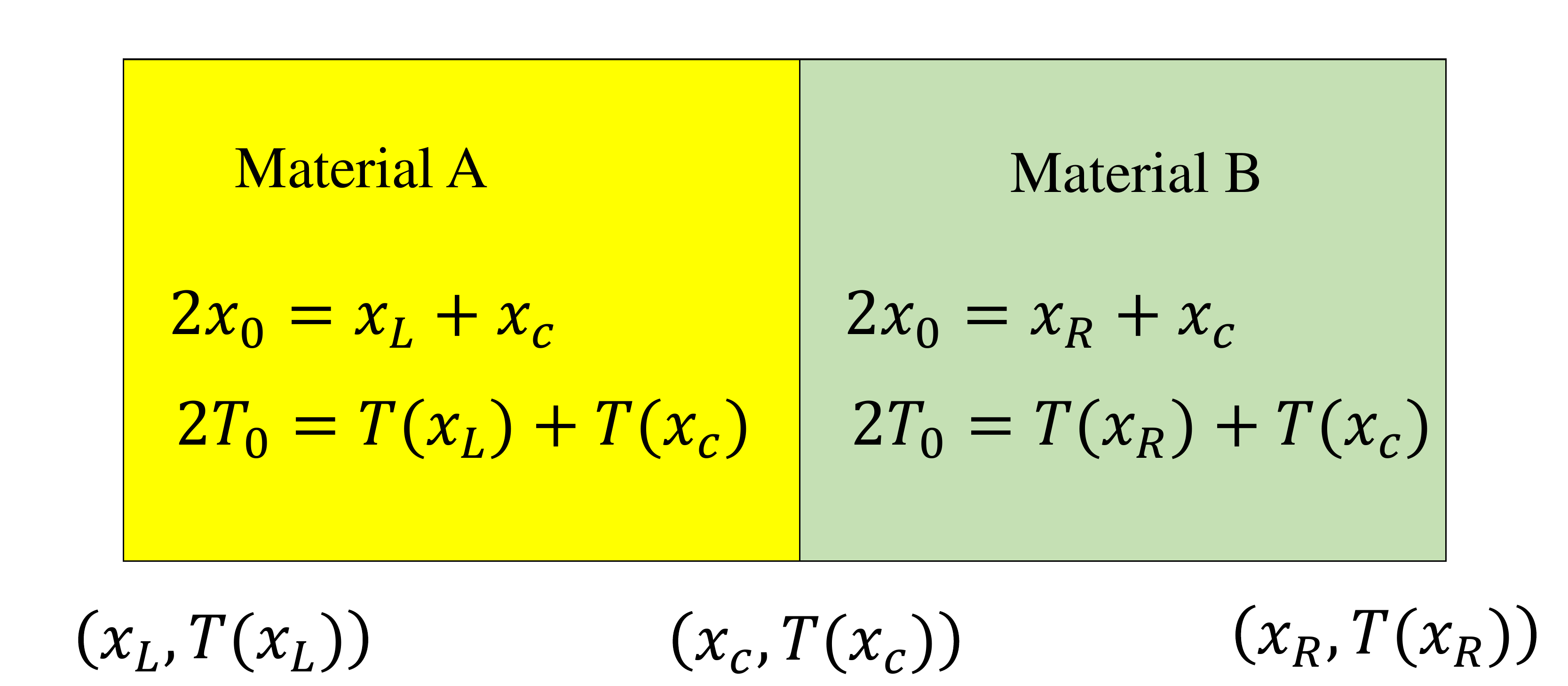}
\caption{The thermal rectification in two-segment bulk materials, where $x_c=(x_L+x_R)/2$ is the position of the interface. }
 \label{schematictwobulk}
\end{figure}
\begin{figure}
 \centering
 \includegraphics[scale=0.45]{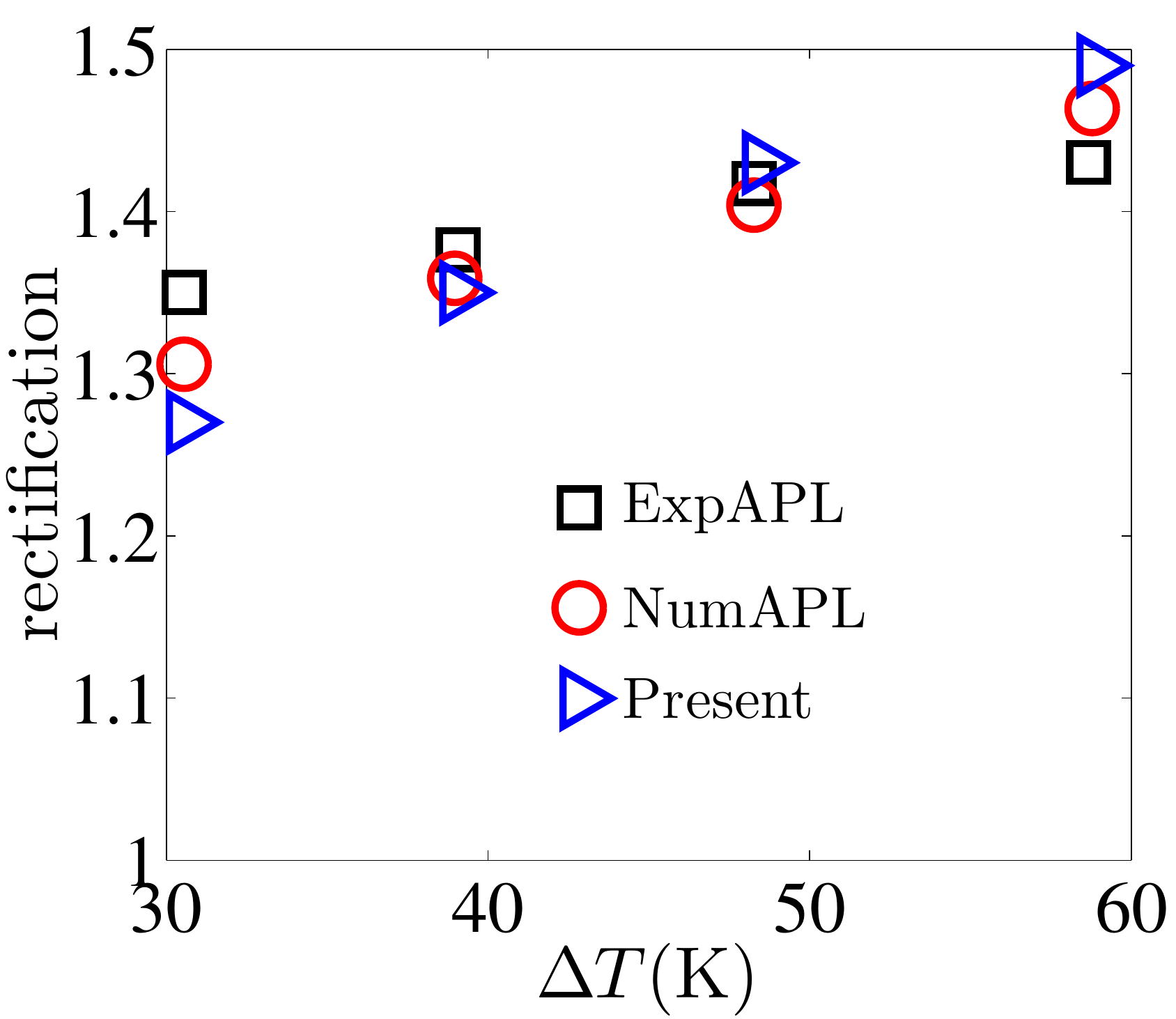}
\caption{A comparison of the thermal rectification ratio among our present numerical results (blue triangle), the measured (black square) and calculated (red circle) shown in Fig. 4(b) in the APL journal~\cite{kobayashi2009}.
 }
 \label{figapl}
\end{figure}
Next, the above equations (Eqs.~\eqref{eq:qzhenglinear},~\eqref{eq:qfulinear}) are used to derive the forward and backward heat flux of a two-segment bulk materials~\cite{dames2009} (\cref{schematictwobulk}).
Here, the thermal conductivity in each segment (labeled as ``A" and ``B") is approximated by a quadratic function of temperature by Taylor expansion.
The forward heat flux is considered first.
For example,
\begin{align}
A: \quad \kappa_e (x,T) &= \kappa_A (T) \approx a_2 T^2+a_1 T +a_0 , &\quad x_L \leq & x < x_c, \\
B: \quad \kappa_e (x,T) &= \kappa_B (T) \approx b_2 T^2+b_1 T +b_0 , &\quad x_c < & x \leq x_R ,
\end{align}
where $a_0,a_1,a_2,b_0,b_1,b_2$ are constants, $x_L,~x_R$ are the spatial position of left and right boundaries with fixed temperatures $T_L,~T_R$, $x_c= (x_L+ x_R)/2$ is the position of the interface between two bulk materials.
$T_c= T(x_c)$ is the temperature at the interface, which is unknown.
There is no temperature jump or thermal resistance at the interface between two bulk materials~\cite{dames2009}.
Note that in each segment, the thermal conductivity is only a quadratic function of temperature, so that $\alpha_x,~\alpha_{xT},~\alpha_{x^2}$ are all zero.

Please note that for each segment, $(x_0,T_0)$, $\kappa_e$, $\Delta T$ and $L$ in Eq.~\eqref{eq:qzhenglinear} are different.
For example (\cref{schematictwobulk}),
\begin{align}
A: \quad x_0= (x_L+x_c)/2, &\quad  T_0=(T(x_L)+T(x_c))/2, \\
B: \quad x_0= (x_R+x_c)/2, &\quad  T_0=(T(x_R)+T(x_c))/2.
\end{align}
For the forward direction, $T(x_L)=T_L,~T(x_R)=T_R,~T_R > T_L$.
Then for each segment of bulk materials, Eq.~\eqref{eq:qzhenglinear} can be used directly, i.e.,
\begin{align}
A: q_{A,+} (T_c) &= -\kappa_A \left(  \frac{T_L + T_c}{2}  \right) \frac{T_c - T_L }{x_c-x_L}  -\frac{T_c - T_L }{x_c-x_L} \frac{a_2  (T_c-T_L)^2  }{24 } , &\quad &x_L \leq  x < x_c, \label{eq:qA} \\
B: q_{B,+} (T_c) &=  -\kappa_B \left(  \frac{T_R + T_c}{2}  \right) \frac{T_c - T_R }{x_c-x_R}  -\frac{T_c - T_R }{x_c-x_R} \frac{b_2  (T_c-T_R)^2  }{24 }, &\quad &x_c <  x \leq x_R , \label{eq:qB}  \\
 q_{A,+} (T_c) &= q_{B,+} (T_c). \label{eq:energys}
\end{align}
Equation.~\eqref{eq:energys} is valid based on the energy conservation, namely, the heat flux across the whole thermal system should be the same.
There is only three unknown variable $T_c$, $q_{A,+}$ and $q_{B,+}$ in above three equations.
So that the temperature at the interface $T_c$ can be calculated as well as heat flux $q_{A,+}$ or $q_{B,+}$.

Similarly, the backward heat flux can be calculated by exchanging $T_L$ and $T_R$ in Eqs.~\eqref{eq:qA},~\eqref{eq:qB} and~\eqref{eq:energys}.
In other words, the thermal rectification can be derived~\cite{dames2009} using our method.
Similar work of thermal rectification in two-segment bulk materials have been done in previous studies~\cite{dames2009,peyrard2006,PhysRevE.98.042131}.

Here is our results for the thermal rectification in the APL paper: An oxide thermal rectifier~\cite{kobayashi2009}.
We set $x_L=0,~x_R=12.4$mm, $x_c=6.3$mm, $T(x_L)=40$K.
The thermal conductivity for each segment in $30-100$K (Fig. 3(a) in Ref~\cite{kobayashi2009}) is approximated as
\begin{align}
&LaCoO_3: &\quad \kappa &= 7.043 \times 10^{-4} T^2 -0.1366 T+8.2058, \\
&La_{0.7}Sr_{0.3}CoO_3: &\quad \kappa &= -0.9278 \times 10^{-4} T^2 +0.03010 T+ 0.07946.
\end{align}
As $T(x_R)=70.6$, $79.1$, $88.5$, $98.9$K, the thermal rectification ratio defined in Ref~\cite{kobayashi2009} is $1.27,~1.35,~1.43,~1.49$, respectively.
A comparison is also made, as shown in~\cref{figapl}.
Present results are in consistent with the results shown in Fig. 4(b) in the paper~\cite{kobayashi2009}.

\bibliography{phonon}